\documentclass[useAMS,usenatbib]{mn2e}
\usepackage{graphicx}
\usepackage{natbib}
\usepackage{times}
\usepackage{float}
\usepackage{epsfig}
\newcommand{\bmth}[1]{\mbox{\boldmath${#1}$}}

\title[Dynamic tides in rotating  stars]{
A unified normal mode approach to dynamic tides and its application to rotating Sun-like stars}
\author[P. B. Ivanov, J. C. B. Papaloizou, S. V. Chernov]{P.B.Ivanov$^{1}$\thanks{E-mail:
pbi20@cam.ac.uk (PBI)}, J. C. B. Papaloizou $^{2}$\thanks{E-mail:
J.C.B.Papaloizou@damtp.cam.ac.uk (JCBP)}  and S. V. Chernov
$^{1}$\thanks{E-mail: chernov@td.lpi.ru (SVCh)}\\
$^{1}$Astro Space Centre, P.N. Lebedev Physical Institute, 84/32
Profsoyuznaya Street, Moscow, 117997, Russia  \\
$^{2}$ DAMTP, Centre for Mathematical Sciences, University of
Cambridge, Wilberforce Road, Cambridge CB3 0WA }

\begin{document}

\date{Accepted. Received; in original form}

\pagerange{\pageref{firstpage}--\pageref{lastpage}} \pubyear{2010}

\maketitle

\label{firstpage}

\begin{abstract}
We determine the response of a uniformly rotating star to tidal perturbations
due to a companion.
General  periodic orbits and parabolic flybys are considered.
We evaluate
energy and angular momentum exchange rates
 as  a sum of contributions from normal modes allowing for dissipative processes.
We consider the case when the response is dominated by the
contribution of an  identifiable regular spectrum of low frequency
modes, such as rotationally modified gravity modes. We evaluate
this response in the limit of very weak dissipation, where
individual resonances can be significant and also when dissipative
effects are strong enough to prevent wave reflection from the
neighbourhood of either the stellar surface or stellar centre,
making radiation conditions more appropriate. The former situation
may apply to Sun-like stars with radiative cores and convective
envelopes and the latter to  more massive stars with convective
cores and radiative envelopes.  We provide general expressions
for transfer of energy and angular momentum  that can be applied to an
orbit with any eccentricity.

Detailed calculations require knowledge of the mode spectrum and
evaluation of the mode overlap integrals that measure the strength
of the tidal interaction. These are evaluated for Sun-like stars
in the slow rotation regime where centrifugal distortion is
neglected in the equilibrium and the traditional approximation is
made for the normal modes. We use both  a WKBJ procedure and
direct numerical evaluation which are found to be in good
agreement for regimes of interest. The former is used to provide
expressions for the mode spectrum and overlap integrals as a
function of mode frequency and stellar rotation rate. These can be
used to find the tidal energy and angular momentum exchange rates
and hence the orbital evolution.

Finally we use  our formalism to  determine the  evolution time
scales for an  object,  in an  orbit of small eccentricity, around
a  Sun-like star in which the tidal response is assumed to occur.
Systems with either  no rotation or synchronous rotation are
considered. Only rotationally modified gravity modes are
taken  into account  under the assumption that wave dissipation occurs close
to the stellar centre.  It is noted that  inertial waves
excited in the convective envelope may produce a comparable amount of tidal
dissipation in the latter case for sufficiently large orbital periods.

\end{abstract}


\section{Introduction}

The tidal interaction between   two bodies
is an important problem in
astrophysics.  It plays a role in close binary stars
leading to synchronisation and
orbital circularisation  (eg. Zahn 1977, Hut 1981),
as well as  stars undergoing a close encounter that may lead to
a tidal capture (eg. Press \& Teukolsky 1977).
In recent years the discovery of many extrasolar planets
orbiting in close proximity to their central stars
has highlighted another situation where tidal interactions
may play an important role in the formation and evolution
of the systems (eg. Barker \& Ogilvie 2009).
In particular hot Jupiters may be formed by a tidal capture into
a highly eccentric orbit followed  by orbital circularisation
 (Weidenschilling \&  Marzari 1996,
Rasio \& Ford 1996, Nagasawa, Ida $\&$ Bessho 2008). During this
process tidal dissipation both in the planet and the star can be
significant (Ivanov \& Papaloizou 2004, 2008, 2011).

The general problem of determining the tidal evolution of a system
involves finding the response of the  rotating stars and/or
planets  due  to the gravitational perturbation of a
companion. This results in the excitation of normal modes and the
exchange of energy and angular momentum with the orbit leading to
its evolution. This may be technically challenging  on account of
the normal mode spectrum being dense and/or continuous (Papaloizou
\& Pringle 1982). Critical latitude phenomena and wave attractors
may also be involved in some cases (Ogilvie \& Lin 2004).
Multidimensional tidal response calculations need to be undertaken
for bodies in a variety of evolutionary states and rates of
rotation. Accordingly it is useful  to  formulate methods for
calculating the tidal interaction that involve the  evaluation of
manageable  analytic expressions  involving  functions of at most
one variable that can be readily tabulated.

In this paper we develop general procedures for expressing tidal
energy and angular momentum exchange rates in  simplified terms
that are applicable to the tidal response that arises from an identifiable
regular spectrum of low frequency modes, comprised of for example  rotationally
modified gravity modes,  for rotating stars with realistic
structure. These are likely to give the dominant tidal
response in bodies with stratification, where the tidal forcing
frequencies significantly exceed the inverse of the convective
time scale associated with any convection zone, making the
associated turbulent viscosity inefficient. The waves associated
with the rotationally modified gravity modes may either
produce a long lasting resonant response through the so-called
'mode locking' mechanism (eg. Witte \& Savonije 1999, 2002) ,
or be dissipated, either near the centre in Sun-like stars with a
radiative core and convective envelope (eg. Barker 2011), or near
the surface in massive stars with a convective core and radiative
envelope (eg. Zahn 1977).

 When decomposition of free stellar pulsations over normal
modes is possible, and the damping mechanism of normal modes is
specified, we give general expressions from which the energy and
angular momentum exchange may be calculated, in principal, for
both bound orbits of arbitrary eccentricity and parabolic
encounters that could lead to tidal captures with an arbitrary
inclination of stellar rotation axis with respect to orbital
plane. These are expressed in terms of contributions from
individual normal modes. In general, they require evaluation of
coefficients characterising the representation of the tidal potential
 as either a Fourier series or a Fourier transform, the spectrum of
normal modes,  normal mode damping rates and
so called overlap integrals that measure the strength of
interaction. It is important to note that in certain regimes of
tidal interaction, such as energy and angular momentum exchange
through  parabolic encounters, or orbital
 evolution in the regime of so-called moderately strong
viscosity (eg. Zahn 1977, Goodman $\&$ Dickson 1998),  the resulting
expressions do not depend on the details of the  mode damping,
 which is poorly
known in many situations, see also the discussion of these regimes
below.

The overlap integrals determine the  coupling of mode eigenfunctions
with the tidal potential, they thus play a central role in our
approach to tidal problem. Note that although we use the
definition of overlap integral first introduced by Press \&
Teukolsky (1977) for non-rotating stars and later generalised by
Ivanov $\&$ Papaloizou (2005, 2007)
 for rigidly rotating stars, they are directly
related to the so-called tidal resonance coefficients first
introduced by Cowling (1941) and studied later in a number of
papers (eg. Zahn 1970, Rocca 1987, Smeyers, Willems $\&$ Van
Hoolst 1998). In order to calculate the overlap integrals one
should be able to evaluate integrals with  highly oscillatory
integrands over the
bulk of the star. This is rather difficult to achieve both
numerically and analytically. In such a situation, numerical and
analytical methods should supplement each other.  In this paper
we give an analytic method for their evaluation,  together with
a calculation of the relevant eigenspectrum and normal modes,
 valid for Sun-like stars having
an inner  radiative region and a convective envelope, that is based
on a WKBJ scheme. In an accompanying paper we also provide values
of overlap integrals for a range of eigenfrequencies in a number
of models of more massive stars of different masses and ages.

It is quite important to stress that  the calculation of the
relevant normal mode spectra for rotating stars is complex,
because unlike the situation for a spherical body, separation of
variables is not in general possible. However, this is rectified
if the well known traditional approximation is used  as is done
below. In the traditional approximation only the radial
component of the angular velocity is retained (Unno et al. 1989).
This expected to be a good approximation for rotationally modified
gravity modes for which motions are mainly horizontal. It has
accordingly been adopted by many authors. For example, it was
used in an asymptotic treatment of the  normal mode problem by
Berthomieu et al. (1978) and in direct tidal response calculations
(Papaloizou \& Savonije 1997, Witte \& Savonije 1999, 2002 and
Ogilvie \& Lin 2004). However,  it is worth mentioning that
 the existence of regular spectra of rotationally modified gravity
modes is not dependent on the use of the traditional
approximation. Such spectra are expected to be associated with the
radiative cores of Sun-like stars and the radiative envelopes of
more massive stars. A spectrum of rotationally modified gravity
modes  in the case of a massive slowly rotating star  was
identified by Savonije et al. (1995). In this case  the
traditional approximation was found to be a good one as long as
their eigenfrequencies placed them outside the inertial regime (
the absolute value of the eigenfrequency exceeded twice the
stellar angular velocity). Similarly Dintrans \& Rieutord (2000)
found a rotationally modified gravity mode spectrum outside the
inertial regime for a $1.5M_{\oplus}$  star with a convective
envelope without making use of the traditional approximation. Note
too that the order of the $l=2$  $g$ mode with eigenfrequency
corresponding to the critical break up rotation  frequency
for a Sun-like star is typically $\sim10$. Accordingly a low
frequency asymptotic analysis should be useful for stars in a
state of modest rotation, at less than about one fifth of the
critical rate, and excited modes with comparable eigenfrequencies.

 Finally,  we demonstrate how our approach can be applied to
the calculation of the orbital circularisation and decay time
scales for objects orbiting Sun-like stars  under the  assumption
of wave dissipation near the centre and small orbital
eccentricity.


 The plan of this study is similar to that of our previous study of
inertial waves in rotating bodies undertaken in Ivanov $\&$ Papaloizou (2011) and
Papaloizou $\&$ Ivanov (2011). In this paper
we mainly concentrate on an analytic approach to the
problem while in an accompanying paper we consider the application of
 numerical methods, extending our calculations to evaluate the tidal response of
stellar models appropriate to stars more massive than the sun.

In section \ref{Basiceq} we give the basic equations governing the
tidal response of a rotating star to tidal forcing. We then use
them to calculate the linear response to harmonic forcing for a
particular azimuthal mode number in section \ref{tidalresp}. To do
this we apply an extension of the method of Ivanov $\&$ Papaloizou
(2007), hereafter IP7,  which provides a solution in terms of a
decomposition in terms of normal modes,  that allows us to
incorporate dissipative forces. Specifically we consider viscous
forces for which the viscosity coefficient may vary with location
but the approach may be generalised to deal with more general
forms of dissipation.

We focus on the case where there is a regular spectrum of normal
modes that can be dense, such as a low frequency gravity mode
spectrum and consider its response. In particular we calculate the
rate of transfer of energy and angular momentum between the star
and the orbit, expressing the results in terms of overlap
integrals that measure  the contribution associated with each
normal mode in terms of the projection of  a related eigenfunction
onto the tidal potential.  We then customise  the expressions for
the tidally induced energy and angular momentum exchange rates to
apply to the case where the response is determined by a dense
regular set of  normal modes, such that  the mode  decay rates are
small compared to their natural frequencies,  but not small
compared to the separation between adjacent  mode frequencies.

As we demonstrate  in an appendix for  slow rotation under the
traditional approximation,  this case  is  expected to  correspond
to the situation where waves associated with the modes propagate
to localised boundary regions, which may be either close to the
stellar centre or surface,  where they are dissipated,  and we
call it the moderately viscous case. This is in contrast to the
situation where dissipation is very weak, and the interaction is
dominated by strong normal mode resonances that are effective only
in very narrow frequency intervals (eg. Savonije  et al.  1995,
Witte \& Savonije  1999, 2002). The expressions we obtain can be
used to determine the orbital evolution of a binary system,  in a
general periodic orbit which may be near circular, or undergoing
encounters in a near  parabolic orbit,
 induced by  tides.  Knowledge of the  adiabatic normal mode spectrum
and the  overlap integrals corresponding to each mode,  in the low frequency regime,
is required for their evaluation in the moderately viscous case, the results then being independent of the
details of the dissipation process.

In section \ref{modgm} we go on to discuss  the linear adiabatic
normal mode problem and the low frequency spectrum provided by
rotationally modified gravity modes. We consider the  low
frequency asymptotic regime  under  the traditional approximation
in which only the radial component of the angular velocity is
retained. When this approximation is made, separation of variables
is possible. Its efficacy  in the stellar case was first
implicitly realised  by Berthomieu et al. (1978), through the
application of an asymptotic procedure,  that effectively  led to
it. We develop expressions for the overlap integrals in the low
frequency limit
 that can be evaluated as  one dimensional integrals
involving the mode angular and radial eigenfunctions.

In section \ref{Numcalc} we perform  direct numerical calculations
of mode eigenfrequencies,  eigenfunctions and their associated
overlap integrals. Rotationally modified  gravity modes under the
traditional  approximation  are found for  models of solar-type
stars.  Models  representing both the present day sun and a young
sun with age $1.66\times 10^8 yr.$ are considered. We also give
power fits to the dependence of the square of the  Brunt - V$\ddot
{\rm a}$is$\ddot {\rm a}$l$\ddot {\rm a}$ frequency inside the
radiative core,  on the distance to the boundary between the
interior radiative zone and the convective envelope (see also
Barker 2011). This dependence was found to have a significant
effect on properties of the normal modes.

In section  \ref{WKBJcalc}
we  calculate  eigenfrequencies, eigenfunctions and  overlap integrals
for   high order adiabatic rotationally modified  gravity modes  for models of
solar-type stars under the traditional approximation following a WKBJ approach that incorporates some corrections
to the formalism of Berthomieu et al. (1978).
We also allow for an arbitrary power law dependence of the Brunt - V$\ddot {\rm a}$is$\ddot {\rm a}$ll$\ddot {\rm a}$
 frequency in the radiative interior  on the distance to the
convective envelope boundary.
We provide expressions for both eigenfunctions and eigenvalues which are found to
provide a good approximation to  the corresponding  numerically determined quantities
for the stellar models considered.

We use these solutions to evaluate the overlap integrals, providing
analytic expressions for them, which together with some quantities represented graphically,
readily enable their evaluation as a function of mode frequency and rotation rate of the star.
These are in very good agreement with the numerically determined quantities
for mode frequencies less than the critical rotation frequency.

Knowledge  of the overlap integrals and mode spectrum enables
calculation of tidal energy and angular momentum exchange rates
with a general orbit. In section \ref{Tidalapp} we  apply  our
formalism  to  the calculation of the tidal evolution timescales
for a  binary with a small eccentricity and with aligned orbital
and spin angular momenta. One of the components is assumed to be a
Sun-like star in which modified gravity modes are excited and
propagate to the centre where they are assumed to be dissipated
corresponding to the moderately large viscosity regime.  The other
component is treated as a point mass and a large range of mass
ratios is considered. We give estimates for the circularisation
time scale when the star is non rotating and when it  rotates
 synchronously with the orbit. In the former case we give the time scale for the semi-major axis
to decrease.

Finally in section \ref{Discuss} we summarize and discuss our results.

\section{Basic equations}\label{Basiceq}

\subsection{Coordinate system and notation}

The basic definitions  and notation adopted in this paper closely
follow IP7. We use
either cylindrical coordinate system $(\varpi, \phi, z)$ or
spherical $(r, \phi, \theta)$ coordinate systems  with origin
at the centre of mass of the star.   When viewed in an inertial
frame with this origin, the unperturbed star is assumed to  rotate
uniformly about the $z$ axis with angular velocity $\Omega.$
Accordingly we adopt the rotating frame in which the unperturbed
star appears at rest as our reference frame.

In this paper the fundamental quantity describing perturbations
of a rotating star will be the Lagrangian displacement vector
${\bmth{ \xi}}$. By definition, this is a real quantity. We represent the
displacement vector as well as any other perturbation quantity in terms of  a
 Fourier series in the azimuthal angle $\phi,$  and
either in terms of a  Fourier series in the time,  $t,$ or a
Fourier integral with respect to  $t,$ depending  respectively on
whether we consider tidal interactions in a binary having a
periodic orbit,  or the orbit is assumed to be  parabolic.

When a discrete Fourier series in   time is appropriate,
for a real quantity, $Q,$  we write
\begin{equation}
Q = \sum_{m, k} \left( Q_{m,k}\exp({-{\rm i}\omega_{m,k} t+im\phi}) + cc
\right ), \label{eq p1}
\end{equation} where $cc$ denotes the complex conjugate,
$\omega_{m,k}=k\Omega_{orb}-m\Omega$,  $\Omega_{orb}$ is the
orbital frequency,  $\Omega $ is  the angular velocity of the star
and $m$ and  $k$ are integers, with only positive values of $m$
included in the summation as in IP7. Note when the orbital plane
is perpendicular to the rotational axis of the star, and only the
quadrupole component of the tidal potential, that is dominant at
large separations  is included, terms with $m=1$ are absent  in
the summation.  It then suffices to consider $m=0, 2$.  In
addition, the  reality of $Q$  implies that we should identify
$Q_{-m,-k} = Q_{m,k}^{*}.$
When forming the sum (\ref{eq p1}), we ensure that terms are not repeated.
For example, when $m=0,$ terms for which $\omega_{m,k}$ changes sign are identical,
they are accordingly combined. After this is done, forcing  for only one sign of $\omega_{m,k}$ has to be considered  .

 For the case involving a  Fourier
transform, we write
\begin{equation}
Q = \sum_{m}\left(  \exp({{\rm i}m\phi})\int^{+\infty}_{-\infty}d\sigma
\tilde Q_{m}\exp({-{\rm i}\sigma t}) + cc \right ), \label{eq p1n}
\end{equation}
where the  Fourier transform is indicated by a tilde. This
requirement   that  $Q$ is real leads to the identification
$\tilde Q_{-m}(-\sigma) = \tilde Q_{m}^*(\sigma).$ The inner
products of two complex scalars $Y_{1}$, $Y_{2}$, and two complex
vectors ${\bmth{ \eta}_{1}}$ and ${\bmth{ \eta}_{2}}$ are
respectively
 defined through
 \begin{eqnarray}
(Y_{1}|Y_{2})&=&\int \varpi d\varpi dz\rho (Y_{1}^{*}Y_{2}) \hspace{1cm} {\rm and}  \\
({\bmth{ \eta}}_{1}|{\bmth{ \eta}}_{2})&=&\int \varpi d\varpi dz\rho
({\bmth{ \eta}}_{1}^{*}\cdot {\bmth{ \eta}}_{2}), \label{eq p2}
\end{eqnarray}
where $({\bmth{ \eta_1}}^{*}\cdot {\bmth{ \eta_2}})$ is the scalar
product, $\rho$ is the density, and $*$ stands for the complex
conjugate.

\subsection{Equations of motion }

In the rotating frame the linearised Navier-Stokes equations take
the form (see IP7)
\begin{equation}
\ddot {\bmth{ \xi}}+2{\bmth {\Omega}}\times \dot {\bmth
{\xi}}+{\bmth {{\cal C}}}{\bmth \xi} =-\nabla \Psi-{\bmth {f}}_{\nu},
\label{eq p3}
\end{equation}
where the dot indicates partial differentiation  with respect to  time and
${\bmth {\cal C}}$ is an integro-differential operator describing the action
of gravity and pressure forces.  For conservative boundary conditions,
it is self-adjoint when the normal product (\ref{eq p2}) is adopted,
 and it can be considered to be a positive operator
 under certain conditions, which are assumed to be
fulfilled in this paper. Its general explicit form is not important for
our purposes and can be found in e.g. Lynden-Bell $\&$ Ostriker
1967.

Retaining only the  assumed dominant  quadrupole component,
the tidal potential,  $\Psi $  can be represented in the
form
\begin{equation}
\Psi=r^{2}\sum_{m,k} \left( A_{m,k}e^{-i\omega_{m,k}t} Y^{m}_{2} +
cc \right) \label{eq4}
\end{equation}
where $Y^{m}_{2}$  are the spherical functions. The coefficients
$A_{m,k}$ are given in Appendix A for the case of coplanar
 orbit with small eccentricity. For orbits having an arbitrary
 value for the  eccentricity,  these coefficients can be expressed in
 terms of so-called Hansen coefficients discussed eg. in Witte
 $\&$ Savonije (1999).
However,  note that the representation of the tidal potential in
terms of  a Fourier series (\ref{eq4}) is not convenient for a
highly eccentric orbit. In that case  it is more appropriate to
 adopt the tidal potential for a parabolic orbit, and its Fourier
transform. The corresponding expressions can be found in e.g.
Ivanov $\&$ Papaloizou (2011).  These authors  as well as Lai
$\&$ Wu 2006  also give expressions, which can be used to
represent the tidal potential in a coordinate system that is
rotated with respect to the one for which the $z$ axis is normal
to  the orbital plane. These expressions  can be used when  the rotation axis
of the star is not perpendicular to the orbital plane.

The viscous force ${\bmth {f}}_{\nu}={\bmth \sigma} \dot{\bmth{
\xi}}$, where the operator ${\bmth \sigma}$ is defined through its
action on a vector ${\bmth{ \eta}}$ as
\begin{equation}
({\bmth \sigma} {\bmth{ \eta}})_{\alpha} = -{1\over \rho}( \rho
\nu \sigma_{\alpha \beta })_{,\beta}, \quad
\sigma_{\alpha\beta}=\eta_{\alpha,\beta}+ \eta_{\beta,\alpha}-
{2\over 3} \eta_{\gamma,\gamma}\delta_{\alpha \beta}
 \label{eq5}
\end{equation}
where $\nu$ is kinematic viscosity,  a comma stands for
differentiation over Cartesian coordinates and summation over
repeating indices is implied. Note that we reserve the indices
$\alpha $, $\beta $ and $\gamma $ to indicate  components of
vectors and tensors  in order to distinguish them from
indices  used in  the summation of  series.  When these indices
are repeated,  summation is implied from now on (the summation convention).

We remark that we have considered a standard form of viscous dissipation
with the possibility that the viscosity coefficient depends on position.
This leads to some simplifying features in the discussion of dissipation
below on account of the viscous stress tensor being symmetric.
However, a corresponding  analysis can also be undertaken for more complicated
forms of dissipation,  including radiative damping, by introducing relevant adjoint operators.
However, as the final results we use do not depend on the details of the dissipation
beyond involving decay rates associated with  individual normal modes
we shall work only  with the standard formulation for viscous dissipation  described above.

\section{The response to  tidal forcing  due to  companions  in periodic
and parabolic orbits }\label{tidalresp}
In this section we calculate the tidal response to  tidal forcing
associated with a general periodic orbit.
Later we extend the discussion to the limiting case  where the
orbit becomes parabolic (see IP7).
We begin by writing the forcing potential  (\ref{eq4}) as a Fourier series of the form
(\ref{eq p1}).  The Fourier coefficients,  $\psi_{m,k},$ are  such that
\begin{equation}
(\nabla \Psi)_{m,k}={A_{m,k}\over 2\pi}\int_{0}^{2\pi} d\phi
e^{-im\phi}\nabla (r^{2}Y^{m}_{2}) \label{eq9n},
\end{equation}
where  $Y^m_l$ indicates the usual spherical harmonic with indices $m$ and $l.$

As was shown by Papaloizou $\&$ Ivanov 2005 and IP7, after
taking the Fourier transform  of equation (\ref{eq p3}),
one can  obtain equations for determining
 the  Fourier coefficients of the Lagrangian displacement
 ${ {\bmth{ \xi}}_{m,k}}.$
 These can be organised
into a  convenient form as described below.
Once determined for periodic orbits,  the ${ {\bmth{ \xi}}_{m,k}}$
can then  be  used to calculate the orbital energy and angular momentum
exchange rates.

\subsection{Determination of the Fourier components of the Lagrangian displacement}
We begin by introducing  a six-dimensional vector
$\vec Z_{m,k} \equiv  ({\bmth {Z}}_{1},{\bmth {Z}}_{2})^{T}$ where
\begin{equation}
{\bmth {Z}}_{1}=\omega_{m,k} { {\bmth {\xi}}_{m,k}} , \quad {\rm and}\quad {\bmth
{Z}}_{2}={\bmth {\cal C}}^{1/2}{ {\bmth{ \xi}}_{m,k}}. \label{eq6}
\end{equation}
Equation $(\ref{eq p3})$ then follows from the pair of equations derived from
\begin{equation}
\omega_{m,k} \vec Z_{m,k} ={\bmth {\cal H}} \vec Z_{m,k}
-i\omega_{m,k}{\bmth{ \cal{D}}}\vec Z_{m,k}+\vec S, \label{eq7}
\end{equation}
where the operator ${\bmth {\cal H}}$ has a matrix structure
\begin{equation}
{\bmth {\cal H}}= \left( \begin{array}{cc} {\bmth {\cal B}} &
{\bmth {\cal C}}^{1/2} \\
{\bmth {\cal C}}^{1/2} & 0 \end{array}\right), \label{eq8}
\end{equation}
with  the operator ${\bmth {\cal C }}^{1/2}$ being  defined through the
condition: ${\bmth {\cal C }}={\bmth {\cal C }}^{1/2}{\bmth {\cal
C }}^{1/2},$  with any ambiguity removed by requirement that it is a positive operator
( see eg. Wouk 1966). In addition
the operator \begin{equation}
{\bmth {\cal B}} {\bmth {\xi}} =-2i{\bmth {\Omega}}\times
{\bmth{\xi}}. \label{eq8n}
\end{equation}
 The operator $\bmath {\cal D}$ is defined through
\begin{equation}
 {\bmath{\cal D}}\vec Z_{m,k}=({\bmth{f_{\nu}}}_{1,k}, {\bmth{f_{\nu}}}_{2,k})^{T}, \label{eq9mm}
  \end{equation}
  where
\begin{equation}
{\bmth{f_{\nu}}}_{1,k}= \frac{1}{\omega_{m,k}} {\bmth { \sigma}} {\bmth {Z}}_{1} \hspace{2mm} {\rm and}
\quad {\bmth{f_{\nu}}}_{2,k}= 0. \label{eq9nn}
\end{equation}

As  we have performed a Fourier decomposition in terms of $m,$
the action of the operator   $ {\bmth {\cal O}}$ being any one of
  ${\bmth {\cal B}} , {\bmth {\cal C}} $ or $\bmth { \sigma} $
on   the Fourier component $ {\bmth {\xi}}_{m,k}$ is obtained through
\begin{equation}
{\bmth {\cal O}} {\bmth {\xi}}_{m,k}=   \exp(-{\rm i}m\phi) {\bmth {\cal O}} ({\bmth {\xi}}_{m,k}\exp({\rm i}m\phi)).
\label{opspec}
\end{equation}
It readily follows  that $\bmth{\cal B}$ is self-adjoint as are  $\bmath {\cal C}$ and $\bmath {\cal C}^{1/2}$
(IP7) .
From the definition (\ref{eq5}) it  also follows that  ${\bmth { \sigma}}$
is self-adjoint.
The  six dimensional source vector $\vec S \equiv ({\bmth{S}}_{1},{\bmth{S}}_{2})^{T} $
 where
\begin{equation}
{\bmth{S}}_{1}=(\nabla \Psi)_{m,k} \hspace{2mm} {\rm and} \quad {\bmth{S}}_{2}=0.
\label{eq9}
\end{equation}

Equation (\ref{eq7}) can be used to define an eigenvalue problem
in the form
 \begin{equation}
\lambda_{j} \vec Z^{j} =({\bmth{\cal H}} -i\omega_{m,k}{\bmth {\cal{D}}})\vec Z^{j}.  \label{eq10}
\end{equation}
Here $\lambda_j$ is the generally complex eigenvalue and
the corresponding  eigenfunction is $\vec Z^{j}= ({\bmth {Z}}_{1j}, {\bmth{Z}}_{2j})^{T},$
the components being related to the corresponding displacement by
\begin{equation}
{\bmth {Z}}_{1j}=\lambda_j { {\bmth {\xi}}_j} , \quad {\rm and}\quad {\bmth
{Z}}_{2j}={\bmth {\cal C}}^{1/2}{ {\bmth{ \xi}}_j}. \label{eq6a}
\end{equation}
Adopting  the inner product defined through
\begin{equation}
<\vec X| \vec Z>=
\int \varpi d\varpi dz\rho ({\bmth {X}}_{1}^*,{\bmth {X}}_{2}^*)({\bmth {Z}}_{1},{\bmth {Z}}_{2})^{T}
, \label{eq11a}
\end{equation}
it follows
from equations  (\ref{eq8}) and (\ref{eq9mm})   that the operators ${\bmth {\cal H}}$
and  ${\bmth {\cal D}}$ are self-adjoint.
Hence
it follows that the operator
${\bmth{\cal H}} +i\omega_{m,k}{\bmth{ \cal{D}}}$
is the adjoint  operator associated with the eigenvalue problem  (\ref{eq10}).   The eigenvalues
will accordingly be  eigenvalues $\lambda_j^*$
with  corresponding eigenfunctions  $\vec X^{j}.$

In addition the eigenfunctions $\vec Z^{j}$  and $\vec X^{j}$ are biorthogonal
with respect to the adopted  inner product so that
\begin{equation}
<\vec X^{j}| \vec Z^{l}>=
  \lambda_{j}^*\lambda_{l}({\bmth {\eta}}_{j}| {\bmth {\xi}}_{l})+
({\bmth {\eta}}_{j}|{\bmth {\cal C}}{\bmth
{\xi}}_{l})=\delta_{jl}N_{j}, \label{eq11}
\end{equation}
where   $\bmth{\eta}_j$  denotes the   Lagrangian displacement
 associated with the eigenfunctions of  the adjoint problem
 defined through (\ref{eq6a}) with ${\bf Z}$ being replaced by ${\bf X}.$
The norm is
\begin{equation}
N_{j}=<\vec X^{j}| \vec Z^{j}>=|\lambda_j|^2({\bmth {\eta}}_{j}|
{\bmth {\xi}}_{j})+({\bmth {\eta}}_{j}
|{\bmth {\cal C}}{\bmth{\xi}}_{j}). \label{eq12}
\end{equation}
We remark that although we have not assumed the viscosity
is small up to now,  when  it is neglected  $\bmth{\xi}_j $ and $\bmth{\eta}_j$
coincide.
\subsection{Solution for  the forced response}
We look for solution of (\ref{eq7}) as a decomposition over the
eigenvectors $\vec Z^{j}$:
\begin{equation}
\vec Z_{m,k}=\sum_{j} \alpha_{j} \vec Z^{j}. \label{eq13}
\end{equation}
Substituting (\ref{eq13}) in (\ref{eq7}), taking the inner product
of the result with the eigenvector $\vec X^l$ and using (\ref{eq10}) and
(\ref{eq11}) we get
\begin{equation}
\alpha_{l}(\omega_{m,k}-\lambda_{l})N_{l}
=<X^l|S>. \label{eq14}
\end{equation}


To proceed further we remark that  an expression for the damping  rate
of the mode with eigenvalue $\lambda_j$  can be
found  by taking the inner product of
(\ref{eq10}) with $\vec Z^{j} .$  Using the fact that ${\cal H}$ is self-adjoint,
we find that
 \begin{equation}
{\cal { I}}m (\lambda_{j}) <\vec Z^{j} | \vec Z^{j}> = -\omega_{m,k} <\vec Z^{j} | {\bmth {\cal{D}}} \vec Z^{j} >,  \label{eq10a}
\end{equation}
where ${\cal { I}}m$ denotes the imaginary part.
Setting ${\cal { I}}m (\lambda_j ) = - \omega_{\nu,kj} $
and making use of (\ref{eq6a}), (\ref {eq9mm}) and (\ref{eq5}) we obtain
\begin{equation}
\omega_{\nu,kj}={|\lambda_{j}|^2\over 2{\cal N}_{j}}(\nu
\sigma^{j}_{\alpha\beta}| \sigma^{j}_{\alpha\beta}),  \label{eq18}
\end{equation}
where the superscript $j$ indicates that
$\sigma^{j}_{\alpha\beta}$ is  evaluated assuming that
$\bmath{\eta}$  as used in equation (\ref{eq5})
    is replaced by $ {\bmth {\xi}}_{j}$ and
   \begin{equation}
{\cal{N}}_{j}=|\lambda_{j}|^{2}({\bmth {\xi}}_{j}|
{\bmth {\xi}}_{j})+({\bmth {\xi}}_{j}
|{\bmth {\cal C}}{\bmth{\xi}}_{j}).\label{norm}
\end{equation}
 Using the above we write
 $\lambda_j =\omega_j-i \omega_{\nu,kj},$
 where $\omega_j$ is the real part of $\lambda_j.$
We remark that for convenience in the above discussion we have suppressed the index
$k$ for $\lambda_j,$ $\omega_j$ and $ {\bmth {\xi}}_{j}.$
Then from equation (\ref{eq14})  we find
\begin{equation}
\alpha_{j}={\lambda_{j} S_{j,k}\over
N_{j}(\omega_{m,k}-\omega_{j}+i\omega_{\nu,kj})}, \label{eq16}
\end{equation}
where
\begin{equation}
S_{j,k}=( {\bmth {\eta}}_{j}| \nabla \Psi_{m,k}),\label{eq17}
\end{equation}
 Now we use equation (\ref{eq13}) to obtain
\begin{equation}
{\bmth {\xi}}_{m,k}=\sum_j {\lambda_{j}^2 S_{j,k}\over \omega_{m,k} N_j
(\omega_{m,k}-\omega_{j}+i\omega_{\nu,kj})} {\bmth
{\xi}}_{j}. \label{eq19}
\end{equation}

\subsection{Energy transfer rate due to tidal forcing}
The canonical energy associated with the perturbations is given by
a standard expression
\begin{eqnarray}
E_{c}&=&\int_{V}d^{3}x \rho {1\over 2}(|{\dot {\bmth {\xi}}}|^{2}+
{\bmth {\xi}}\cdot {\bmth {\cal C}}{\bmth {\xi}})\hspace{2mm} {\rm  which\hspace{1mm}
 leads\hspace{1mm}  to\hspace{1mm}  }\nonumber \\
\langle E_{c}\rangle&= &2\pi \sum_{m,k}
\lbrace  \omega_{m,k}^2({\bmth {\xi}}_{m,k}|  {\bmth {\xi}}_{m,k})+ ({\bmth
{\xi}}_{m,k}| {\bmth {\cal C}}{\bmth {\xi}}_{m,k})\rbrace,
\label{eq20}
\end{eqnarray}
where we integrate over the volume of the star.  Recall that
${\bmth {\xi}}$ is real.  The last expression, being a time average,  displays
explicitly  contributions corresponding to different azimuthal
numbers $m$ and forcing frequencies $\omega_{m,k}.$ The
contribution of an equal  response to the complex conjugate of the
forcing potential in (\ref{eq4}) has also been included.  The rate of
change of $E_{c}$ with time  as a result of the action of
dissipative forces follows from (\ref{eq p3}) and (\ref{eq20}) as
\begin{equation}
\dot E_c=-\int d^3x\rho (\dot {\bmth {\xi}}\cdot {\bmth
{f}_{\nu}})=-{1\over 2} \int d^3x\rho \nu [\dot \sigma_{\alpha
\beta} \dot \sigma_{\alpha \beta}], \label{eq21}
\end{equation}
where we use (\ref{eq5})   with ${\bmth {\eta}}$ replaced by  ${\bmth {\xi}}$ and integrate by parts.
We remark that when, as here a sum of steady periodic responses
is excited by a forcing potential, the rate of energy loss resulting from dissipative forces
is balanced by input from the forcing potential.
We now go on to obtain an  expression for $\dot E_c$ in the form of a
 sum of contributions corresponding to different
$m$ and $\omega_{m,k}.$

To do this we substitute (\ref{eq19})
in the appropriate
form of
(\ref{eq p1}) and  the result
 into  (\ref{eq21}).  An average of the resulting expression over a time period
 that is large in comparison to the periods of eigenmodes
that contribute significantly to the sum is then taken.
 We  thus obtain
\begin{equation}
\dot E_c=-\pi \sum_{m,k,j,j^{'}}\left(
{\lambda_j^{2}\lambda_{j'}^{*2}S_{j,k}S^{*}_{j^{'},k}\over
N_{j}N^{*}_{j^{'}}D_{k,j,j^{'}}} (\nu \sigma_{\alpha \beta
}^{j^{'}}|\sigma_{\alpha \beta }^{j})+cc \right), \label{eq22}
\end{equation}
where
\begin{equation}
D_{k,j,j^{'}}=(\omega_{m,k}-\omega_{j}+i\omega_{\nu,kj})
(\omega_{m,k}-\omega_{j^{'}}-i\omega_{\nu,kj^{'}}).
\label{eq23}
\end{equation}

When the mode energy does not grow with time on average, and,
accordingly, the stationarity condition is implied as in the bulk of the
paper\footnote{We briefly  consider the situation when the stationarity
 condition may be invalid in the end of Discussion.}
the rate of energy dissipation by the action of viscosity must
be negative of the work done by tidal forces, $\dot E_{c}=- \dot
E_{T}$, where $\dot E_{T}=-2\pi(\dot {\bmth {\xi}} | \nabla \Psi)$, see
equation (\ref{eq p3}) and averaging over a time  larger than a
characteristic time scales associated with the forcing frequencies is
implied.  From equations (\ref{eq p1}), (\ref{eq6}) and (\ref{eq13})
it is easy to see that
\begin{equation}
\dot E_{T}=2\pi i\sum_{m,k,j}\omega_{m,k}(\alpha_{j}((\nabla \Psi)_{m,k}|{\bmth
{\xi}}_{m,k})-cc).
\label{eq23n}
\end{equation}
Note that when viscosity is small the complex conjugate of $({\bmth
{\xi}}_{m,k}|(\nabla \Psi)_{m,k})$ is
approximately given by equation (\ref{eq17}).

\subsubsection{The case of very small viscosity and near normal mode resonances}
The expression (\ref{eq22}) can be further simplified by assuming
that a strong resonance exists in the vicinity of a particular forcing
frequency, $\omega_{m,k},$
 on account of very  weak dissipation.
If  we assume  as a consequence of this that, for a given forcing frequency,  the dominant contribution to the sum comes from
a particular mode that  is nearly resonant,  only  the diagonal terms  in  the sum over $j$ and
$j^{'}$  need to be retained as they provide this  contribution.
But note that the appropriate  damping rates due to viscosity should be sufficiently small,
as indicated  further below.
 Then  we have
\begin{eqnarray}
\dot E_c&=&-2\pi \sum_{m,k,j}\left( {|\lambda_{j}|^4|S_{j,k}|^2\over
|N_{j}|^2D_{k,j,j}}(\nu \sigma_{\alpha \beta }^{j}|\sigma_{\alpha
\beta }^{j}) \right)\nonumber \\
 & =&-4\pi
\sum_{m,k,j}{\omega_{\nu,kj} |\lambda_{j}|^2{\cal {N}}_j}{
|S_{j,k}|^2\over| N_j |^2D_{k,j,j}}, \label{eq24}
\end{eqnarray}
where we have used equation (\ref{eq18}) to obtain the last equality.

\subsubsection{The case of moderately large viscosity}\label{Modvisc}
In order to make the above procedure valid, the effect of
viscosity should be small. But note that  we shall be  interested
in the case when the viscous decay rate of a mode, although being
small compared  to its frequency, may be large compared to the
frequency separation between modes. This situation is less clear
as  many modes could  interact and treatment of viscosity
 by straightforward perturbation theory
is not possible.
However, taking the inner  product of equation (\ref{eq10}) with $\vec Z^{j'}$
and subtracting the complex conjugate of the corresponding equation
with $j$ and $j'$ interchanged, we obtain

\begin{equation}
  (\lambda_{j} - \lambda_{j'}^*)<\vec Z^{j'}| \vec Z^{j}>=
-2{\rm i}\omega_{m,k}<\vec Z^{j'}|{\cal{D}} \vec Z^{j}>
\label{eqee1}
\end{equation}
which leads to the following expression
\begin{equation}
 \frac {\lambda_{j}\lambda_{j'}^*
 (\nu \sigma_{\alpha \beta }^{j'}|\sigma_{\alpha
\beta }^{j})}{(\lambda_{j} - \lambda_{j'}^*)}
={\rm i}\left(\lambda_{j}\lambda_{j'}^*({\bmth {\xi}}_{j'}| {\bmth {\xi}}_{j})+
({\bmth {\xi}}_{j'}|{\bmth {\cal C}}{\bmth
{\xi}}_{j})\right).
\label{eqff1}
\end{equation}

From this we can argue that, when the mode separation frequencies
are less than the characteristic damping rate (assumed to be
comparable for neighbouring modes), the contribution of off
diagonal terms, as compared to diagonal terms  in (\ref{eq22}) is
measured by the magnitude of  an overlap expression between the
displacements of the two modes that approaches zero in the limit
of zero viscosity when $j\ne j'.$ Provided such overlaps remain
relatively small,  it should be  possible to neglect the off
diagonal terms in the sum. This would be satisfied if the form of
the eigenfunctions is preserved when the  level of viscosity of
interest is introduced. This is physically reasonable when the
damping rate  is small compared to the actual mode frequency and
viscosity is important only in  localised  regions of the
star/planet where the mode wavelength is small, that  for example,
could be modelled using a boundary layer treatment, in such a way
that the form of the  normal modes of interest is  preserved in
most of the star/planet. Such a situation may occur for stars with
radiative envelopes when $g$ modes are effectively damped only
near the surface. We shall assume it to be the case from now on so
that (\ref{eq24}) applies  and  inviscid modes can be used to
evaluate the terms in the summation.

\subsection{Angular momentum transfer rate and energy transfer rate as viewed in the inertial frame}

 Equation (\ref{eq24}) gives the transfer rate  of canonical energy
defined in the rotating frame. The angular momentum transfer rate,
$\dot L_c$, and the  energy transfer rate as defined in the inertial
frame, $\dot E_{I}$, follow by using the  fact that for forcing with a harmonically varying potential
with pattern speed $\omega_{m,k}/m,$ the rates of change of angular momentum and canonical energy are related by
${\dot L_{c}}=m {\dot E_{c}} /\omega_{m,k}.$  In addition, for a harmonically varying forcing potential, the
 rate of change of  energy in the
inertial frame is related to the rates of change of
canonical energy and angular momentum through
\begin{equation}
{\dot E_I}={\dot E_c}+\Omega {\dot L_c}= {\dot E_c}(\omega_{m,k}+  m\Omega)/\omega_{m,k}.\label{ECLCEI}
\end{equation}
Applying the above results separately to the contributions to
  the sum in equation (\ref{eq24}) arising from distinct forcing frequencies, we obtain
\begin{eqnarray}
\dot L_c&=&
-4\pi
\sum_{m,k,j}{m\omega_{\nu,kj} |\lambda_{j}|^2{\cal {N}}_j
|S_{j,k}|^2\over\omega_{m,k}| N_j |^2D_{k,j,j}}, \label{eq250} \\
\dot
E_I&=&
-4\pi
\sum_{m,k,j}
{\omega_{\nu,kj} |\lambda_{j}|^2 (\omega_{m,k}+  m\Omega){\cal {N}}_j
|S_{j,k}|^2\over\omega_{m,k}|N_j |^2D_{k,j,j}}
.\label{eq25}
\end{eqnarray}
 Note that in the limiting case of a  small stellar angular
velocity  directed perpendicular to the orbital plane,
and under the assumption that the effect of rotation can be dealt with
 by simply Doppler shifting forcing frequencies,
analogous equations have been derived by,  e.g. Kumar et al.  (1995).

\subsection{A simplification in the case of dense spectrum of normal modes}

The expressions (\ref{eq24}) (\ref{eq250}) and (\ref{eq25}) can be significantly
simplified in the case when the spectrum of eigenmodes
contributing most significantly to the sums
becomes  dense with a typical distance between two neighbouring
modes $ | \omega_{j+1}-\omega_j|$  approaching  a smooth function
that we can identify with  $|{d\omega_j / dj}| \ll |\omega_{j}|$ for  $ j \rightarrow \infty.$
 An example of  this situation
occurs  for  non rotating  cool stars with radiative regions that support internal
gravity modes.  These  have the required property in the high order/low
frequency limit. In that case $|{d\omega_j / dj}| \rightarrow  |\omega_{j}/j|$
for $j \rightarrow \infty.$ In general,  a high degree of regularity of the spectrum is required
for it to have this property.

 As above we suppose that the main contribution to the series
comes from  modes having near resonance with  the external forcing
frequencies. These are such that  $\omega_{j}\approx \omega_{m,k}$  making  the
magnitude of the denominator,  $D_{k,j,j},$ a small quantity  for such modes.
Let us assume
that for a particular forcing frequency, $\omega_{m,k},$ the
 resonance condition is most closely satisfied for a particular
mode with  index $j=j_{0},$ (which we recall will vary with $m$ and  $k$).
Accordingly  we have
\begin{equation}
\omega_{j_{0}}=\omega_{m,k}+\Delta \omega_{j_{0}}, \label{eq26}
\end{equation}
where   the offset, $\Delta \omega_{j_{0}}, $ is such that
 $|\Delta \omega_{j_{0}}| < |{d\omega_{j_{0}} /
dj_{0}}|$.
 We approximate values of neighbouring mode frequencies by  representing
$\omega_j$ as a Taylor series in $j-j_0$ up to the linear term, thus
$\omega_j\approx \omega_{j_{0}}+ ({d\omega_{j_{0}} / dj_{0}})l$, where
$l=j-j_{0}$ \footnote{It is assumed hereafter that
${|d\omega_{j_{0}} / dj_{0}|}$ is not so small that
 the quadratic term in the Taylor expansion in $l$
has to be retained.}.
This is then substituted  in $D_{k,j,j}$ and
the result used in  equations (\ref{eq24}) (\ref{eq250}) and (\ref{eq25}).
In addition quantities other than  $D_{k,j,j}$ in
eg. (\ref{eq24}) are assumed to depend smoothly on $j$. As the
dominant contribution to the sums comes from values of $j$ close
to $j_0,$  such quantities can be evaluated for $j=j_0$ and
$\omega_j=\omega_{m,k}$ and taken out of  sums  over $j$. In
addition as the viscous decay rates are assumed small we set
$\lambda_j=\omega_{m,k}.$   Following the discussion in section \ref{Modvisc},
we assume the normal
mode spectrum is approximately independent of dissipation so that
the inviscid case may be adopted, we set $N_j$ and ${\cal
{N}}_j$ equal to ${\cal {N}}_{j0}\equiv N_{j0}.$

In this way from (\ref{eq24}) we obtain
\begin{eqnarray}
\dot E_c&=&-4\pi \sum_{m,k} {\cal A}_k
{\omega_{\nu,kj_0}\omega^{2}_{m,k}\over |{d\omega_{j_{0}}/ dj_{0}}|^{2}}
{|S_{j_{0},k}|^2\over N_{j_{0}}}, \hspace{2mm}  {\rm where}\nonumber \\
{\cal A}_k&=&\sum^{+\infty}_{-\infty}{1\over (l+\delta )^{2}+\kappa^2}
 \label{eq27}
\end{eqnarray}
and  we formally extend the sum over the index $(l)$ to $\pm \infty, $
\begin{equation}
\delta = \left |{\Delta \omega_{j_{0}}\over {d\omega_{j_{0}}/
dj_{0}}}\right |, \hspace{2mm}    {\rm and}  \hspace{2mm}
\kappa = \left |{\omega_{\nu,kj_0} \over  d\omega_{j_{0}}/ dj_{0}} \right |.
\end{equation}
The factor ${\cal A}_k$ can be evaluated by  standard  complex variable
techniques to give
\begin{equation}
{\cal A}_k={\pi \sinh \pi \kappa \cosh \pi \kappa \over \kappa (\sinh^2
\pi \kappa + \sin^2 \pi \delta )}. \label{eq28}
\end{equation}

\subsubsection{ The case of moderately large viscosity or an
 averaged dissipation }\label{MLV}
Let us consider the case of large $\kappa \gg 1$  which occurs when the
characteristic  viscous decay  rate of a resonant
mode is much larger than the  frequency difference between the relevant  mode and a neighbouring
mode.  On the other hand the viscous decay time should also be much longer
than the inverse mode frequency so that strong resonances with one particular
mode are still possible.

As discussed by Goodman $\&$ Dickson (1998), physically this situation
occurs when the  decay time
of a mode due to viscosity is much smaller than the time for wave propagation
with the group velocity,  associated with the  potentially resonant modes, which is identified as
$|d\omega_{j_0}/ dj_{0}|^{-1}.$
To see that the latter identification is reasonable,
suppose that  mode $j$ is associated with an appropriately defined  mean wavenumber $k_j $
which, for large $j,$  asymptotically approaches $j/\Delta,$ where $\Delta$ is a
 length scale that is characteristically of the order of the size of the region containing the modes.
We have
$|d\omega_{j_{0}}/ dj_{0} |= |d\omega_{j}/ dk_{j}|_{j=j_0}/\Delta.$
It is seen that the right hand side of the above is the inverse of the time to propagate
across $\Delta$ with the notional group velocity   $ |d\omega_{j}/ dk_{j}|_{j=j_0}.$

In the limit $\kappa \gg 1,$
${\cal A}_k\approx \pi/\kappa$ and the expression for the energy transfer
\begin{equation}
\dot E_{c}=-4\pi^2 \sum_{m,k}{\omega^{2}_{m,k}\over
|{d\omega_{j_{0}}/ dj_{0}}|} {|S_{j_{0},k}|^2\over N_{j_{0}}}.
\label{eq29}
\end{equation}
does not depend on the value of viscosity.
This is physically reasonable because propagating disturbances decay
before they reach at least one boundary. Accordingly,  the calculation of the tidal
response can be performed in a subdomain of the star with
boundary conditions
on the state variables corresponding to  outgoing waves where these leave the subdomain.

In Appendix B we support this conclusion by  showing  how to
obtain (\ref{eq29})  for a model of a non-rotating star, for which
there is  potential resonance  with modes in a $g$ mode spectrum,
by a standard technique for solving the tidal response problem
with appropriate radiation boundary conditions.

 It is important to note that the expression (\ref{eq29}) as well
as the expressions (\ref{eq34}) below  apply  not only to the case of
 moderately large viscosity, but they  also  give  mean energy and angular
 momentum transfer rates obtained by averaging the factor ${\cal A}_k$
over the interval
$0 \le \delta \le 1.$
 Indeed, from equation (\ref{eq28})  it is easy to see that
$\int^{1}_{0}d\delta {\cal A}_k= \pi/\kappa$. Thus, these expressions
are formally valid when  orbital or stellar  parameters
change at uniform rates such that  a large  range of $\delta $
is  covered in the  course of orbital and/or stellar evolution.

Let us stress, however, that when the energy dissipation is very weak
the orbital evolution may become  stalled at some fixed value of $\delta$
with the system located  between resonances,
see eg. Terquem et al (1998).

\subsubsection{The case of very low viscosity}\label{lv0}

Now let us consider the case of small $\kappa \ll 1$. In this case
the factor ${\cal A}_k\approx 1/( \kappa^2 +\sin^2 \pi \delta
/\pi^2)$, and we have from (\ref{eq27})
\begin{equation}
\dot E_c=-4\pi \sum_{m,k}{\omega_{\nu,kj_0}\omega^{2}_{m,k}\over
(\omega^2_{\nu,kj_0}+  ({d\omega_{j_{0}}/ dj_{0}})^2\sin ^2 \pi
\delta / \pi^2)} {|S_{j_{0},k}|^2\over N_{j_{0}}}.
 \label{eq30}
\end{equation}
When $\delta $ is not very close to either zero or one the second
term in the denominator of (\ref{eq30}) is much larger than the
first one. In this case the energy exchange is proportional to the
value of viscosity as expected. However, close to resonances when
$\omega_{\nu,kj_0} > |{d\omega_{j_{0}}/ dj_0}||\sin  \pi \delta /
\pi|$ we can neglect the second term. In this case the energy
transfer is strongly amplified being inversely proportional to
$\omega_{\nu,kj_0}$.

\subsection{Energy and angular momentum transfers expressed in terms of overlap integrals}\label{Olap}

It is convenient to represent the quantities  $S_{j,k}$ entering
into expressions for the energy and angular momentum transfer and
defined through equation (\ref{eq17}) as an inner  product of  a quantity
determined by the forcing potential  and  an eigenmode.
Following the discussion above,
the latter from now on will be  taken to be those appropriate to the inviscid case $\nu=0$
for which we recall  ${\bmth {\eta}}_{j}={\bmth {\xi}}_{j}.$

Using  equation
(\ref{eq9n})  for the Fourier components of the gradient of the forcing potential, we can write
\begin{eqnarray}
&\hspace{-3mm} S_{j,k}&={A_{m,k}\over 2\pi}Q_{j},
{\rm  with}\hspace{1mm}
Q_{j}=\left({\bmth\xi}_{j}{ |}\int_{0}^{2\pi} d\phi e^{-im\phi}\nabla
(r^{2}Y^{m}_{2})\right)\nonumber \\
&\hspace{-3mm}=&\int d^3x\rho e^{-im\phi}{\bmth
\xi}^{*}_{j}\cdot \nabla (r^{2}Y^{m}_{2}). \label{eq31}
\end{eqnarray}
The quantities $Q_j,$ so defined,  coincide with the so-called
overlap integrals considered by e.g. Press $\&$ Teukolsky 1977,
Ivanov $\&$ Papaloizou 2004, IP7 and others. Additionally, instead
of using the norm given by the  expression (\ref{eq12}) evaluated in the inviscid limit,
 it is more convenient to
consider another norm
\begin{equation}
n_{j}=\pi (({\bmth {\xi}}_{j}| {\bmth
{\xi}}_{j})+({\bmth {\xi}}_{j}|{\bmth {\cal
C}}{\bmth{\xi}}_{j})/\omega_{j}^2), \label{eq32}
\end{equation}
which  reduces to a standard  form in the limit of a
non-rotating  star.
Let us express the expressions for the energy and angular momentum
transfer in terms of $Q_j$ and $n_j$. Substituting (\ref{eq31})
and (\ref{eq32}) into  the  expressions (\ref{eq24}), (\ref{eq250})  and
(\ref{eq25}) we obtain
\begin{eqnarray}
\dot E_c&=&-\sum_{m,k,j}{\omega_{\nu,kj}
{|A_{m,k}\hat Q_{j}|^2\over  D_{k,j,j}}}, \nonumber \\
\dot L_c&=&-
\sum_{m,k,j}m{\omega_{\nu,kj}\over \omega_{m,k}}{ |A_{m,k}\hat
Q_j|^2\over D_{k,j,j}}   {\hspace{1cm}} {\rm and}\nonumber \\
  \dot E_I&=&-
\sum_{m,k,j}{\omega_{\nu,kj}(\omega_{m,k}+m\Omega)\over \omega_{m,k}}{
|A_{m,k}\hat Q_j|^2\over D_{k,j,j}}. \label{eq33}
\end{eqnarray}
where
\begin{equation}
\hat Q_j=Q_j/\sqrt{n_j}, \label{eq33a}
\end{equation}
does not depend on normalisation of eigen functions, and we use
equation (\ref{eq18}) to obtain the last equality.
\subsubsection{A dense spectrum of modes and moderately large viscosity}
 In this case equation
(\ref{eq29}) together with (\ref{ECLCEI})  leads to
\begin{eqnarray}
\dot E_{c}&=&-\pi \sum_{m,k}{|A_{m,k}\hat Q_{j_{0}}|^2\over
|{d\omega_{j_{0}}/ dj_0}|},  \nonumber \\
 \dot L_{c}&=&-\pi \sum_{m,k}{m\over
\omega_{m,k}}{|A_{m,k}\hat Q_{j_{0}}|^2\over |{d\omega_{j_{0}}/
dj_0}|}, \hspace{1cm}{\rm and}   \nonumber \\
\dot E_{I}&=&-\pi \sum_{m,k}\left(1+{m\Omega\over
\omega_{m,k}}\right ){|A_{m,k}\hat Q_{j_{0}}|^2\over |{d\omega_{j_{0}}/
dj_0}|}.
 \label{eq34}
\end{eqnarray}
\subsubsection{ A dense spectrum of modes and very small viscosity}\label{lv1}
In this case  equation
(\ref{eq30}) similarly leads to
\begin{eqnarray}
\dot E_c&=&-\sum_{m,k}{\omega_{\nu,kj_0}\over D_*}
|A_{m,k}Q_{j_{0}}|^2, \nonumber \\
\dot L_c &=&-\sum_{m,k}{m\omega_{\nu,kj_0}\over \omega_{m,k}D_*}
|A_{m,k}Q_{j_{0}}|^2,  \hspace{1cm}  {\rm and} \nonumber   \\
\dot E_I&=&-\sum_{m,k}\left (1+{m\Omega\over
\omega_{m,k}}\right){\omega_{\nu,kj_0}\over D_*}
|A_{m,k}Q_{j_{0}}|^2, \label{eq35}
\end{eqnarray}
where $D_{*}=(\omega^2_{\nu,kj_0}+ ({d\omega_{j_{0}}/ dj_0})^2\sin
^2 \pi \delta/ \pi^2)$.

\subsection{Energy and angular momentum transfer in the case of
nearly parabolic orbit}
The transfers of
energy, $\Delta E_{c}$, $\Delta E_{I}$ and angular momentum,
$\Delta L_{c}$, between the orbital motion and normal modes of the
star during a periastron flyby on a formally parabolic orbit can
also be expressed in terms of overlap integrals. For completeness we give
expressions for these here.
As
mentioned above, in this case we should represent the results in
terms of the Fourier transform of the tidal potential (see equation (\ref{eq p1n})) instead of the
Fourier series used in the previous sections.
Expressions for $\Delta E_{c}$, $\Delta E_{I}$ and
$\Delta L_{c}$ in terms of this were
derived by IP7
\footnote{
Note that  in equation (5) of IP7 there should be a $(-)$  sign in
front of the expression on the right hand side and  in
equation (28) the imaginary unit should  be inside the
summation.}. As in the case discussed above they also can
be represented as sums over quantities determined by
the eigenmodes, which for a given $m$  have eigenfrequency $\omega_k,$  that  take the form
\begin{eqnarray}
\Delta E_{c}&=&{8\pi^3}\sum_{m,k}{\omega_k^2|S_{m}(\omega_k)|^2\over
N_k}, \nonumber \\
\Delta L_{c}&=&{8\pi^3}\sum_{m,k}{m\omega_k
|S_{m}(\omega_k)|^2\over N_k}  \nonumber \\
\Delta E_{I}&=&{8\pi^3}\sum_{m,k}{\omega_k(\omega_k+m\Omega)|S_{m}(\omega_k)|^2\over
N_k}. \label{eq36}
\end{eqnarray}
Representing again the quantities $S_{m}(\sigma)$ as a product of
two factors determined by the orbit and  by the eigenmodes,
respectively,  we obtain a relation corresponding to  (\ref{eq31}):
$S_{m}(\sigma)=(1/ 2\pi) A_m(\sigma)Q_{j}.$  Using the
normalised overlap integrals defined in (\ref{eq33a}) we can
rewrite (\ref{eq36}) in a more convenient way as
\begin{eqnarray}
\Delta E_{c}&=&{2\pi^2}\sum_{m,k}{|A_{m}(\omega_k)\hat Q_k|^2}, \nonumber \\
\Delta L_{c} &=&{2\pi^2}\sum_{m,k}{m \over \omega_k}
{|A_{m}(\omega_k)\hat Q_k|^2}, \nonumber \\
\Delta E_{I} &=&
{2\pi^2}\sum_{m,k}\left (1+
{m\Omega\over \omega_k}\right )
{|A_{m}(\omega_k)\hat Q_k|^2}. \label{eq37}
\end{eqnarray}
Note that the quantities $A_m(\sigma)$ entering (\ref{eq37}) are
discussed in e.g. Ivanov $\&$ Papaloizou (2011) for the  general
case of a parabolic orbit inclined with respect to equatorial
plane of a rotating star\footnote{Note that analogous expressions
of Ivanov $\&$ Papaloizou 2011 differ by factor of two from  those
given in (\ref{eq37}). This is due to the fact that in Ivanov $\&$
Papaloizou 2011,  the summation over the azimuthal  mode number is
formally performed  over positive and negative values of $m$,
while in this paper we consider  only positive azimuthal mode
numbers. Note also that in this paper we take all quantities
related to eigenmodes to be defined in the rotating frame,  while
in Ivanov $\&$ Papaloizou (2011) the inertial frame was used.} .

\section{Rotationally modified gravity modes}\label{modgm}

\subsection{Use of the traditional approximation}
In what follows we are going to apply the general theory discussed
above to the situation when the
 energy and angular momentum exchange between
a  rotating Sun-like star  and its orbit  occurs as
a result of the excitation of  internal  rotationally modified  gravity (g)
modes.  These are assumed to be associated with
 discreet low frequency spectrum.

For the  cases of a periodic orbit and of a nearly
parabolic flyby, the energy and angular momentum exchanges between
the orbit and the star are characterised  by two quantities
associated with the normal modes.  These are  their eigenfrequencies
$\omega_{j}$ and  the corresponding overlap integrals $Q_j.$
 Thus for given
orbital parameters and the rate of rotation of the star they
determine the energy and angular momentum transfer rates in the case
of  'moderately large viscosity'  discussed above. In this regime the decay
rate of a characteristically excited mode is small compared to
its frequency as seen in the rotating frame. However, it is large
compared to the frequency separation of neighbouring modes.
Then the  waves associated with the rotationally modified (g) modes
are dissipated near a boundary, preventing the setting up of standing
waves. This leads to a radiation boundary  condition for the forced problem.
However,  these disturbances are solutions of the adiabatic linearised problem
away from localised dissipative regions.
 When dissipative
effects are very weak, an alternative
prescription for their action  on stellar perturbations
should be used (see sections \ref{lv0} and \ref{lv1}.

The rotational angular frequency $\Omega $,  can be expressed in
units of  the natural frequency of the primary star,
$\Omega_*=\sqrt{GM_*/ R_*^3}$, where $M_*$ and $R_*$ are the
stellar mass and radius, respectively. As the non spherical
distortion of the star is  $\propto (\Omega/\Omega _*)^2$ it may
be assumed to be spherically symmetric   when $(\Omega/\Omega _*)$
is small with internal structure  the same as in the non-rotating
state.

The problem on hand has a symmetry property $\omega_j \rightarrow
-\omega_j$, $\Omega \rightarrow -\Omega$. In the preceding
sections it was assumed that $\Omega$ is positive while the
eigenfrequencies $\omega_j$ can have either sign depending on
direction of their propagation with respect to rotation of the
star. In what follows it is convenient to adopt the alternative
convention assuming that the eigenfrequencies are always positive
while the case $\Omega  > 0$ and $\Omega < 0$ correspond to the
modes propagating in the prograde and retrograde sense with
respect to rotation of the star, respectively.

When the orbital period is sufficiently large,  tidal perturbations  mainly
interact  modes of large radial order.
 In this case one can make use of
 the Cowling approximation which  neglects variations of  self-gravity when
describing free normal modes of the star and the
'traditional' approximation, where only components of the Coriolis
force perpendicular to the radius vector from the centre of the
star  are included.
 When the traditional approximation is adopted and the star is assumed to be
spherical the equations
describing the stellar pulsation admit
 separation of variables (see below).
The problem of finding $\omega_j$ in such a setting has been
considered by many authors, either by numerical means or using the
WKBJ technique ( e.g. Lee $\&$ Saio 1987, Bertholomieu et al
1978 among others ).

\subsection{Calculation of the overlap integrals}
 However, less attention has been payed to  the calculation of
 the overlap integrals,  $Q_j,$
 Rocca (1987) attempted to calculate  quantities  called tidal
resonant coefficients that are proportional to $Q_j$, following the
earlier work by Zahn (1970) who studied the same quantities
 in the case of non-rotating star.

However,   significant potential
 inaccuracies  remain,
for example,  Berthomieu et al (1978) assumed that the WKBJ
approach is approximately valid in the  convective regions of the
star bounding a radiative region
where the waves associated with the  modes propagate.
 Although they did not consider the overlap integrals,
 this is not appropriate for the modes of low azimuthal order that  are
important for the tidal problem.
In addition, when calculating the eigenfrequencies
and eigenmodes,  it has been generally  assumed that,
in a radiative zone,  the square of the
Brunt - V$\ddot {\rm a}$is$\ddot {\rm a}$l$\ddot {\rm a}$ frequency behaves approximately linearly  with
the  distance to  the base of convection zone. Since
for some  models  solar mass  stars,  this approximation is not
accurate  (see Fig \ref{Fig3} below) we allow for a more
general power law dependence\footnote{Note that Provost $\&$
Berthomieu (1986) give WKBJ expressions for the displacement
vector in a radiative region valid for our power law dependence.
However,   solutions in the convective
are obtained  numerically precluding the
development of an analytic asymptotic representation of the overlap integrals,
which were not considered,  at low forcing frequency.}

Our approach to calculation of $Q_{j}$ is similar to the approach
of Rocca (1987). However, in
contrast  to that work,  we consider a general
power law dependence,  on the distance to the
convection zone boundary,  for  the Brunt - V$\ddot {\rm a}$is$\ddot {\rm a}$l$\ddot {\rm a}$ frequency.
We revisit the  calculation of $\omega_j$ and
$Q_j,$ mainly concentrating on calculation of $Q_j,$ since
calculation of the eigenfrequencies is relatively straightforward.
 We  find two
contributions to the overlap integrals, the first   comes  from the radiative  region
close to the base of convection zone and other comes  from the
convection zone itself.  We correct some errors in  earlier work,
and find the asymptotic form at low forcing frequency.
 We compare the results we  obtain with numerical
calculations for solar mass  models in a regime where these can be performed.

\subsection{Linearised equations for normal modes}
Using the Cowling and traditional approximation in (\ref{eq
p3}), while neglecting dissipative effects
and the
external forcing potential,  we get

\begin{equation}
-\omega^2\mbox{\boldmath$\xi$}  -2i\omega \left({\bmth {\Omega}}\cdot {\hat {\bf r}}\right){\bf {\hat r}}\times {\bmth{\xi}}
+{\bmth {\cal C}}\mbox{\boldmath$\xi$} = 0
\label{eomt}
\end{equation}
where
\begin{equation}
{\bmath {\cal C}}\mbox{\boldmath$\xi$}=\frac {1}{\rho}\nabla P'
-   \frac{P'}{\rho\Gamma_{1} P }\nabla P
+N^2(\mbox{\boldmath$\xi$}\cdot {\bf\hat{ r}}) {\bf\hat{ r}},\label{linop}
\end{equation}
\begin{equation}P'(\mbox{\boldmath$\xi$})
=-\Gamma_1 P\nabla\cdot(\mbox{\boldmath$\xi$}) -      \mbox{\boldmath$\xi$}
\cdot\nabla P \label{ep1}\end{equation} and
the square of the Brunt - V$\ddot {\rm a}$is$\ddot {\rm a}$l$\ddot {\rm a}$ frequency is given by
\begin{equation} N^2= \frac{1}{\rho}\frac{dP}{dr}
\left( \frac{1}{\rho}\frac{d\rho}{dr}-\frac{1}{\Gamma_{1} P
}\frac{dP}{dr}\right),
\label{BVF}\end{equation}
where $\Gamma_{1}=({\partial \ln P / \partial \ln
\rho})_{ad},$ with the subscript $ad$ indicating that the derivative is taken at constant entropy.

We adopt  spherical polar coordinates $(r, \theta, \phi)$ with $\xi_{r}$,
$\xi_{\theta}$ and $\xi_{\phi}$ being the  components of the vector ${\bmth
{\xi}}$ in these
coordinates,
the component of
${\bmath{\Omega}}$ acting in the radial direction is
${\bmath{\Omega}}~\cdot~{\bf {\hat r}}~\equiv~(\Omega\cos\theta, 0, 0)$
where ${\bf {\hat r}}$ is the unit vector in the radial direction. It will be taken as read
that all quantities of interest  will be taken to be expressed in terms of
Fourier series as specified as in  (\ref{eq p1})
with the equations  applying to separate coefficients.
Here $\rho$
is the density and $P$ is the pressure in the background state.
  The gravitational
acceleration at  radius $r$
 is $g={GM(r)/ r^{2}}$
 and $M(r)$  is the mass enclosed within a
sphere of radius $r.$

 The  density and pressure perturbations are $\rho^{'}$ and
$P^{'}$
respectively.
We note that  $\rho^{'},$  $P^{'}$ and $\xi_r$
satisfy the condition for adiabatic perturbations
which gives
\begin{equation}
P^{'}=c_{s}^2\rho^{'}- {c_{s}^{2}\rho N^{2}\over g}\xi_{r},
\label{eq39}
\end{equation}
where $c_{s}=\sqrt{\Gamma_{1}P/\rho}$ is the adiabatic sound speed.
Note that if $N^2\ge 0$ everywhere as assumed
in this paper, the operator  ${\bmath {\cal C}}$  is positive and
the linear eigenvalue problem  can be brought into the self-adjoint form (\ref{eq10}).
For simplicity we suppress the mode index (j ) below,  wherever
this  doesn't lead to any ambiguity.


 As is well known (e.g. Unno et al 1989),  both the forced response problem  and
 the  unforced $(\Psi=0)$
 problem considered  in this section
 can be solved by adopting the separation of variables anzatz:
\begin{equation}
\xi_r={\xi(r) H(\theta)\over \sqrt{2\pi}},\hspace{1mm}
\rho^{'}={\hat \rho(r) H(\theta)\over \sqrt{2\pi}},
\hspace{1mm}{\rm and}\hspace{1mm} P^{'}={\hat P(r) H(\theta)\over
\sqrt{2\pi}} \label{eq42}
\end{equation}


\noindent together with
\begin{eqnarray}
\xi_{\phi}&\hspace{-2mm}=&\hspace{-2mm}
{1\over \sqrt{2\pi}}{\xi^{S}(r)\over (1-\nu^2 \cos^2 \theta)}\left({im\over \sin
\theta }H(\theta)-i\nu \cos \theta {dH(\theta)\over d\theta}\right), \nonumber \\
\xi_{\theta}&\hspace{-2mm}=&\hspace{-2mm}{1\over
\sqrt{2\pi}}{\xi^{S}(r)\over (1-\nu^2 \cos^2 \theta)}\left
({dH(\theta)\over d\theta}-m\nu \cot \theta H(\theta )\right).
\label{eq43}
\end{eqnarray}
Here quantities with a {\it hat} are functions only of $r,$
 $\nu~=~{2\Omega /~\omega}$ and the factor
 ${1/ \sqrt{2\pi}}$ is inserted for  convenience (see below).
We remark  that the separation of the $\phi$ dependence through the usual  factor $\exp({\rm i}m\phi$)
is here taken as read.

 Substituting equations
(\ref{eq42}) and (\ref{eq43}) into equations  (\ref{eomt}) -(\ref{eq39})
we verify separability provided that
the function $H(\theta)$ is determined through solving the  eigenvalue  problem
 specified by
\begin{eqnarray}
\hat L H&=&-\Lambda H,\hspace{1mm}
{\rm where} \hspace{1mm} {\rm the}\hspace{1mm} {\rm operator}
\hspace{1mm}} \hat L \hspace{1mm} {\rm  is \hspace{1mm} defined \hspace{1mm} by \label{Laplace0}\\
\hat L&=&{d\over d\mu}\left({(1-\mu^2)\over (1-\nu^2
\mu^2)}{d\over d\mu}\right)\nonumber \\
&+&\hspace{-3mm} {1\over (1-\nu^2 \mu^2)}\left ({m\nu
(1+\nu^2\mu^2)\over (1-\nu^2 \mu^2)}-{m^2\over (1-\mu^2)}\right ),
\label{eq47}
\end{eqnarray}
with $\mu=\cos \theta $.  The eigenvalues, $\Lambda,$ specified by
(\ref{Laplace0}), which defines  the Laplace tidal equation,  are  also separation
constants. The associated eigenfunctions are Hough functions
( see e.g. Longuet-Higgins, 1968, for a discussion of the
eigenvalues and eigenfunctions).

After separating  the variables
by the means as indicated above,  it is found that the  angular components of (\ref{eomt})
require that
\begin{equation}
\xi^S={\hat P \over r\omega^2 \rho}. \label{eq46}
\end{equation}
The radial component of (\ref{eomt}) together with (\ref{ep1})
then lead to  two ordinary differential equations
for the radial dependence
of $\xi(r)$ and ${\hat P}.$  These equations take  the form
\begin{equation}
{d\xi \over dr}=-\left ({2\over r}+{1\over \Gamma_1 P}{dP\over dr}\right )\xi
+{\hat P  \over \rho}\left ({\Lambda \over \omega^2 r^2}-{1\over c_s^2}\right ), \label{eq44}
\end{equation}
and
\begin{equation}
{d\hat P \over dr}=\rho (\omega^2 - N^2)\xi +{\hat P  \over \Gamma_1
P}{dP\over dr}. \label{eq45}
\end{equation}
In general $\Lambda $ is a function of
$\nu $, and, accordingly, $\omega$. Therefore, when $\Omega \ne 0$
the eigenvalues  $\Lambda$ and $\omega$  defined through the solution of
 equations (\ref{eq44}),
(\ref{eq45}) and   (\ref{Laplace0}) must be determined  together.

 In the degenerate case $\Omega=0,$ $\nu=0,$  (\ref{Laplace0})
 reduces to Legendre's  equation. Its solutions are
accordingly are associated Legendre functions $P^{m}_{l}(\mu)$ and
$\Lambda = l(l+1)$, where $l$ is an natural number. Since in this limit the
tidal problem is such that only eigenmodes corresponding to $l=2$ are coupled
to the dominant quadrupole component of the tidal potential,
$\Lambda $ can be set to $6.$
The  eigenvalue
problem for the determination of $\omega$ is accordingly decoupled from the angular one in the
non-rotating case.  Equations (\ref{eq44})-(\ref{eq45}) then  coincide
with standard equations describing the free modes of pulsation of a
non-rotating star under  the Cowling approximation, see e.g.
Christensen-Dalsgaard (1998).

\subsection{Properties of the Laplace tidal equation}\label{Laplace}

We here summarize  properties of equation   (\ref{Laplace0})  required for our purposes.
When the parameters $\nu $ and $m$ are fixed the operator $\hat L$
is self adjoint with respect to the natural inner product defined
by integration  over $(-1,1)$ with respect to $\mu.$  There is a  countable infinite set of
solutions corresponding to different characteristic values of $\Lambda,$
$\Lambda_i$.  The corresponding eigenfunctions, $H_i$,  are
orthogonal and normalised  such that
\begin{equation}
\int ^{1}_{-1}d\mu H_i(\mu)H_k(\mu)=\delta_{ik}.
\label{eq48}
\end{equation}

In general, the eigenvalues $,\Lambda_i,$  can be either positive or negative. In this paper
we assume below that they are non-negative,  which is necessary  for the associated solutions
to   correspond
to wave-like
perturbations in stably stratified regions, as required for potentially resonant
rotationally modified gravity modes.
Thus the  $\Lambda_i, i=1,2...,$ will  be arranged  such that
$\Lambda_{i+1} > \Lambda_i$ with $\Lambda_1$ corresponding to the  minimum  non-negative eigenvalue
 for a given $\nu $ and $m.$

As  stated above, when $\nu =0,$ the  $H_i$ are proportional
to  associated  Legendre functions, $P^m_l.$
In the general case, $\nu \ne 0,$
 the functions $H_i$ may be represented as  an  infinite series of
 associated Legendre functions through
\begin{equation}
H_i=\sum_{n \ge 0 }\alpha_{i,n}\tilde P^m_{(2n+2)}, \hspace{1mm}
{\rm with} \hspace{1mm}\tilde P^m_{(2n+2)}=\sum_{i >0}\alpha_{i,n}H_{i}.
\label{eq49}
\end{equation}
Here the scaled associated Legendre functions
\begin{equation}
{\tilde P}^{m}_{l}(\mu)= \sqrt {\frac{(2l+1) (l-m)!}{ 2(l+m)!}}P^{m}_{l}(\mu)
\end{equation}
satisfy the condition
$\int^{1}_{-1} d\mu \tilde P^m_{l}(\mu)\tilde P^m_{l'}(\mu)= \delta_{l,l'}$.
The second equality in (\ref{eq49}) follows from orthogonality and normalisation  of each
of the  sets of functions
 $H_i$ and $\tilde P^m_{(2n+2)}$. We calculate
the decomposition of the tidal potential
 as a series of Hough functions using equation (\ref{eq49}) truncating the
summation  at sufficiently large value of,  $n,$ with help of the
procedure described in Ogilvie $\&$ Lin  (2004).

As the dominant tidal potential has the angular dependence of a
quadrupole, it follows from equation (\ref{eq31}), that the
overlap integrals involve a product of $\tilde P^m_{2}$ and $H_i$,
and therefore, the coefficients
\begin{equation}
\alpha_i(\nu)\equiv \alpha_{i,0}={\int^{1}_{-1} d\mu H_i \tilde
P^m_2} \label{eq50}
\end{equation}
play an important role. It describes the 'angular'
coupling of the rotationally modified  g-modes with the  tidal field within the framework of the
traditional approximation.

\begin{figure}
\begin{center}
\vspace{8cm}\includegraphics{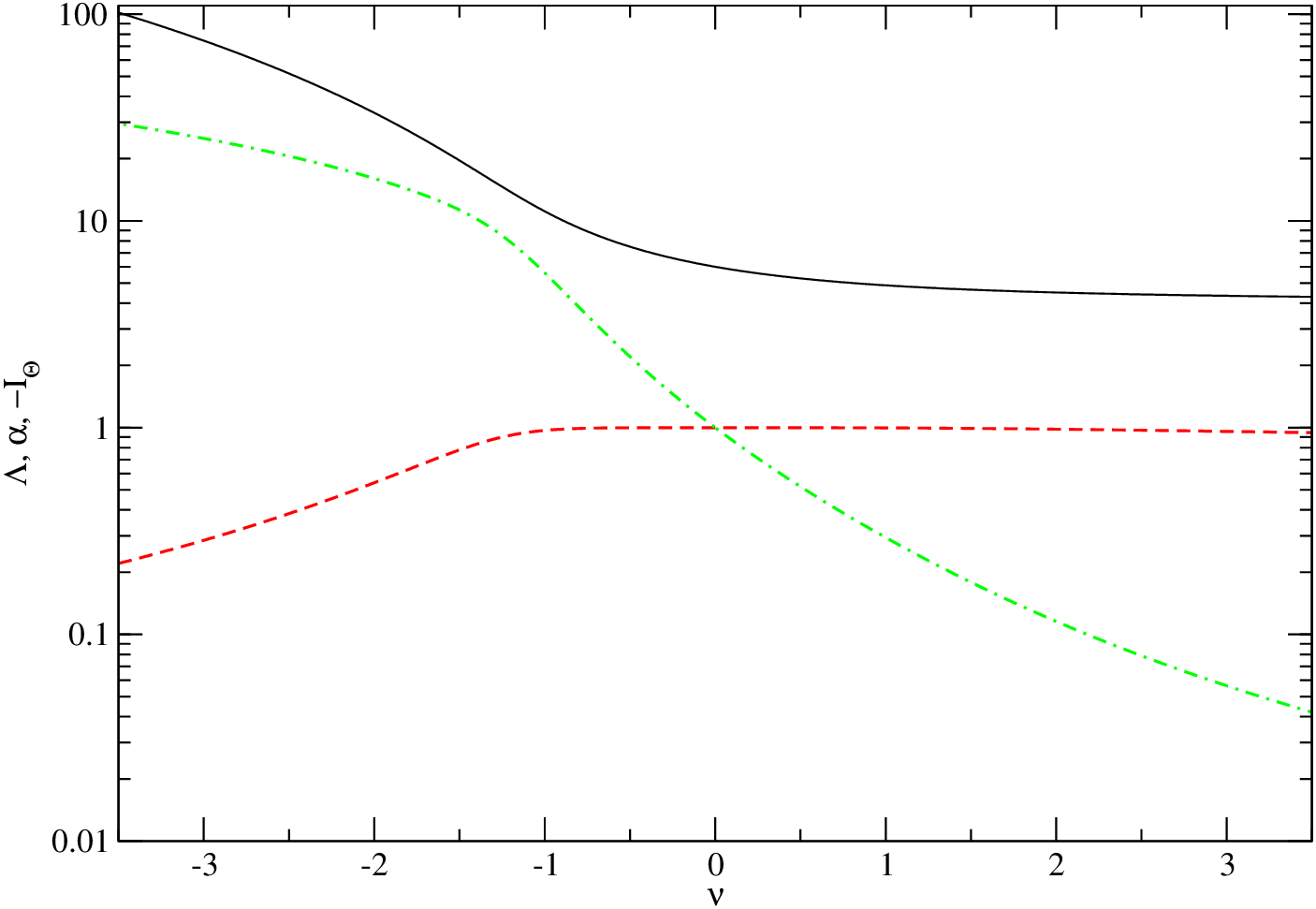}
\end{center}
\vspace{0.5cm} \caption{The values of $\Lambda_1$ (solid line),
$\alpha_1$ (dashed line) and  $-I_{\theta}$, where
$I_{\theta}$ is defined  through equation (\ref{eq56}) (dot-dashed line) for $m=2$
as functions of the parameter $\nu $.} \label{Fig1}
\end{figure}

\begin{figure}
\begin{center}
\vspace{8cm}\includegraphics{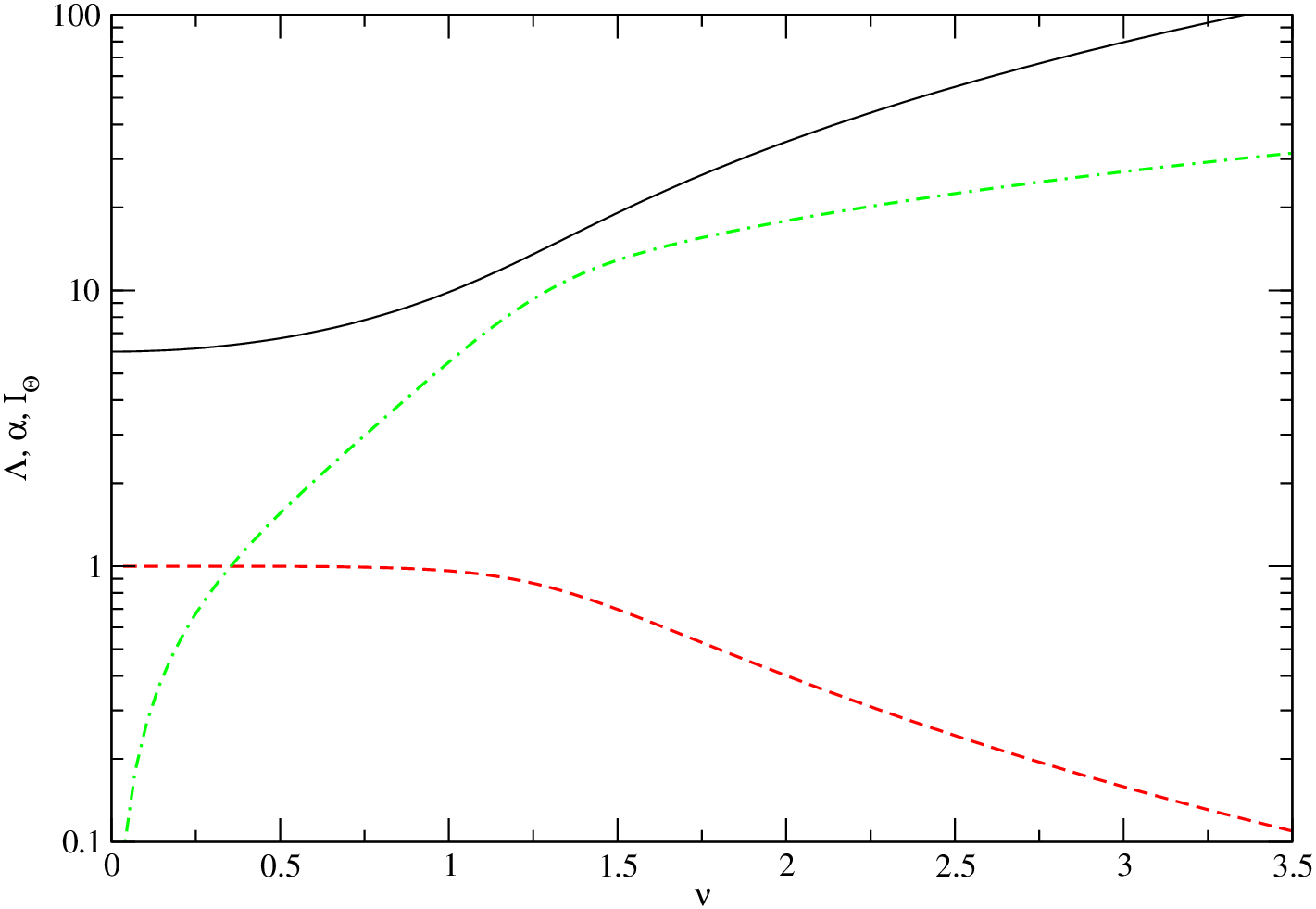}
\end{center}
\vspace{0.5cm} \caption{Same as Fig. \ref{Fig1} but the case $m=0$
is shown and the dot-dashed line shows $I_{\theta}$. }
\label{Fig1a}
\end{figure}

For moderate values of $\nu$ ( for which, $|\nu | \le 3,$ say)
only the quantities corresponding to $i=1$ are important. We
calculate $\Lambda_{1}(\nu)$ and $\alpha_1(\nu)$ numerically as
well as an additional quantity, $I_{\theta}(\nu),$ determined by
the Hough functions and defined below (see equation (\ref{eq56})
and plot  them in Fig. \ref{Fig1} for $m=2$ and in Fig.
\ref{Fig1a} for m=0.  Note that in the latter case from the
definition of $\Lambda_{1}(\nu)$ and $\alpha_1(\nu)$ it follows
that they are even functions of $\nu$ while $I_{\theta}$ is an  odd
one. Therefore, we show only positive values of $\nu $ in Fig.
\ref{Fig1a}.

The case $m=2$ is the most important for  tidal interaction.
We see from Fig. \ref{Fig1} that $\Lambda_{1}(\nu)$ and
$-I_{\theta}(\nu)$ increase as $\nu$ decreases.  On the other
hand $\alpha_1(\nu)$ increases with $\nu$ until $\nu=-1,$
levelling off at larger values. As $\alpha_1(\nu)$ measures the
strength of coupling to the tidal field, this latter behaviour
results in the stronger excitation of modes associated with
positive values of $\nu$.

The values of $\nu$ are in general different for different modes since
these  are obtained from the  simultaneous solution of  eigenvalue  problems in the
radial and  $\theta$ directions. Nonetheless, we
classify the eigenmodes using  two indices, the index $(i)$
corresponding to a particular $\Lambda_i(\nu_i),$  which accordingly
indicates the order of the angular eigenvalue, and the index
$(n)$ describing the number of nodes in the radial direction, for
a given $m$.  Thus our mode index $j,$  used in previous sections,
should be understood as a pair $(i,n)$.

Note that when the star is non-rotating, and, accordingly, $\nu=0$, from
 equation(\ref{eq43}) it follows that ${\bmth {\xi}}= \xi Y^m_l{\bf e}_r+\xi^{S}r\nabla Y^m_l$,
where ${\bf e}_r$ is the unit vector in
the radial direction.

\subsection{ An expression for the overlap $ Q_j$ in the low frequency limit
in the traditional approximation}
We now reduce the expression  (\ref{eq31}) for the overlap integral  $Q_j$
to a form involving an integral over the radial coordinate only in the traditional approximation.
This is useful for either performing numerical calculations or undertaking further asymptotic analysis
(see below). We work on the expression for $Q_j$ given by equation (\ref{eq31}).

 It is convenient to
express  the term $\nabla (r^2Y^m_2)$ that appears there  as
 $2rY^{m}_2{\bf
e}_r+r^2 \nabla Y^m_2.$ We then  split the integral for  $Q_j$
into two parts, the first containing $2rY^{m}_2{\bmath
\xi}^{*}_j\cdot {\hat {\bf r}}$ and the second containing
$r^2{\bmath \xi}^{*}_j \cdot \nabla Y^m_2$. We remark that the
forms  of both  the  free and forced problems specified by
(\ref{eomt}) are such that after separation of variables,  the
functions $\xi$  and $\xi^{S}$ written as ${\xi}_j$ and
$\xi^{S}_j,$  when these correspond to a free normal mode $j,$ may
be taken to be real. Noting that $Y^m_2=\tilde P^m_2\exp({\rm
im}\phi) /\sqrt{2\pi}$ and  using (\ref{eq50}), the first part of
(\ref{eq31})  can be transformed to  become
\begin{equation}
2\alpha_i\int_0^{R_s} r^3dr \rho \xi_j.\nonumber
\end{equation}
To deal with  the second
part we first note that equation (\ref{eq47}) is equivalent to the
relation
\begin{equation}
 r\nabla_{\perp}(e^{{\rm i}m\phi} {\bmath \xi}_j)=-\Lambda_i \xi^S_j H_ie^{{\rm i}m\phi}/\sqrt{2\pi},
\label{eq51}
\end{equation}
where
$\nabla_{\perp} \equiv \nabla - {\hat {\bf r}}  {\hat {\bf r}}\cdot\nabla.$
 Making use of (\ref{eq51}), performing an integration by parts, and noting
(\ref{eq49}) and  (\ref{eq50}),   the second part can be reduced to the form
\begin{equation} \alpha_i \Lambda_i\int_0^{R_*} r^3dr \rho \xi^S_j. \nonumber
\end{equation}
Summing
the contributions of the two parts,  we  obtain
\begin{equation}
Q_j=\alpha_i Q_{rj}, {\rm with}  \quad Q_{rj}=\int^{R_*}_{0}r^3dr \rho (2\xi_j
+\Lambda_i \xi^S_j). \label{eq52}
\end{equation}
 When the star is
non-rotating $\alpha_i=1,$ $\Lambda=6$, and the expression (\ref{eq52})  reduces to
its standard form, see e.g. PT, IP4.

We remark that, when considering which regions of the star  contribute to
the overlap integrals, the original expression
 (\ref{eq31}) can also, after a straightforward  integration by parts, be
reduced to an integral for which the integrand  is proportional to the
mode density perturbation. Viewed in this way, only regions with
non vanishing density perturbation contribute,  consistent
with expectation for a tidal interaction controlled by the
gravitational potential due to a perturber.

The expression for $Q_{rj}$ can be   transformed into a form that is
explicitly $\propto \omega_j^2$ which is thus
very convenient in the low frequency limit, see also Rocca (1987).  In order to  proceed
we first use (\ref{eq46}) to express $\hat P$ in terms of $\xi^S$
and substitute the result in  equations  (\ref{eq44}) and (\ref{eq45}).
These then yield the two relations:
\begin{eqnarray}
\xi_j&=&{\omega_j^2\over N^2}\left (\xi_j +{rN^2\over g}\xi_j^S-(r\xi_j^{S})^{'}\right),
\label{eq520} \\
\xi_j^{S}&=&{1\over \Lambda_i}\left({1\over \rho r}(r^{2}\rho
\xi_j)^{'}+{rN^2\over g}\xi_j+{\omega_j^2r^2\over c_s^2}\xi_j^S\right),
\label{eq52a}
\end{eqnarray}
where a prime indicates  differentiation with respect to  $r.$
 Now we substitute the right hand side of equation(\ref{eq52a}) for
$\xi^S_j$  in (\ref{eq52})
and then integrate  the term
proportional to $(r^{2}\rho \xi)^{'}_j$ by parts, assuming that
$\xi_j$ is smooth, but
we allow for  the density $\rho $ to  undergo a discontinuous  jump at
some point $r=r_*$ as is found to occur in some stellar models.
  In addition we assume  that there are
no boundary contributions coming from the surface or centre of the
star.  We accordingly  obtain
\begin{equation}
Q_{rj}=\int^{R_*}_{0} r^4dr\rho \left ({N^2\over
g}\xi_j+{\omega_j^2r\over c_s^2}\xi_j^S\right) - r^4\xi_j\left[ \rho
\right ]^{\pm},
\label{eq52b}
\end{equation}
where the term in  square brackets in (\ref{eq52b}) should be understood as
being equal to the difference between the expression in square brackets taken at
radii $r_*+\epsilon$ and $r_*-\epsilon$ and then taking the limit  $\epsilon \rightarrow
0$.

In order to get an expression for $Q_r$ that is  explicitly proportional to
$\omega_j^2$ we substitute the right hand side of equation (\ref{eq520})
for $\xi_j$
into (\ref{eq52b})
and then
integrate  the term proportional to
$(r\xi^{S}_j)^{'}$ by parts assuming the same conditions as above.
We then  get
\begin{equation}
Q_{rj}=\omega_j^2 \int^{R_*}_{0} r^4 dr\rho \left({\xi_j\over g}+{\xi_j^S\over r^3}\left({r^4\over
g}\right)^{'}\right)-S^{\pm}
\label{eq52c}
\end{equation}
where the surface term
\begin{equation}
S^{\pm}=r^4\xi_j\left[\rho \right]^{\pm}  -\omega^2_j{r^5\over g}\left[\rho\xi^{S}_j\right]^{\pm}. \label{eq52n}
\end{equation}
In order to evaluate $S^{\pm}$ we  take equation (\ref{eq520}),
multiply  both sides by $\rho N^2$ and substitute the expression given by
(\ref{BVF}) for the square of the
 Brunt - V$\ddot {\rm a}$is$\ddot {\rm a}$l$\ddot
{\rm a}$ frequency.
 Integrating the result
between $r_*+\epsilon$ and $r_*-\epsilon,$
  making use of the fact that the  displacements
are are finite at $r=r_*$ and that
the pressure gradient is equal to $-\rho g,$
 we readily obtain  $S^{\pm}=0$.


So far the expressions for $Q_{rj},$ though adapted to low frequencies, are exact
and can be used  for numerical evaluation.
To facilitate analytic discussion in the relevant limit of low frequency,  we  express the integrand in (\ref{eq52b})
in terms of $\xi_j$ only,  and in so  doing,  obtain  a simplified expression for $Q_{rj}$
that neglect  terms of order  $\omega_j^4$ and higher.

In order to do this,  we substitute the right hand side of
equation (\ref{eq52a}) into (\ref{eq52c}), first  noting that
  the last term may be neglected to the order we are working.
 Integrating
again by parts, assuming no boundary contributions or internal jumps in $\xi_j^S,$  we obtain
\begin{eqnarray}
Q_{rj}&=&\omega_j^2\int^{R_*}_0 r^4 dr\rho h(r)\xi_j + O(\omega_j^4), \hspace{1mm} {\rm where} \nonumber \\
h(r)&=&{1\over g}\left(1-{(12-4\upsilon)\over \Lambda_i}\right)\nonumber \\
&+&{1\over
\Lambda_i}\left({r^{2}g^{''}\over g^2}-{2(rg^{'})^{2}\over
g^3}+{(8-\upsilon)rg^{'}\over g^2}\right). \label{eq52d}
\end{eqnarray}
Here $\upsilon= N^2r/g.$

\subsection{An expression for the norm}
In order to calculate energy and angular momentum exchange rates
resulting from tidal interaction, in addition to the overlap
integrals $Q_j,$ we also require the norms $n_j$  ( see eg.
section \ref{Olap}). Here we develop a form for $ n_j$ in terms of
one dimensional angular and radial integrals that can be readily
evaluated. To do this we  transform  the expression (\ref{eq32})
for the norm, $n_j,$ so that it involves integrals only over the
radial coordinate. Using the  that ${\bmath \xi}_j$ satisfies,
$\omega^2{\bmath \xi}_j~=~{\bmath{{\cal
C}\xi}_j}~+~\omega~{\bmth{\cal B}}{\bmath \xi}_j,$
  (\ref{eq32})  can also be written as
\begin{equation}
n_j={2\pi \over \omega_j^2}(({\bmath \xi}_j|{ {\bmath {\cal C}} }{\bmath
\xi}_j)+{\omega_j\over 2}({\bmath \xi}_j|{\bmath {\cal B}}{\bmath \xi}_j)).
\label{eq53}
\end{equation}
  Using  equation (\ref{linop}) to specify (\ref{eq53})
  in terms  of  the  linear perturbations,  making use of (\ref{ep1})-(\ref{eq45}),
  performing  an integration over the angular coordinates by parts,
 and then making  use of
(\ref{eq51}), we obtain
\begin{equation}
({\bmath \xi}_j|{\bmth {\cal C}}{\bmath\xi}_j)={\omega_{j}^2\over 2\pi}\int_0^{R_*} r^2dr \rho
(\xi_j^2+\Lambda_i (\xi^S_j)^2). \label{eq54}
\end{equation}
This expression  reduces to the standard one in the non-rotating
case when $\Lambda = 6$. Using (\ref{eq8n}),
with $\bmath\Omega$ replaced by its radial component
$(\bmath\Omega\cdot {\hat {\bf r}}) {\hat {\bf r}} $ as appropriate for the traditional approximation,
 together with
(\ref{eq43}),    we obtain
\begin{equation}
({\bmath \xi}_j|{\bmath{\cal B}}{\bmath \xi}_j)={2\over \pi}\Omega
I_rI_{\theta}, \label{eq55}
\end{equation}
where
\begin{eqnarray}
&\hspace{-5mm}I_r& = \int_0^{R_*} r^2dr \rho (\xi^S_j)^2,
\hspace{3mm}{\rm and} \hspace{3mm}I_{\theta} \hspace{2mm}= \int_{-1}^{1}{\mu d\mu
\over (1-\nu^2\mu^2)^2}\times\nonumber \\
& &\hspace{-9mm}\left[\nu \mu(1-\mu^2)\left[{dH\over
d\mu}\right]^2+m(1+\nu^2\mu^2)H{dH\over d\mu}+{m^2\nu\mu H^2\over
1-\mu^2}\right]. \label{eq56}
\end{eqnarray}
Substituting (\ref{eq54}) and (\ref{eq55}) in (\ref{eq53}) we get
$n_j=n_{st}+n_r, $
\begin{equation}
{\rm  where } \hspace{1mm}  n_{st}=\int_0^{R_*} r^2dr \rho (\xi_j^2+\Lambda_i
(\xi^S_j)^2), {\rm and}\hspace{1mm} n_r=\nu I_r I_{\theta}. \label{eq57}
\end{equation}
We remark that the integrals with respect to $r$ can be straightforwardly
evaluated once the eigenfunctions are known. The angular integral $I_{\theta}$
is a function of $\nu$ only. The   dependence  of $-I_{\theta}$ on  $\nu$
is illustrated in  Fig. \ref{Fig1}.

\section{Numerical  calculation of $\omega_{\lowercase{j}},$  $Q_{\lowercase{j}}$ and $\lowercase { n}_{\lowercase{j}}$
for   rotationally modified  gravity modes for models of
solar-type stars}\label{Numcalc}

In  this section we  describe the numerical evaluation of
the eigenfrequencies $\omega_j$ and the overlap integrals $Q_j$
for rotationally modified $g$ -modes
under the traditional approximation for stars with solar-type structure.
We also highlight aspects of the models that are relevant to
a subsequent asymptotic analysis,
\subsection{Stellar models considered}
 Our calculations are for
a rotating star with a
radiative core and extended convective envelope. The structure is assumed  not
 to be modified by rotation,  such modifications being a second order effect.
This has the advantage that  standard spherical models can be used.

\begin{figure}
\begin{center}
\vspace{8cm}\includegraphics{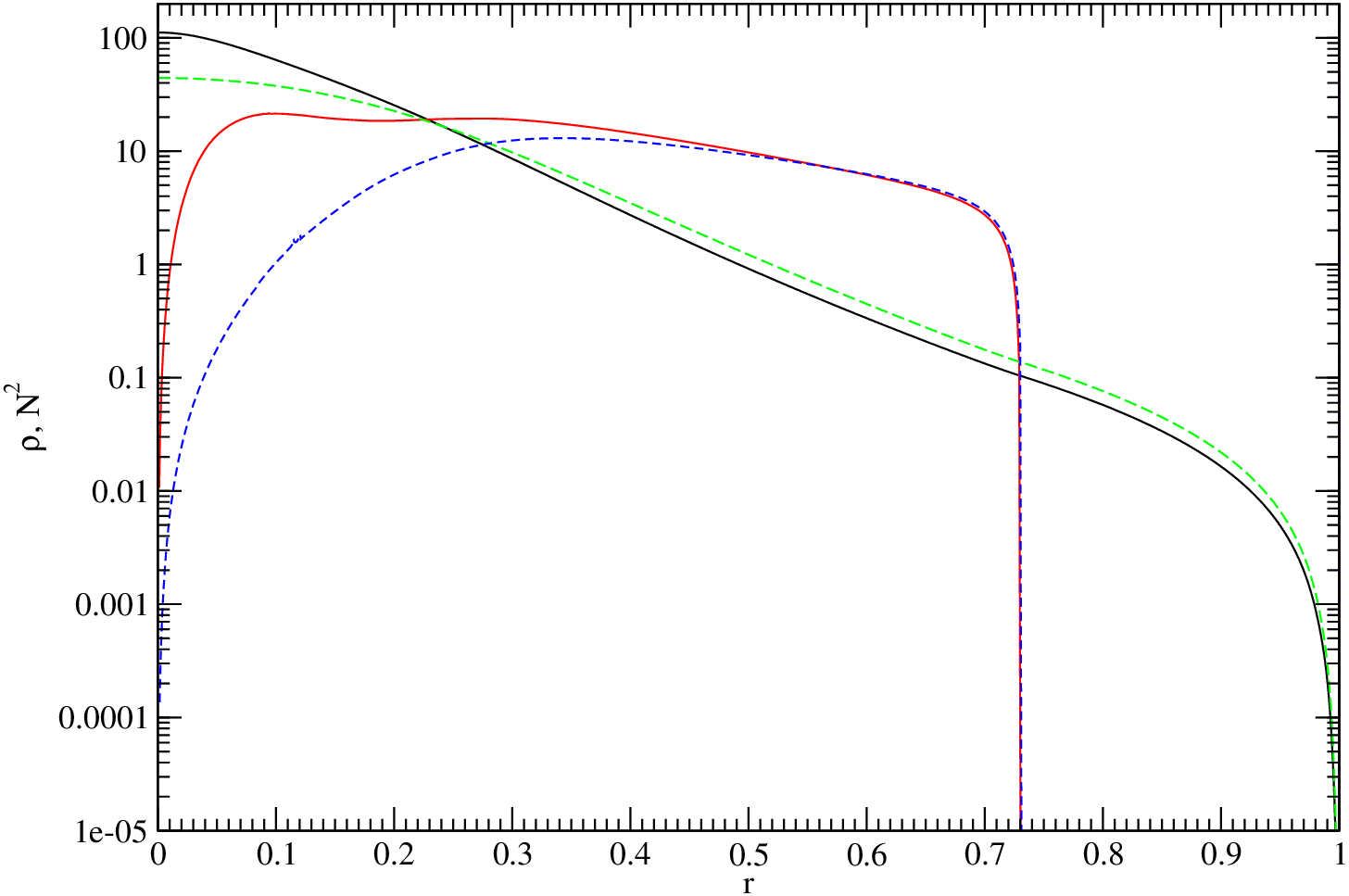}
\end{center}
\vspace{0.5cm} \caption{The density $\rho $ (in units of the
mean density $3M_*/ (4\pi R_*^3)$ and square of the Brunt - V$\ddot {\rm a}$is$\ddot {\rm a}$l$\ddot {\rm a}$
frequency,  $N^2$ (in units of $GM_*/ R_*^3$) as functions of
the radius $r$ expressed in units of $R_*.$  The solid curves correspond to the CD model while
the dashed ones are for the IR model. The curves monotonically
decreasing with $r$ are for the density distributions,  while the
curves having maxima at some values of $r$ are for the
Brunt - V$\ddot {\rm a}$is$\ddot {\rm a}$l$\ddot {\rm a}$
frequencies.} \label{Fig2}
\end{figure}

We have considered an almost zero age model of the star with
$M_*=M_{\odot}\approx 2\cdot 10^{33}g$ and age $t\approx 1.66\cdot
10^8yr$ provided to us by I.W. Roxburgh (see Roxburgh 2008 for a
description of his numerical code) referred hereafter to as IR
model, and also a model of the present day Sun
(Christensen-Dalsgaard et al. 1996) with age $t=4.6\cdot 10^9yr,$
referred hereafter to as the CD model. Plots the density $\rho $
and the square of the
 Brunt - V$\ddot {\rm a}$is$\ddot {\rm a}$l$\ddot {\rm a}$
 frequency, $N^2,$  against radius,  for these
models are given in Fig. \ref{Fig2}.
Note that the  total stellar radius of the IR model, $R_*$ is smaller
than that  of the present day Sun,  taken to be
$R_*=R_{\odot}=7\cdot 10^{10}cm$.  For the IR model we have
 $R_* = 6.3\cdot 10^{10}cm$ approximately.
\subsection{Numerical approach}

The eigenfrequencies are obtained by
finding appropriate solutions of
 (\ref{eq44}) and
(\ref{eq45}),
in the core and in the envelope and applying
matching conditions at the base of convection zone.
Since we consider low frequency modes we can neglect
the last term in the brackets multiplying $\hat P$ in (\ref{eq45})
as this is important only very close to the surface in
the limit $\omega^2 \rightarrow 0$ of interest.
The radial eigenvalue problem is solved together with the azimuthal eigenvalue  problem.
To solve the former the standard fourth order Runge-Kutta method was used.
 To solve the latter for a given eigenfrequency,  we adopted the matrix method of Ogilvie and Lin 2004.
We employed   $\sim 10^6-10^7$ grid points in the radial direction
in order to  ensure numerical convergence for eigenfrequencies
$\omega  < \sim 0.1$ in natural units. Since the CD and IR models
were provided with a much smaller number of grid points,  $1282$
and $2001$ respectively, linear interpolation was employed to
supply the data for our finer grid.
 Having obtained the eigenfunctions,
we then use  the expressions (\ref{eq52c}) and
(\ref{eq56}) to calculate  the overlap integrals and norms.
\subsection {Asymptotic  analysis in the limit $\omega \rightarrow 0.$}

In subsequent sections we shall discuss
the asymptotic form of the normal modes, overlap integrals and norms
in the low frequency limit following a WKBJ approach.
In order to proceed we first discuss the behaviour of $N^2$ in the vicinity
of the interface between the radiative and convective regions of the stellar model,
this being  important  consideration for the asymptotic analysis .

\subsection{Behaviour of the Brunt - V$\ddot {\rm a}$is$\ddot {\rm a}$l$\ddot {\rm a}$
frequency near the base of the convection zone} 
\begin{figure}
\begin{center}
\vspace{8cm}\includegraphics{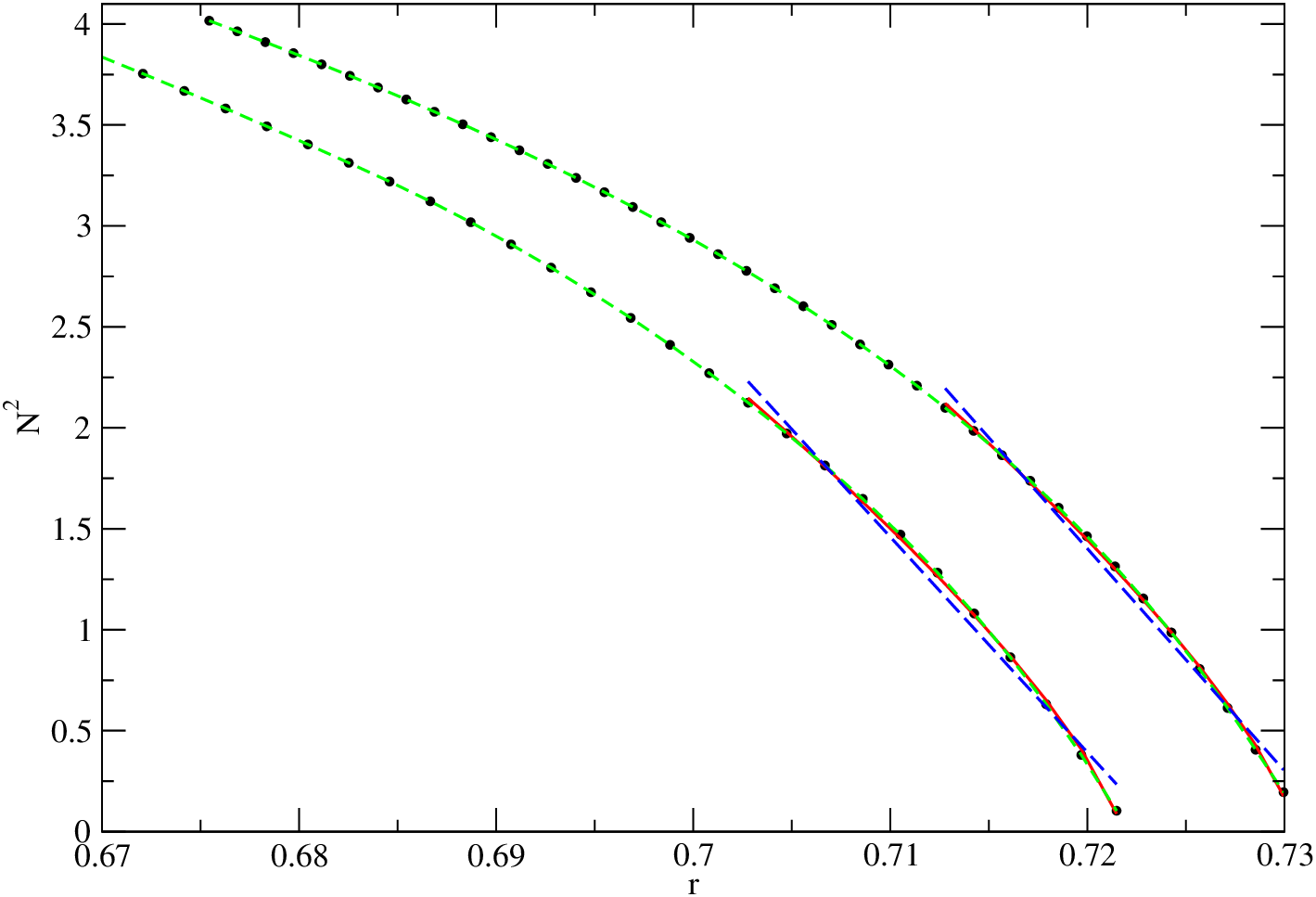}
\end{center}
\vspace{0.5cm} \caption{$N^2$ close to the base of
convection zone plotted as a function of radius,  $r,$ expressed
in units of $R_* .$  The filled circles show the  data points used to make the fits.
Those  with larger values of $N^2$ at a given $r$
correspond to the IR model. The solid and dashed curves show
the CD1, IR1,  and the  CD2,  IR2 fits  respectively. The dashed
straight
lines show the fits  CDL and IRL which had $q=1,$  see table \ref{tab:t1} for the
values of the parameters $A$ and $q$  appropriate to these fits.
\label{Fig3}}
\end{figure}

\noindent In
previous studies (e.g. Zahn 1970,  Berthomieu et al 1978, Rocca
1987) it was assumed that near the  boundary between convective
and radiative zones,  but on the radiative side,  $N^2 \propto $
the difference $x=|r_c-r| \ll r$, where $r_c$ is the radius of the
boundary ,  defined by the condition $N^2(r_c)=0$. However, this
is  not to be  the case for many stellar models (see Barker 2011).
In particular, as seen from  Fig. \ref{Fig3}, for the solar models
we use,  $N^2$ decreases more slowly  with $x$ than expected
for a  linear dependence. Here we consider  more general possibilities,  by
assuming  that when  $ r_c > r$, so that $x=r_c-r >0$,
for   $x \ll r$ we may write
\begin{equation}
N^2\approx Ax^{q}, \label{eq61}
\end{equation}
where $q$ is a constant, see also Provost $\&$ Berthomieu (1984).
 Note that as is seen from Fig. \ref{Fig3} the dependence of $N^2$
  on $x=r_c-r$ is not strictly given by a single  power law.
   Local fits result in smaller values of  $q$ at
  larger values of $x.$
 However, this feature is not very important for our
  purposes.  For our estimates of the overlap integrals below,
 only  the local value of $q$ that is found to be applicable to the range of $x$ such that $1> N^2 >  0.05$  is
 important.  This is  needed to find the contribution to  $Q_{r}$ given by
  equation (\ref{eq85}), which plays a role only when eigenfrequencies
  are in the range $\omega \sim 0.2-1$, see Figs. \ref{Fig6}-\ref{Fig7}.

We use equation (\ref{eq61}) to fit the dependence of the
Brunt - V$\ddot {\rm a}$is$\ddot {\rm a}$l$\ddot {\rm a}$ frequency
on $r$ in the CD and IR models in the vicinity of the base of
convection zone using the Levenberg-Marquardt algorithm.

\noindent  Since the dependence is
not, in fact, an exact power law we  made fits using
 two different  data sets in order to explore
the range of the resulting values of $A$ and $q.$
 For the first
data set the number of data points is constrained approximately by
the condition $N^2 < 2,$ while for  the second, and more
extended data set,   the number of points is constrained
by the condition  $N^2 < 4$.
Additionally, we made fits to the data fixing the power index  $q=1$
for the data sets with $N^2 < 2.$  The results are illustrated in Fig.
\ref{Fig3} and  tabulated in table \ref{tab:t1}.

The results labelled  CD1, CD2 and CDL
are respectively for  the fits using the data sets with $N^2 <2$, $N^2 <
4,$  and  with assumed  linear dependence of $N^2$ on $x,$
for the CD model of the star.
 The labels  IR1, IR2 and IRL
describe the same type of fits  made for the IR model.
\begin{table}
\begin{tabular}{llll}
 Model & A & q & $r_c$ \\
 CD1   & 41.54 & 0.747 & 0.722 \\
 CD2   & 24.32 & 0.617 & 0.721 \\
 IR1   & 41.4 & 0.737 & 0.731 \\
 IR2    & 23.1 & 0.564 & 0.73  \\
 CDL   & 106 & 1 & 0.724 \\
 IRL   & 110 & 1 & 0.733 \\
\end{tabular}
\caption{Results of fitting of  the dependence of $N^2$
on $r$ for the CD and IR models. Different entries in the table
correspond to either general fits made with different data sets,  or fits assuming $q=1,$  as described in the text
(see also Fig. \ref{Fig3}).  Note that $A$ and $r_c$ are given in natural units
such that  $r$ and $N^2$ are expressed in the units of $R_*$ and the
natural stellar frequency $GM_*/ R_*^3$,
respectively.\label{tab:t1}}
\end{table}

From  Fig. {\ref{Fig3} we see that the fits obtained
 with a smaller number of data points give somewhat  larger value of $q\approx 0.74$ whereas
the larger data sets are fitted quite well by
(\ref{eq61}) with $q\approx 0.6$.
This indicates that  the value $q$
 required  for the most accurate fit may decrease with the radial
 extent of the region of interest.
 This in turn may be a function of frequency when normal modes are considered.
We discuss below  how
 quantities obtained in our analysis depend on the
assumed fit.

\section {Calculation of eigenfrequencies, eigenfunctions, overlap integrals and norms
for   high order rotationally modified  gravity modes for models of
solar-type stars following a WKBJ approach}\label{WKBJcalc}
In order to provide a check on numerical calculations and to
extend results to frequencies below which the numerical approach
can run into difficulties on account of inadequate resolution,  we perform an an asymptotic
analysis following a WKBJ approach. We first obtain expressions for eigenfrequencies
and eigenmodes and then go on to use them to evaluate overlap integrals and norms.

\subsection{Asymptotic  solution of the radial eigenvalue problem providing
 expressions for the radial eigenfunctions  and
associated eigenfrequencies}
We now discuss the asymptotic form of the eigenfrequencies and  eigenfunctions
found from solving the radial eigenvalue problem in the low frequency limit
adopting a WKBJ approach.
In what follows, for simplicity  we make the assumption that
$\Gamma_1=const$ throughout  the regions of the star of interest,
  setting  $\Gamma_1=5/3$ in
numerical expressions. Note that these assumptions can be easily
relaxed if needed.
Similarly the term involving $1/c_s^2$ in (\ref{eq44})
becomes important only very close to the surface
it be neglected in the same limit.
This term  can also be taken into
account in a more advanced variant of our scheme.

\subsubsection{The WKBJ solution in the radiative core}
Under the assumptions mentioned above  equations
(\ref{eq44}) and (\ref{eq45}) can be combined to yield the  second
order differential equation
\begin{equation}
\frac{P^{1/\Gamma_1}}{\rho} {d\over dr}\left(\frac{\rho }{P^{2/\Gamma_1}}{d\over
dr}(r^2P^{1/\Gamma_1}\xi)\right)+{\Lambda }\left(\frac{ N^2 }{ \omega^2}-1\right)\xi=0.
\label{eq58}
\end{equation}
An approximate WKBJ  solution in the radiative core,  sufficiently
far from the centre that  $N^2/\omega^2 - 1$ may be replaced by
$N^2/\omega^2 $ in (\ref{eq58}),  has the standard form (see e.g.
Christensen-Dalsgaard 1998)
\begin{eqnarray}
\xi&=&C_{WKBJ}r^{-3/2}(\rho N)^{-1/2}\sin ({W/
\omega}+\phi_{WKBJ}), \hspace{1mm}{\rm where}\nonumber \\
 W&=&\Lambda^{1/2}\int^{r}_{0}{dr\over
r}N, \label{eq59}
\end{eqnarray}
 $C_{WKBJ}$ is a constant and the angle $\phi_{WKBJ}$ is
fixed by the condition that the solution matches onto a  regular one  as
$r\rightarrow 0$ as
\begin{equation}
\phi_{WKBJ}=-{l\pi\over 2}, \hspace{1mm} {\rm with} \hspace{1mm}  l={1\over
2}(\sqrt{1+4\Lambda}-1), \label{eq60}
\end{equation}
see e.g. Vandakurov (1968)\footnote{It can be shown that close to
the centre we have approximately: $\xi_{core}\propto
r^{-3/2}J_{l+1/2}(C_{core}r)$, where $J_{\nu}(z)$ is a Bessel
function and $C_{core}$ is a constant. This solution is regular
when $l \ge 1$. At sufficiently large $r$ the solution should be
matched to (\ref{eq59}). This determines the phase
$\phi_{WKBJ}$.}. Note that $l$ is not, in general, an integer.
When the star is non-rotating we recall that $\Lambda=6,$
and, accordingly, $l=2$.

\subsubsection{The solution close to the base of convective envelope}
We look for a solution that is valid in a radial region of extent
$r  \ll r_c$ in the radiative region close to its boundary with the
convection envelope. To do this we substitute (\ref{eq61}) in
(\ref{eq58}),  neglecting unity in the brackets in the last term,
the variation of $P$ and $\rho,$  and recalling that $x=r_c-r.$
Equation (\ref{eq58})  then reduces to
\begin{equation}
{d^2 \xi \over d x^2}+{\Lambda A x^q\over r^2_c\omega^2}\xi=0.
\label{eq62}
\end{equation}
The general  solution of (\ref{eq62}) is  a linear combination of
Bessel function $J_{\nu}(z)$ such that
\begin{eqnarray}
\xi&=&\sqrt{x}(C_1J_{{1\over q+2}}(z)+C_2J_{-{1\over q+2}}(z)),\hspace{1mm}
{\rm where }\nonumber \\
 z&=&{K\over \omega} x^{{{q+2}\over{ 2}}}, \quad
 K={2\sqrt {\Lambda A}\over (q+2)r_c}.
\label{eq63}
\end{eqnarray}
Note that the expression (\ref{eq63}) is equivalent to what is
obtained in Provost $\&$ Berthomieu (1984) in the limit of small
$x$ provided that one discards
 higher order  terms as discussed in their paper.

The solutions (\ref{eq59}) and (\ref{eq63}) must be matched in a
region where $x/r \ll 1$ but $z \gg 1$. This is easily
accomplished  by the standard method of comparing the former with
the latter after utilising  the asymptotic expansion of the Bessel
functions for large $z.$
 The matching procedure
allows us to express the constants $C_1$ and $C_2$ in (\ref{eq63})
in terms of  $C_{WKBJ}$ through
\begin{eqnarray}
C_1&=&-\sqrt{{\pi K\over 2\omega}}C_*{\cos(\phi_c-{\pi q/
(4(2+q))})\over \sin({\pi/( 2+q)})}, \nonumber \\
 C_2&=&\sqrt{{\pi K\over
2\omega}}C_*{\cos(\phi_c-{\pi (4+q)/( 4(2+q))})\over
\sin({\pi/(2+q)})}, \label{eq65}
\end{eqnarray}
where $C_*=C_{WKBJ}\rho_{c}^{-1/2}r_c^{-3/2}A^{-1/4}$, and
\begin{equation}
\phi_c=\phi_{WKBJ}+{\sqrt{\Lambda}\over
\omega}\int_{0}^{r_c}{dr\over r}N. \label{eq66}
\end{equation}

But note that the approximations that lead to   equation (\ref{eq62}) imply that this  equation  must
be modified when $N^2\le \omega^2$, which corresponds to
\begin{equation}
x < x_{crit}=A^{-{1\over q}}\omega^{{2\over q}}. \label{eq67}
\end{equation}
On the other hand from (\ref{eq63}) it follows that
when $z \ll 1$ the solution approximately satisfies   ${d^2\xi /dx^2}=0$
and the neglected term cannot produce significant changes over  the small shrinking domain $(0,x_{crit})$
as $\omega \rightarrow 0.$

\subsubsection{The solution in the convective envelope}

 Extending  the solution into the convective region,
 we can set approximately $N^2=0$.  The stellar gas, therefore,
can be approximated as having barotropic equation of state with $P=K_c\rho^{\Gamma_1},$
which  corresponds  to  a polytrope  with
polytropic index equal to $1.5$.   Assuming that the convective
envelope has a negligible  mass compared to that  of the star,
of mass $M_*,$
 the hydrostatic balance equation takes the form
\begin{equation}
{dP\over dr}=-{GM_*\rho \over r^2}.
\label{eq68}
\end{equation}
Solving (\ref{eq68})
to  find the dependencies of  the pressure and density on $r,$ assuming these vanish at $r=R_*,$
we obtain
\begin{equation}
\rho=\rho_*\left({R_*\over r}-1\right)^{{1\over \Gamma_1-1}}, {\rm with}\hspace{2mm}
 \rho_*=\left({(\Gamma_1 -1)GM_*\over K_c\Gamma_1 R_*}\right)^{{1\over \Gamma_1-1}}.
\label{eq69}
\end{equation}
Setting $N^2=0$ in equation (\ref{eq58}), using the polytropic  equation of state and equation
(\ref{eq69}) we obtain
\begin{equation}
{d^2 \Phi_c\over dr^2}+{1\over (\Gamma_1-1)r(1-r/R_*)}{d\Phi_c \over dr}
-{\Lambda \over r^2}\Phi_c=0,
\label{eq70}
\end{equation}
where $\Phi_c=r^2\rho \xi$. We look for a solution to (\ref{eq70}), which tends to zero when
 $r\rightarrow R_*$. This solution is expressed through
the hypergeometric function, $F(\alpha,\beta,\gamma,z)$, as
\begin{equation}
\Phi_c=C_cr^ax^{{\Gamma_1\over \Gamma_1-1}}F(b_1,b_2,b_3,x),
\label{eq71}
\end{equation}
where $x=1-r/R_*$,
\begin{eqnarray}
a&=&{1\over 2}\left(-{2-\Gamma_1\over \Gamma_1-1}+\sqrt{\left({2-\Gamma_1\over \Gamma_1-1}\right)^2+
4\Lambda}\right), \nonumber \\
b_{1,2}&=&{1\over 2}(2a-1\pm\sqrt{1+4\Lambda})+{\Gamma_1\over \Gamma_1 -1},\hspace{2mm} {\rm and} \nonumber \\
b_3&=&{2\Gamma_1 -1\over \Gamma_1-1}
\label{eq72}
\end{eqnarray}
and $C_c$ is an arbitrary constant. As we discuss in the next section,
for our purposes it is important to know the quantity
\begin{equation}
B_c=-\left({{d\xi / dr}\over \xi}\right)_{r=r_c},
\label{eq73}
\end{equation}
calculated for the solution discussed  above, at the base
of the convective  envelope,  as a function of $\Lambda$. This
quantity  can be  calculated  from (\ref{eq71}).
It will be seen from the above analysis that this quantity depends on the
eigenfrequency only through its dependence on $\Lambda.$
The  dependence $B_c$  on $\Lambda,$   for
$\Gamma_1=5/3$ and $r_c/R_*=0.7,$  is illustrated in Fig. \ref{Fig4}.
\begin{figure}
\begin{center}
\vspace{8cm}\includegraphics{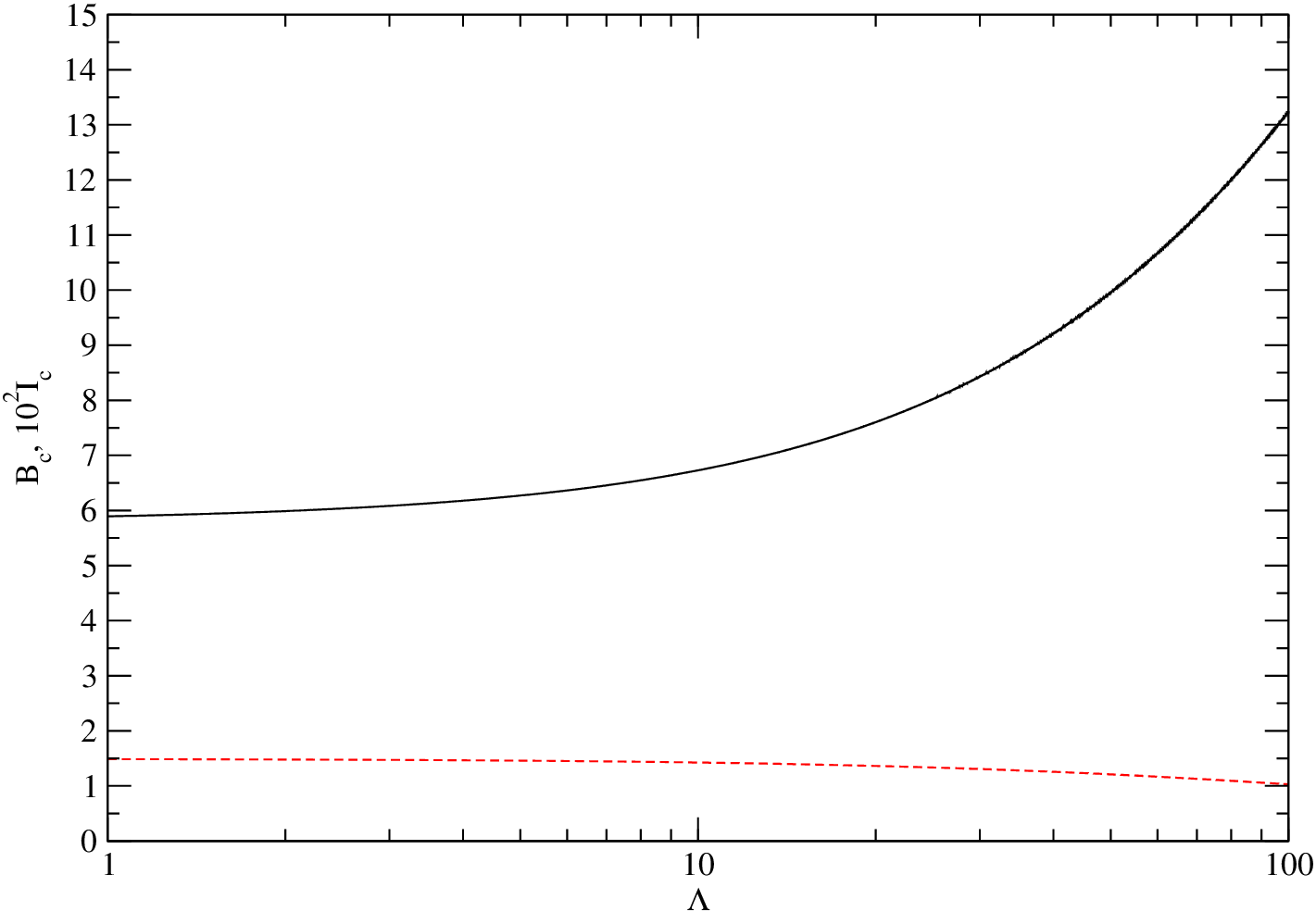}
\end{center}
\vspace{0.5cm} \caption{The quantities $B_c$ (\ref{eq73}) and
$10^2 I_c$ (see (\ref{eq91}))  determined from  the solution (\ref{eq71}) as
functions of $\Lambda $ for  $r_c  = 0.7R_*$. $B_c$ is
shown as the solid curve.  Note that
the values of $B_c$ and $I_c$ found from the asymptotic
analysis are in quite good agreement
with  those  obtained  using numerically calculated  eigenfunctions.}
\label{Fig4}
\end{figure}

\subsubsection{Matching conditions and eigenfrequencies}

In order to mach the solutions in the radiative zone and  the convective envelope we use
the solution (\ref{eq63}) in the limit $z\rightarrow 0$   to calculate
$-(d \xi / dr)/ \xi$  as the base of the convective envelope is approached from below.
 Then, we equate the resulting expression to (\ref{eq73}) which gives the same
quantity as the base of the convective envelope is approached from above.
This  gives  a compatibility relation, which will be satisfied
only when $\omega$ is an allowed eigenfrequency.
In the limit $z\rightarrow 0$ equation  (\ref{eq63}) gives
\begin{eqnarray}
&\xi& \approx C_1 ({K/( 2\omega) })^{{1/( 2+q)}}{x\over \Gamma({(3+q)/(2+ q)})}+\nonumber  \\
&C_2&\hspace{-3mm}({K/( 2\omega) })^{-{1/(2+q)}}{1\over \Gamma({(1+q)/( 2+q)})},
\label{eq74}
\end{eqnarray}
where $\Gamma(z)$ denotes  the gamma function. We then find
$-(d\xi /dr)/ \xi$  in the limit $z\rightarrow 0$ from
(\ref{eq74}) and equate  it to the same quantity determined from
the solution in the convective envelope given by equation
(\ref{eq73}).

Equation  (\ref{eq65}) can be used to specify the
ratio   $C_1/C_2$ which, after its elimination from the expression,
yields   a compatibility condition from which the eigenfrequencies
can be determined. This compatibility condition can be written down as
\begin{equation}
\hspace{-1mm}\frac{\cos(\phi_c-{\pi q/(8+4q)})}{\cos
(\phi_c-{\pi(4+q)/(8+4q)}} =-{{\Gamma ((3+q)/( 2+q))}\over {\Gamma
((1+q)/(2+q ))}} B_c {\bar \omega}^{{2\over  2+q}}, \label{eq75}
\end{equation}
where ${\bar \omega}=   2\omega / K$,  and $\phi_c$ is given by
equation (\ref{eq66}).

In the limit $\omega \rightarrow 0$ the term on the right hand side of (\ref{eq75}) is
small. This allows us to solve equation (\ref{eq75})
iteratively. First  we set  the term on the right hand side to zero.
\footnote{This   is equivalent to setting
 $C_1=0$ which   corresponds to  the radial eigenfunction
having its derivative close to zero at the base of convection zone}.
 Then we find that we must have
$\phi_c=\pi (n+1/2+q/(4(2+q)))$, where $n$ is an integer.
 Using  this  result in  equations (\ref{eq60}) and (\ref{eq66}) we obtain
\begin{equation}
\omega\approx \omega_{0}={\sqrt \Lambda I\over \pi (n+(1+l)/ 2+q/( 4(2+q)))},
\label{eq76}
\end{equation}
where
\begin{equation}
I=\int^{r_c}_0{dr\over r}N.
\label{eq77}
\end{equation}

Now let us
take account of a small non zero $B_c$
by setting $\phi_c=\pi (n+1/ 2+q/ 4(2+q))+\Delta \phi$, where $\Delta \phi $ is assumed
to be small. We substitute  this expression into (\ref{eq75})
and solve the resulting equation  for  $\Delta \phi $ using perturbation theory
to linear order,   thus  obtaining
\begin{equation}
\Delta \phi =B_c({2\omega / K})^{2/ (2+q)}{\Gamma ((3+q)/(
2+q))\over \Gamma ({(1+q) /(2 +q)})}\sin ({\pi /(2+q)}),
\label{eq78}
\end{equation}
and accordingly
\begin{equation}
\omega ={\sqrt \Lambda I\over \pi (n+(1+l)/ 2+q/ (4(2+q)))+\Delta \phi(\omega_0)}.
\label{eq79}
\end{equation}
We recall   that for rotating stars equation (\ref{eq79}) gives
only an implicit equation for eigenfrequencies since in that case
$\Lambda $ is a function of $\omega $.

\begin{figure}
\begin{center}
\vspace{8cm}\includegraphics{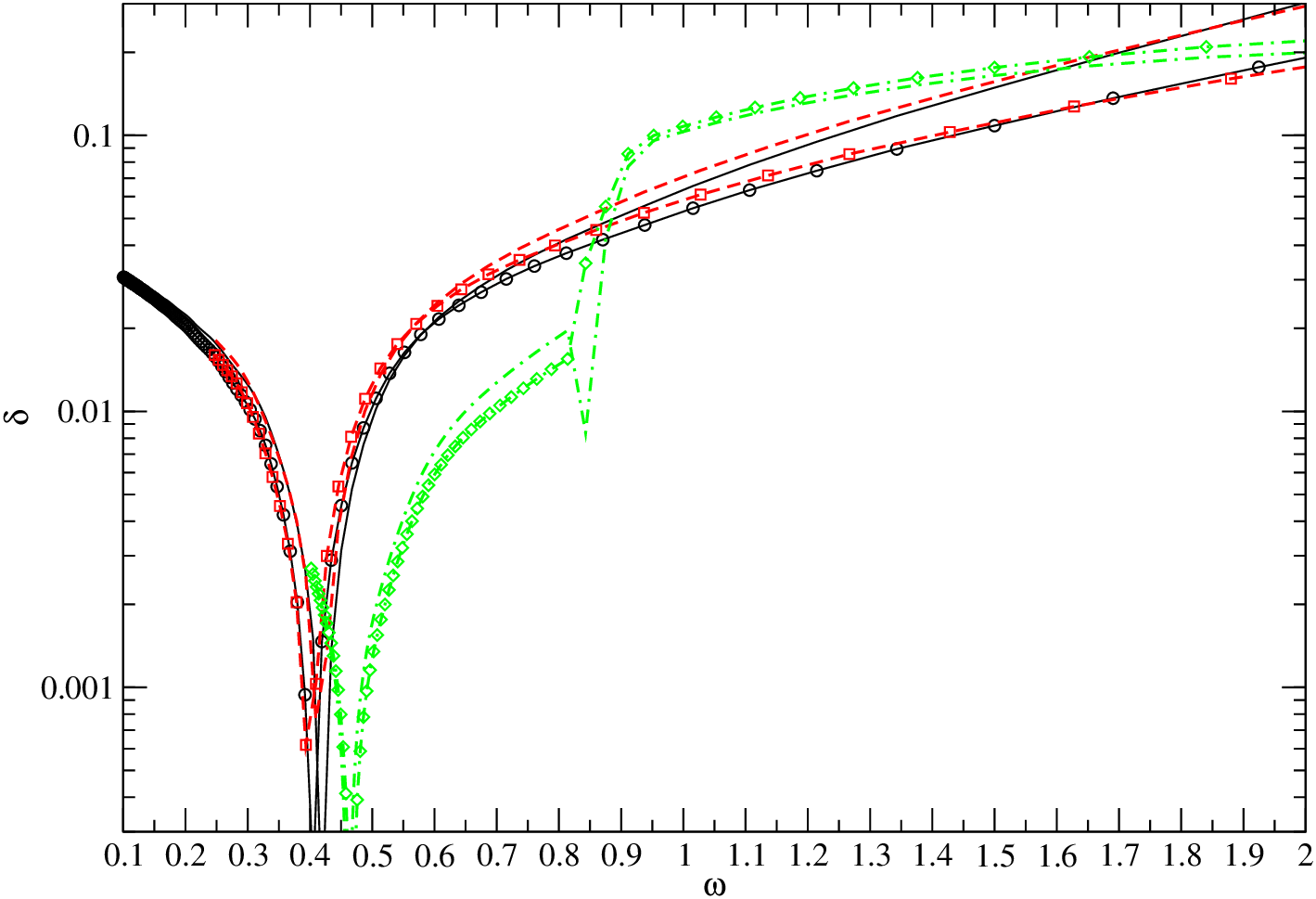}
\end{center}
\vspace{0.5cm} \caption{The difference $\delta
=|\omega_{WKBJ}-\omega_{n}|/\omega_{n}$ between the WKBJ
eigenfrequencies $\omega_{WKBJ}$ and the corresponding frequencies determined numerically $\omega_n.$
The results are plotted as a function of $\omega_n$ calculated for the CD model of the
present day Sun and expressed in units of $\sqrt{ GM_*/R_*^3}.$
 The value of $I$ defined in equation (\ref{eq77})
is approximately equal to $14.91\sqrt{{CM_*\over R_*^3}}$. Since
all the  fitting models shown in table \ref{tab:t1} give quite similar
curves we show only the results corresponding to the CD2 model.
 See the text for the description of different
curves.} \label{Fig5}
\end{figure}
We remark that for low frequencies the influence of the convective
envelope solution on the spectrum, through $B_c,$   provides only a small correction
to the eigenspectrum. This is consistent with the idea that the properties of
the rotationally modified $g$ mode spectrum are almost completely  determined
by the radiative zone.

We now  compare results obtained from  our analytic expression for the eigenfrequencies
given by equation (\ref{eq79}) with those obtained
 from an  expression for $\omega$ given
in Berthomieu et al (1978) and also  with those obtained from numerical calculations
 for the CD model of the Sun in Fig.
\ref{Fig5}. The
simple expression obtained by Berthomieu et al (1978) differs
from equation (\ref{eq79}) by replacing  the  denominator
 in that equation with  $\pi(n+{1\over 2}(1+\sqrt{\Lambda + {1\over 4}}))$.

In Fig. \ref{Fig5} we show the quantity $\delta
=|\omega_{WKBJ}-\omega_{n}|/\omega_{n}$, where $\omega_{WKBJ}$ and
$\omega_n$ stand for the eigenfrequencies determined from the WKBJ
analysis and numerical calculation respectively, as a  function of
$\omega_n$ expressed in units of $\sqrt{GM_*/R_*^3}.$  The curves
with symbols showing positions of the eigenfrequencies are
calculated from our analysis,  while the curves with no symbols
are for eigenfrequencies determined from the expression of
Berthomieu et al (1978).
The solid, dashed and dot dashed curves are for $\Omega=0,$
$\Omega=0.42\Omega_*$ and $\Omega=-0.42\Omega_*$, respectively.  It is
seen that the deviation between the numerical and
analytical results does not vary  monotonically with   $\delta $ increasing as
 $\omega $ becomes very small. This feature is most probably a consequence of numerical errors.
The curves corresponding to $\Omega=0$ and $\Omega=0.42\Omega_*$
are similar to each other, this can be explained by  the rather
weak dependence of $\Lambda $ on $\nu$ when $\nu
={2\Omega\over \omega}$ is positive, see Fig. \ref{Fig1}.
Since the WKBJ approximation gets better with increasing
numbers of nodes $n$, which is, in its turns, gets larger with
increasing  $\Lambda$, the deviation $\delta $ corresponding to
those curves is appreciably larger than the deviation corresponding to
 retrograde rotation at $\omega < 0.8$ because  $\Lambda $
increases significantly with decreasing  $\omega $ in that case, see Fig.
\ref{Fig1}.

As seen from Fig. \ref{Fig5} our approach gives  deviations from
the numerical results  that are smaller than those of  Berthomieu et al
(1978) for  prograde rotation and no rotation.  However,
the difference is typically rather moderate being of the order of
several tens per cent.
For  the case of retrograde rotation and
large values of $\omega$ the Berthomieu et al. (1978) expression
gives somewhat better agreement though the difference is small.
When $\omega < 0.847\Omega_*$ our approach is closer to the
numerical one than that of Berthomieu et al. (1978). Note a
non-monotonic behaviour of the curve corresponding to the
Berthomieu et al (1978) expression close to $\omega =
0.847\Omega_*$. This value of $\omega$ corresponds to $\nu=1$,
which is a special value of this parameter delimiting the modes
dominated by the gravity ($\nu < 1$) and Coriolis forces ($\nu >
1$). We suspect that this feature may be purely numerical because of
 the  behaviour of  the singularities of  equation  (\ref{Laplace0}),  determining
$\Lambda,$ as $ \nu$ passes through unity.

For our estimates of the overlap integrals given  below it is convenient
to  substitute the explicit expression for the phase $\phi_c$
calculated above into equations (\ref{eq65})  for the constants
$C_1$ and $C_2.$  Making use of  (\ref{eq78}) to do this,  we
easily obtain
\begin{eqnarray}
C_1\hspace{-3mm}&=&\hspace{-3mm}(-1)^n\sqrt{{ {\pi K\over 2\omega}}}
B_c \left( {2\omega  \over K}\right)^{2/ (2+q)}{\Gamma ({(3+q)/ (2+q)})\over \Gamma ({1+q)/(2+q)})}C_* , \nonumber \\
C_2\hspace{-3mm}&=&\hspace{-3mm}(-1)^n\sqrt{{{\pi K\over
2\omega}}} \left(1-B_c\left({2\omega \over K}\right)^{2/
(2+q)}\right.
\times  \nonumber \\
&& \left. {\Gamma ((3+q)/ (2+q))\over
 \Gamma ((1+q)/( 2+q))} \cos({\pi / ( 2 +q)})\right)C_*.
\label{eq80}
\end{eqnarray}

\subsection{Asymptotic expressions for the normalised overlap integrals $\hat Q_j$ }
Having determined eigenfrequencies and eigenfunctions, we can use them
to estimate the normalised overlap integrals determining the strength of  tidal interactions. We
evaluate the unnormalised   $Q$ derived from  (\ref{eq31}),  and the norm $n$
given by (\ref{eq57})( from now on we  omit the mode  subscripts
 $(i, j)$ below unless absolutely necessary).

The overlap integral is written as $Q=\alpha Q_r.$
The factor $\alpha,$ coming from the angular integration  is  calculated numerically
and, as it does not depend on the radial form of the eigenfunctions,
 it is assumed to be  known function of frequency (see Fig. \ref{Fig1}).
As  we are working in the low frequency limit,   we  use the reduced form   (\ref{eq52d}) to evaluate
the factor  $Q_r.$
It is important to note that the evaluation of this  expression
 can be simplified for  solar type
stars.  For these,  the  main contribution to the integral comes from,   either the transition region between
the radiative core and  the convective envelope,  or from  the convective envelope itself. We discuss these
contributions separately below.  Note that in view of various integrations by parts,  these contributions
do not necessarily relate to physical contributions from the regions concerned.

In what follows we adopt  natural units, expressing
radii in units of the  stellar radius $R_*$, mass in units of
$M_*$, frequencies  in units of $\sqrt{GM_*/ R_*^3}$ and density
in units of the stellar mean  density $\rho_*={3/(4\pi)}(M_*/
R_*^3)$. In these natural units the factor $GM_*$ in (\ref{eq81})
can be set equal to $1$. The normalised  overlap integrals have dimension of
$\sqrt{M_*}R_*.$  Accordingly we  express them in these units.
From equation (\ref{eq31}),  given the fact that we express the density in
units of the mean stellar density,  it follows that expressions for the
normalised overlap integrals obtained with quantities
expressed in our  units, must be additionally
multiplied by factor of $\sqrt{3 /( 4\pi)}$. This will be
assumed implicitly.

\subsection{An expression for the norm}

 We consider the expression (\ref{eq57}) which gives the norm $n$
 written as the sum of two contributions  such that $n=n_{st}+n_{r}.$
 In the limit  $\omega \rightarrow 0,$ the  main contribution to the
radial integrals in (\ref{eq57}) is expected to come from  the region in the radiative
core where the radial wavelength and group velocity are smallest
and accordingly radial displacement is given by the WKBJ expression
(\ref{eq59}). In the same limit and in the same region of the
star,  $|\xi^{S}|\gg |\xi|$ as expected for low frequency modes in a
stratified medium. Thus it  follows from equations
(\ref{eq46}), (\ref{eq44})  and (\ref{eq59}) that
\begin{eqnarray}
\hspace{-30mm}&\xi^S&\hspace{-3mm}\approx {r\over \Lambda }\xi^{'}\approx {C_{WKBJ}\over\Lambda^{1/2}
 \omega}\rho^{-1/2}r^{-3/2}N^{1/2}
\cos(\Psi_{WKBJ}) \hspace{2mm} {\rm with} \hspace{-3cm}\nonumber \\
&&\Psi_{WKBJ}={W\over \omega}+\phi_{WKBJ}.
\label{eq82}
\end{eqnarray}
Substituting this expression into the integrand  determining $n_{st}$ in (\ref{eq57}),
 neglecting the contribution
proportional to  $\xi$ there and adopting  the average value of $\cos^2 (\Psi_{WKBJ})=
{1\over 2}$ we get
\begin{equation}
n_{st}\approx {C^{2}_{WKBJ}\over 2\omega^2}I, \label{eq83}
\end{equation}
where $I$ is given by equation (\ref{eq77}).
Since $n_r= \nu I_r I_{\theta},$ with the integral $I_r,$
defined through (\ref{eq56}),  readily being seen to be approximately equal to $n_{st}/\Lambda $
on account of $|\xi^{S}|\gg |\xi|,$ we finally obtain
\begin{equation}
n=n_{st}+ n_r \approx  {C^{2}_{WKBJ}\over 2\omega^2}I\left(1+{\nu
I_{\theta}\over \Lambda}\right), \label{eq84}
\end{equation}
where $I_{\theta}$ is defined through equation (\ref{eq56}). We recall  that in general the latter has to be
calculated numerically.

\subsection{The estimate of the overlap integral  and  $Q_r$}

In order to calculate the integral   in the expression for $Q_r,$ we use the
approximate expression (\ref{eq52d}).  The integrand in that expression is proportional to  the function
 $h(r)$ defined there.  We remark
 that the main contribution to the integral
  comes from  regions  close to the boundary
 between the radiative zone and the convective envelope where the eigenmodes
 have their largest scale. The term
proportional to the Brunt - V$\ddot {\rm a}$is$\ddot {\rm a}$l$\ddot {\rm a}$  frequency in (\ref{eq52d}) is small in
these regions and we set in to zero. Also, as  indicated above,
the mass of the star
 may be taken  to be constant
in these regions so that $g\approx GM_*/r^2.$ Substituting this
into the expression for  $h(r)$ in (\ref{eq52d}), we obtain the
useful and simplified form
\begin{equation}
h(r)\approx {r^2\over GM_*}\left(1-{30\over \Lambda }\right).
\label{eq81}
\end{equation}

 The integral may
be regarded as being  composed of   two contributions,  which should be evaluated separately,
 the first comes from  the radiative region close to the base of
 convective envelope,  and the second from the envelope
itself.

\subsubsection{The contribution  to the integral from  the  radiative region close to the
base of the convective envelope}

In order to calculate this contribution we set all quantities in the
integrand,  multiplying the radial displacement $\xi,$ equal to their
values at $r=r_c$ and assign the subscript $(c)$ to them,
e.g. $\rho_c=\rho(r_c)$.
We assume that the distance from the base,
$x=r_c - r > 0,$ is small and change the integration variable from $x$ to  $z$  as defined
in equation (\ref{eq63}).
 We substitute
$\xi$ as given by  (\ref{eq63}) into the integral  in (\ref{eq52d}). Taking the lower limit
to be zero   and formally
extending the upper limit of integration to infinity gives the contribution from the radiative region.
 Proceeding in this way we get a
linear combination of two integrals containing Bessel functions
and powers of $z$. These integrals can be evaluated by standard
methods  using   Gradshteyn $\&$  Ryzhik (2007) giving the
contribution to $Q_r$ from the radiative zone, $Q_{rr}$ as
\begin{eqnarray}
&Q_{r,r}&=\left(1-{30\over \Lambda}\right){2^{3/(2+q)}\over
  (2+q)}\rho_c r_c^6D_c^{3/2}\omega^2\times\nonumber \\
 & &\left ( C_1{\Gamma ({2/( 2+q)})\over
  \Gamma({(1+q)/( 2+q)})}+C_2{\Gamma ({1/( 2+q)})\over
  \Gamma({q/( 2+q)})}\right),
\label{eq85}
\end{eqnarray}
where $D_c=({\omega / K})^{{2/( 2+q)}}$ and we recall  that $K$
is defined in (\ref{eq66}).
By construction this  expression does not include the contribution
to the integral from the convective envelope
which is considered separately below.
 Next we substitute
$C_1$ and $C_2$ as given by (\ref{eq80}) and use (\ref{eq65})  to
replace  $C_*$ by $C_{WKBJ}$ in the  expression for $Q_{r,r}.$  In this
way we  obtain
\begin{eqnarray}
&\hspace{-8mm} Q_{r,r}&\hspace{-8mm}=\left(1-\frac{30}{
\Lambda}\right)Q_{b}\left ({\Gamma ({1/( 2+q)})\over
  \Gamma({q/( 2+q)})}\left (1-\Delta_{corr}\cos (\pi /( 2+q)\right)\right.\nonumber\\
&\hspace{-3mm}+\Delta_{corr}&\hspace{-3mm}\left.{\Gamma ({2/ (2+q)})\over
  \Gamma({(1+q)/( 2+q)}) }\right),
\label{eq86}
\end{eqnarray}
where
\begin{eqnarray}
\Delta_{corr}&=&{\Delta \phi \over \sin
(\pi/(2+q))}\nonumber \\
&=&B_c\left({2\omega \over K}\right)^{2/( 2+q)}{\Gamma
({(3+q)/( 2+q)})\over \Gamma ({(1+q)/ (2+q)})}, \label{eq87}
\end{eqnarray}
see equation (\ref{eq80}), and
\begin{equation}
Q_b=\frac{(-1)^{n}(K/2)^{-{4-q\over
2(2+q)}}\sqrt{\pi}}{(2+q)A^{1/4}}
\rho_c^{1/2}r_c^{9/2}\omega^{{3(4+q)\over 2(2+q)}}C_{WKBJ}.
\label{eq88}
\end{equation}
Expressions (\ref{eq86}-\ref{eq88}) determine the contribution of the
radiative region close to the convective base to the overlap integral. Note that
$\Delta_{corr}\ll 1$ when $\omega $ is small. It thus  represents a
correction to the leading term.

\subsubsection{Contribution to the overlap integral from the convective envelope}

In order to find this contribution we first  determine
the constant $C_c$   that scales the solution (\ref{eq71}),
applicable in the convective envelope,
in terms of  the constant $C_2$ that scales
the solution in the radiative region.
This can be done by  applying the condition
that  the  radial displacement $\xi $ be continuous  at
$r=r_c$.
 Using (\ref{eq71}) and (\ref{eq74}) evaluated at $x=0$
to obtain the displacements just outside and just inside $r=r_c$
respectively, we find that the continuity condition gives
\begin{equation}
C_c=C_2({K/ 2\omega })^{-{1\over 2+q}}{1\over \Gamma ((1+q)/( 2+q))\zeta},
\label{eq89}
\end{equation}
where $\zeta=(r_c^{a-2}/ \rho_c)x_{c}^{{\Gamma_1\over
\Gamma_1-1}}F(b_1,b_2,b_3,x_c)$ and $x_c=1-r_c$.
After using (\ref{eq89})  to substitute for $C_c$  in (\ref{eq71})
 we substitute the result  into  (\ref{eq52d}).  Making use of
(\ref{eq65}), (\ref{eq80}) and (\ref{eq81}) we  evaluate the integral to  obtain
\begin{eqnarray}
&Q_{r,c}&=(1-\frac{30}{ \Lambda})(1-\Delta_{corr}\cos({\pi\over
2+q})) Q_{b_c}, \hspace{2mm} {\rm where}  \nonumber \\
& & \hspace{-9mm} Q_{b,c}= (-1)^{n}{\sqrt{\pi \rho_c}{({K/
2})}^{q\over 2(2+q)}I_c \omega^{{8+3q\over 2(2+q)}}\over
\Gamma({(1+q)/(2+q)})r_{c}^{3/2}A^{1/4}}C_{WKBJ},  \label{eq90}
\end{eqnarray}
with
\begin{equation}
I_c={\int_{r_c}^{1}drr^{4+a}x_{c}^{{\Gamma_1\over
\Gamma_1-1}}F(b_1,b_2,b_3,x)\over
r_{c}^{(a-2)}x_{c}^{{\Gamma_1\over
\Gamma_1-1}}F(b_1,b_2,b_3,x_c)}. \label{eq91}
\end{equation}
The integral $I_c(\Lambda )$ is plotted as a function of $\Lambda$
in Fig. \ref{Fig4}. It will be  seen that $I_c$  depends rather weakly
on $\Lambda $ in the considered range,  being approximately equal to
its value for a non-rotating star, namely  $1.45\cdot 10^{-2}$.

Note that when $\Lambda $ is assumed to be constant (as  in
the non-rotating case) and $q=1,$  from (\ref{eq90}) it follows that
$Q_r\propto \omega^{{11\over 6}}$. This scaling  was obtained by
Zahn (1970) using  a different formalism.

\subsubsection{Analytic expressions for the  normalised overlap integrals}

The
normalised overlap integrals entering the expressions for the tidal
energy and angular momentum transfer given by (\ref{eq33a})
are related to $Q_r, \alpha,$ and $n,$ through
\begin{equation}
\hat Q_j=\alpha \hat Q_r, \quad {\rm where} \hspace{2mm} \hat Q_r={Q_{r}\over \sqrt n}.
\label{eq92}
\end{equation}
Note that  $\alpha $ is defined through equation (\ref{eq50}) and
its form for the dominant mode $(i=1)$ is shown in
Fig. \ref{Fig1}.   The quantities  $Q_r$ can be specified as
$Q_r=Q_{r,r}+Q_{r,c},$
where $Q_{r,r}$ and $Q_{r,c}$ are given in equations (\ref{eq86})
and (\ref{eq90}), respectively and we recall that $n$ is given by (\ref{eq84}).

In order to evaluate these expressions,  along with $\alpha,$
we require $ \Lambda,$  both of which are shown in Fig. \ref{Fig1},  $ B_c,$
and $I_c,$ the latter two quantities being
plotted in Fig. \ref{Fig4}.  In addition to these we require the stellar
model dependent quantities, $I, A, q, r_c,$ and $\rho_c.$
These are specified below and in table \ref{tab:t1}.
The above information enables straightforward calculation of
 analytic estimates for the normalised overlap integrals for rotating stars,  which
match well the numerically determined ones,  for forcing frequencies
less than the critical rotation period.


The values of the integral $I$ are taken to be
equal to $14.91$ and $5.898$ for the CD and IR models,
respectively. The density $\rho_c,$
in standard units,
at the base of convection zone
is equal to $0.1048$ in the CD model and to $0.1375$ in the IR
model. It is assumed that the base of the convection zone is
situated at $r_c=0.729$ in all cases.

\subsection{Quantitative comparison of the the overlap integrals
obtained numerically and analytically}

\begin{figure}
\begin{center}
\vspace{8cm}\includegraphics{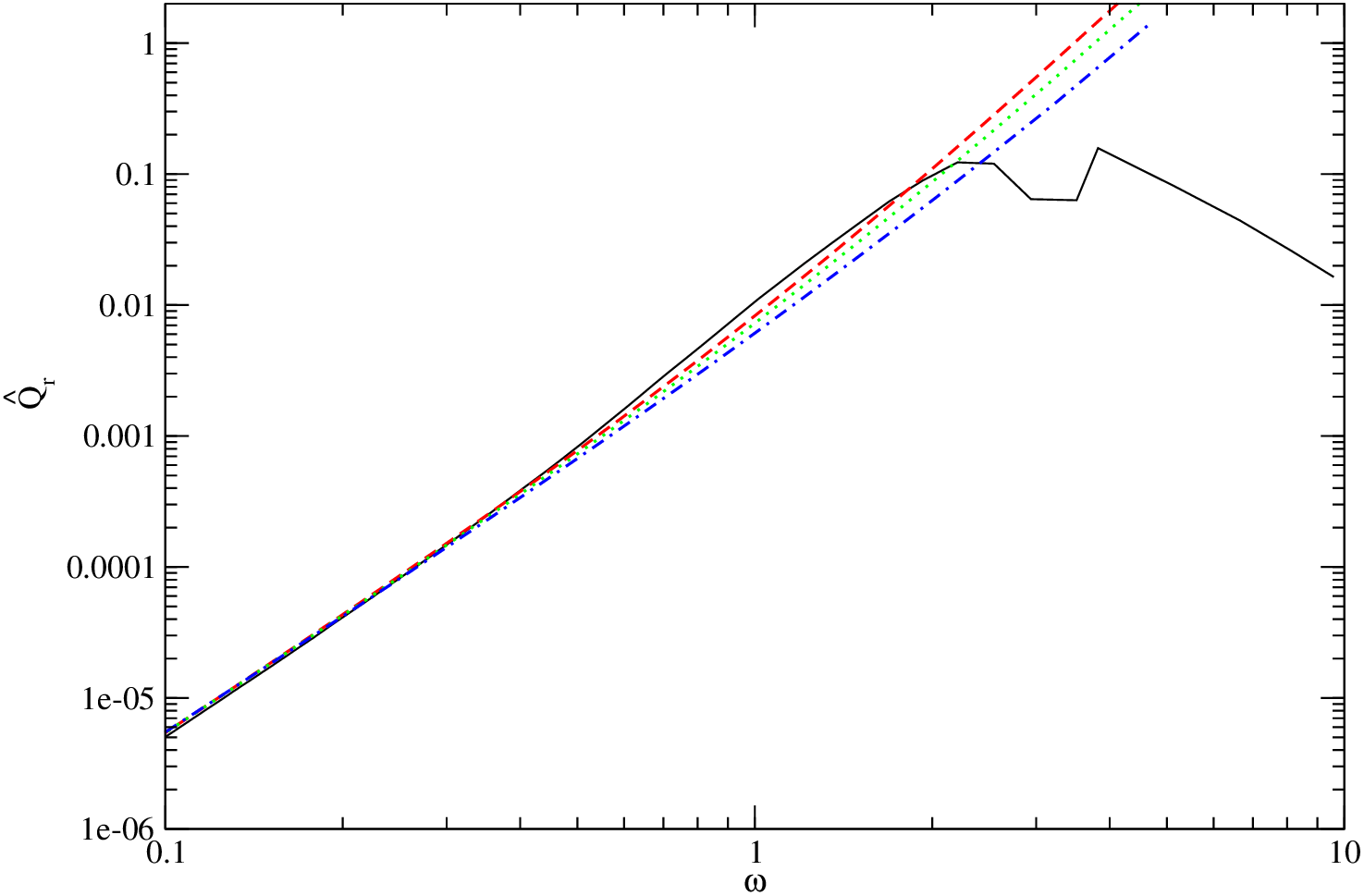}
\end{center}
\vspace{0.5cm} \caption{The dependence of $\hat Q_r$ on $\omega $
for the non-rotating CD model with  different fitting models for the radial  dependence of the
Brunt - V$\ddot {\rm a}$is$\ddot {\rm a}$l$\ddot {\rm a}$  presented
in table \ref{tab:t1}. See the text for a description of the
different curves. } \label{Fig6}
\end{figure}

\begin{figure}
\begin{center}
\vspace{8cm}\includegraphics{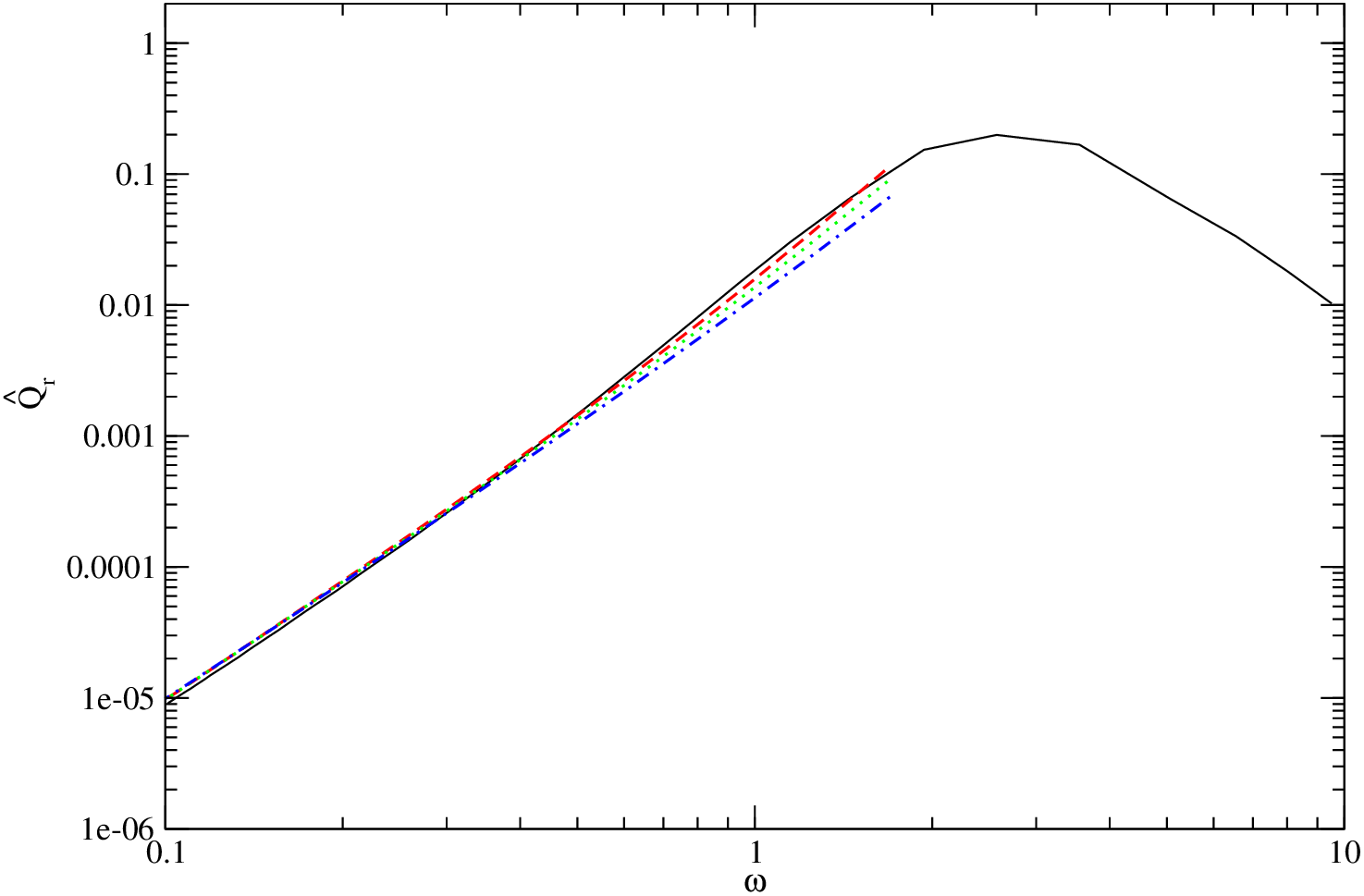}
\end{center}
\vspace{0.5cm} \caption{As in Fig. \ref{Fig6} but for the  IR model.}
\label{Fig6a}
\end{figure}

\begin{figure}
\begin{center}
\vspace{8cm}\includegraphics{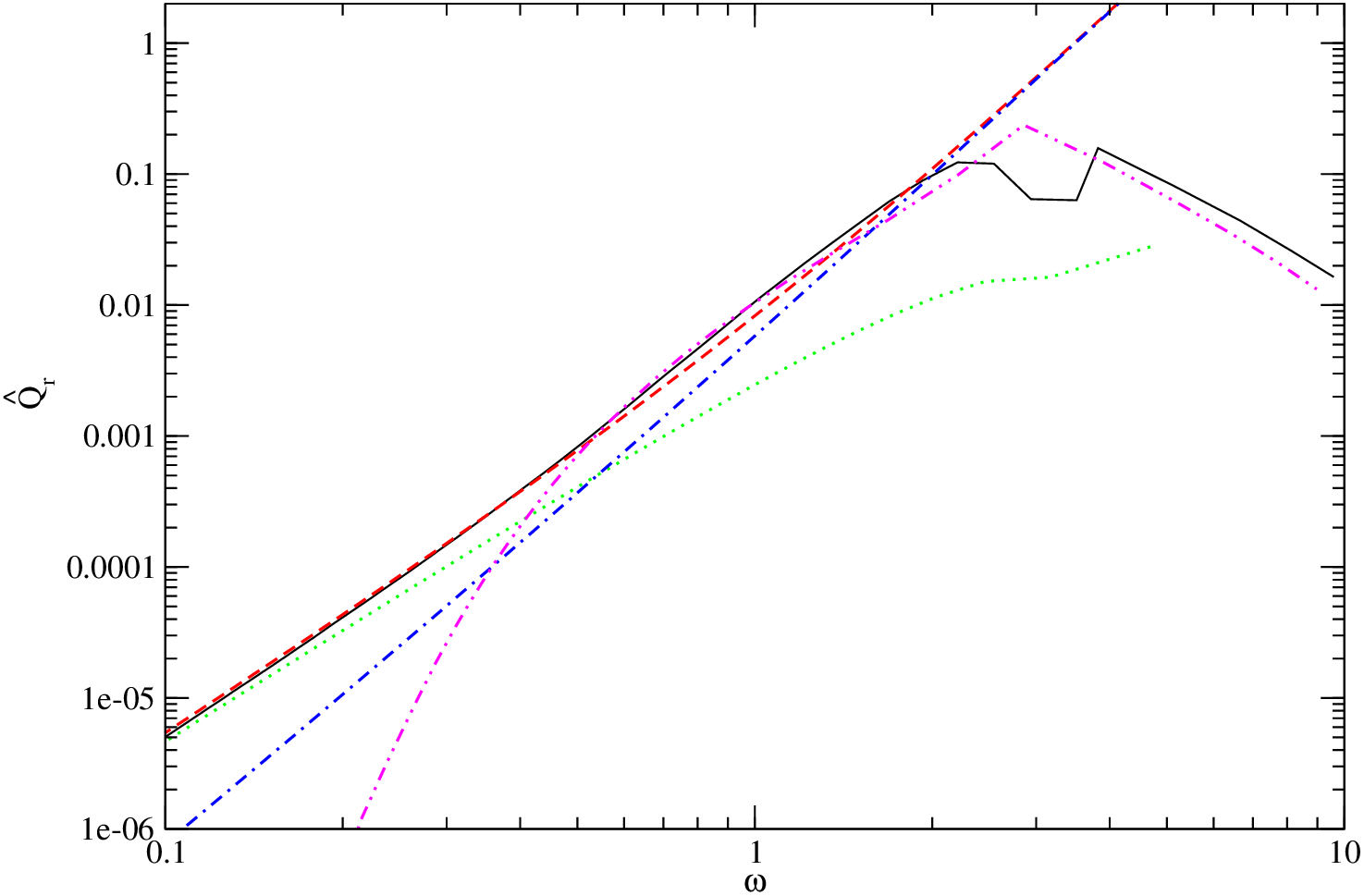}
\end{center}
\vspace{0.5cm} \caption{As in Fig. \ref{Fig6},  but showing
the contribution of different terms to  $\hat Q_r$. All curves are for
the CD model with  the CD2 fitting model for  the radial
dependence of the  Brunt - V$\ddot {\rm a}$is$\ddot {\rm a}$l$\ddot {\rm a}$  frequency.} \label{Fig7}
\end{figure}

\begin{figure}
\begin{center}
\vspace{8cm}\includegraphics{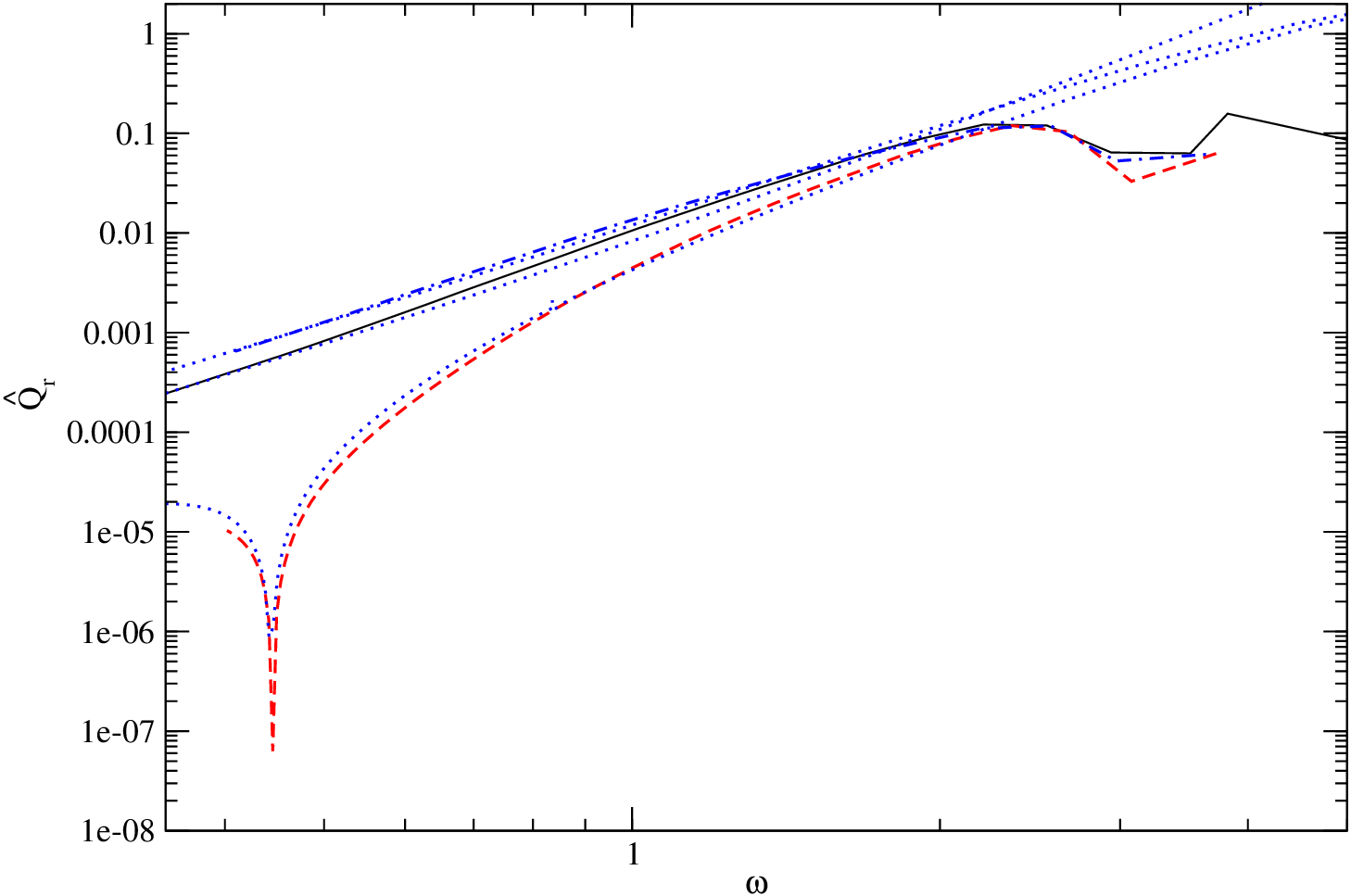}
\end{center}
\vspace{0.5cm} \caption{As in Fig. \ref{Fig6} but showing
$\hat Q_r$  for a rotating  star with $\Omega_r = \pm 0.42.$  All
curves are  for the CD model with the CD2
fitting model for the radial dependence of the Brunt - V$\ddot {\rm a}$is$\ddot {\rm a}$l$\ddot {\rm a}$  frequency. } \label{Fig8}
\end{figure}

\begin{figure}
\begin{center}
\vspace{8cm}\includegraphics{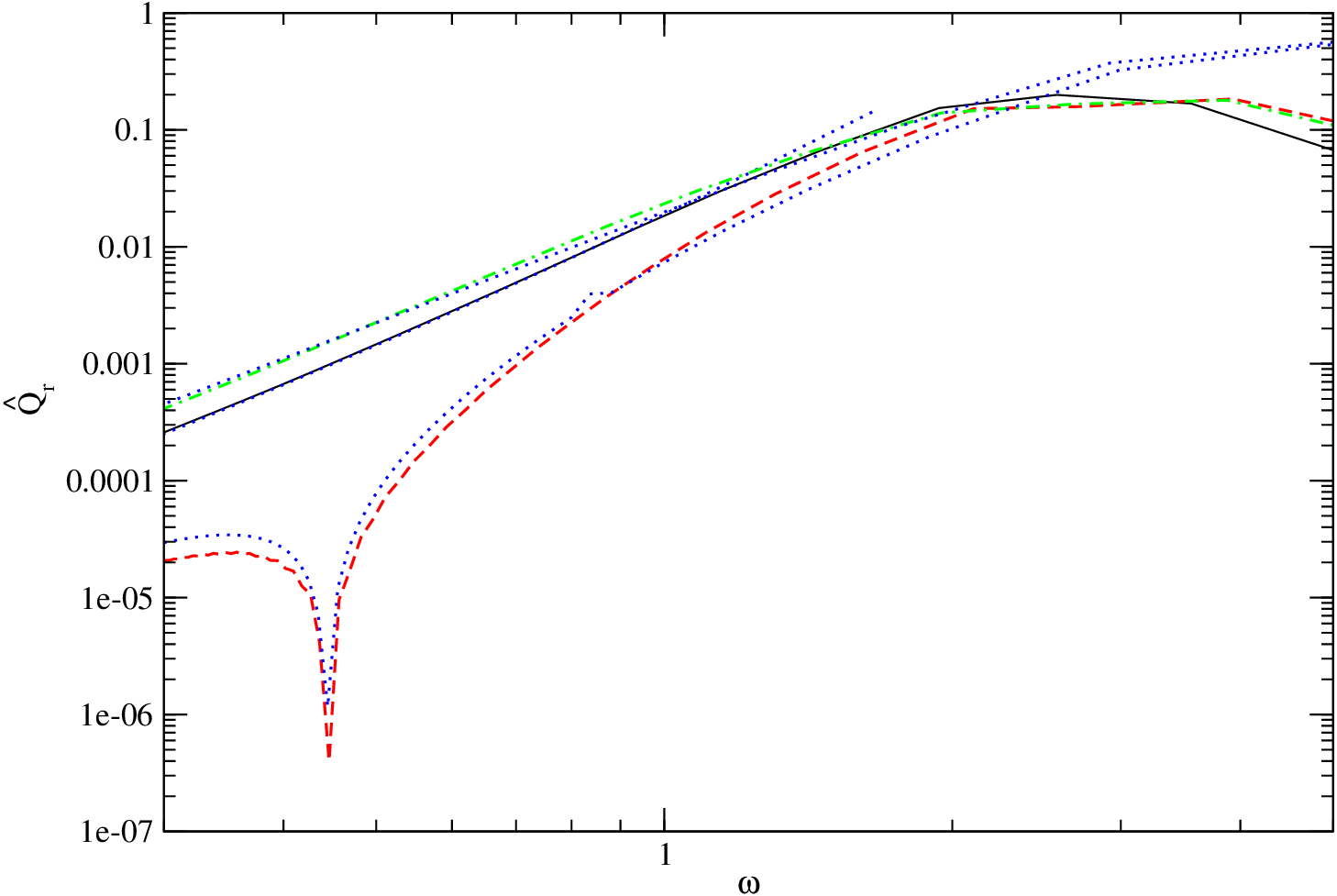}
\end{center}
\vspace{0.5cm} \caption{As in  Fig. \ref{Fig8} but for the IR
model of the star with the IR2 fitting model for  the  radial dependence Brunt - V$\ddot {\rm a}$is$\ddot {\rm a}$l$\ddot {\rm a}$  frequency.}
\label{Fig9}
\end{figure}
The  comparison of the analytic estimates  for the overlap
integrals  determined from  equation (\ref{eq92}) with values
obtained numerically is illustrated  in Figures
\ref{Fig6}-\ref{Fig9}.

 In Figs \ref{Fig6} and \ref{Fig6a} we show  $\hat Q_r$ as
a function of  $\omega$ for the non-rotating CD  and IR models,
respectively.  For non-rotating models we straightforwardly
have $\hat Q=\hat Q_r$. The solid curves give numerically
evaluated overlap integrals with the uppermost one
 corresponding to the IR model.
  For the non-rotating models of the star the Cowling
approximation is relaxed. The dashed, dotted and dot dashed curves
give our analytically  determined  overlap integrals
for the CD2 and IR2 fitting models
(dashed curves), for the CD1 and IR2 fitting models (dotted
curves) and for the CDL and IRL models (dot dashed curves),
respectively. As seen from this Fig. when $\omega $ is small our
analytic results are  an excellent agreement with the numerically determined
ones, especially for the CD2 and IR2 fitting models, which will be
used later in other Figures. For the CD2 model the disagreement is
smaller than or of the order of $10$ per cent when $\omega \le
0.5$ and smaller than $30$ per cent when $\omega \le 1$. For the
IR model the disagreement is even smaller being of the order of or
smaller than $10$ per cent when $\omega \le 1.5$.

It is very important to stress that the IR model,  describing a
young zero age main sequence star of one solar mass,  has larger
overlap integrals. This means that tidal interactions are stronger
in that  case. Our analytic approach to the calculation of  overlap
integrals allows us to explain this  as follows. One
can check that the main difference in the values of the overlap
integrals is determined by the the integral $I$ entering in the
expression for the norm (\ref{eq84}).
 As  mentioned above,
we have $I=14.91$ and $5.83$ for the CD and IR models,
respectively. Our analytically determined overlap integrals are
inversely proportional to square root of $I$. Therefore, the
overlap integrals of the IR models are  larger than
the ones of the CD models by a factor approximately
given by  $\sqrt{14.91/5.83}\approx 1.6.$
Since the energy and angular momentum exchanges through
tidal interactions are proportional to square of the overlap
integrals,  in the case of  the zero age MS star,  these are
approximately $2.55$  times larger in comparison with
those obtained for the present day solar
model.

In Fig \ref{Fig7} we show different contributions to
$\hat Q_r$, for the CD2 model, separately. The solid and dashed curves are
 as in Fig. \ref{Fig6}.  The dot dashed curve shows the
 analytically  determined  $\hat Q_r,$ but with with $Q_{r,c}$ set to zero
in (\ref{eq92}). The dotted curve shows $\hat Q_r,$ but with $Q_{r,r}=0$
in (\ref{eq92}). Thus, the dot dashed and dotted  curves show
the separate contributions of the radiative region close to the base of
convection zone and the  contribution from convection zone i to $\hat Q_r$.
As
seen from Fig. \ref{Fig7},   $Q_{r,r}$ is more
important for relatively large values of $\omega, $ while the
$Q_{r,c}$  gives a contribution that becomes increasingly important in the
limit of $\omega \rightarrow 0$. However, for the range of
$\omega $ shown, both contributions must be taken into account  in order to obtain
good agreement between the numerical and analytical results.
 Additionally,
we show the overlap integrals of a star modelled as an  $n=3$
polytrope,   plotted  using  the dot dot dashed curve. Unlike the
overlap integrals for  the the realistic star,   the polytropic
ones decrease exponentially with decreasing  $\omega.$ Therefore,
when  g modes of sufficiently high order  determine the tidal
exchange of energy and angular momentum,   this  is much smaller
for a  polytropic star.

In Fig. \ref{Fig8} we  compare  numerical and
analytic  results for a rotating CD2 model. In order to  delineate  the
effects of rotation, the rotational frequency is taken to be
large with  $|\Omega_r| = 0.42$. The solid, dashed, and
dot dashed curve show the numerical results corresponding to
$\Omega_r=0$, $-0.42$ and $0.42$, respectively. The dotted curves
correspond to the analytic calculations and  are closest to their corresponding
numerical   curves.  One can see that the agreement
between analytical and numerical results is again very good for
$\omega < 1$ for  the  non-rotating star and the star with  prograde
rotation. The analytic and numerical curves can hardly be distinguished.
 For the case of retrograde rotation the curves disagree
by   up to 30 per cent in the  range shown. The curves
corresponding to  prograde rotation are qualitatively similar
to the curves corresponding to  a non-rotating star, the only
difference being  that they indicate somewhat  larger values of $\hat Q_r$ for a given
$\omega$. The curve corresponding to retrograde rotation behaves
in a non-monotonic manner. It has a strong minimum at $\omega =
0.44$, which corresponds to $\nu =-1.9$. From Fig. \ref{Fig1} it
follows that $\Lambda=30$ for this value of $\nu$ and from
equation (\ref{eq81}) it follows that the factor $h(r)$ entering
our analytic expression for $\hat Q_r$ is then  equal to zero.
 This explains the non-monotonic character of the
behaviour of $\hat Q_r$ for retrograde rotation. In general, in
this case the radial overlap integral is smaller than for
the non-rotating star. This as well as  the decrease
of the 'angular overlap integral' $\alpha $ for retrograde
rotation (see Fig. \ref{Fig1}) suppresses  the tidal transfer of
energy and angular momentum for  sufficiently large retrograde
rotation. Note, however, that for  retrograde rotation  there
are other factors, such as a  match between forcing
and   mode frequencies that tends to occur for  lower order modes
in comparison to the prograde case,
that may  amplify the transfer, see e.g. Ivanov $\&$
Papaloizou (2011) .
In Fig. \ref{Fig9} the corresponding results are given  for the
rotating IR2 model. The results are qualitatively  similar to the
previous case.

\begin{figure}
\begin{center}
\vspace{8cm}\includegraphics{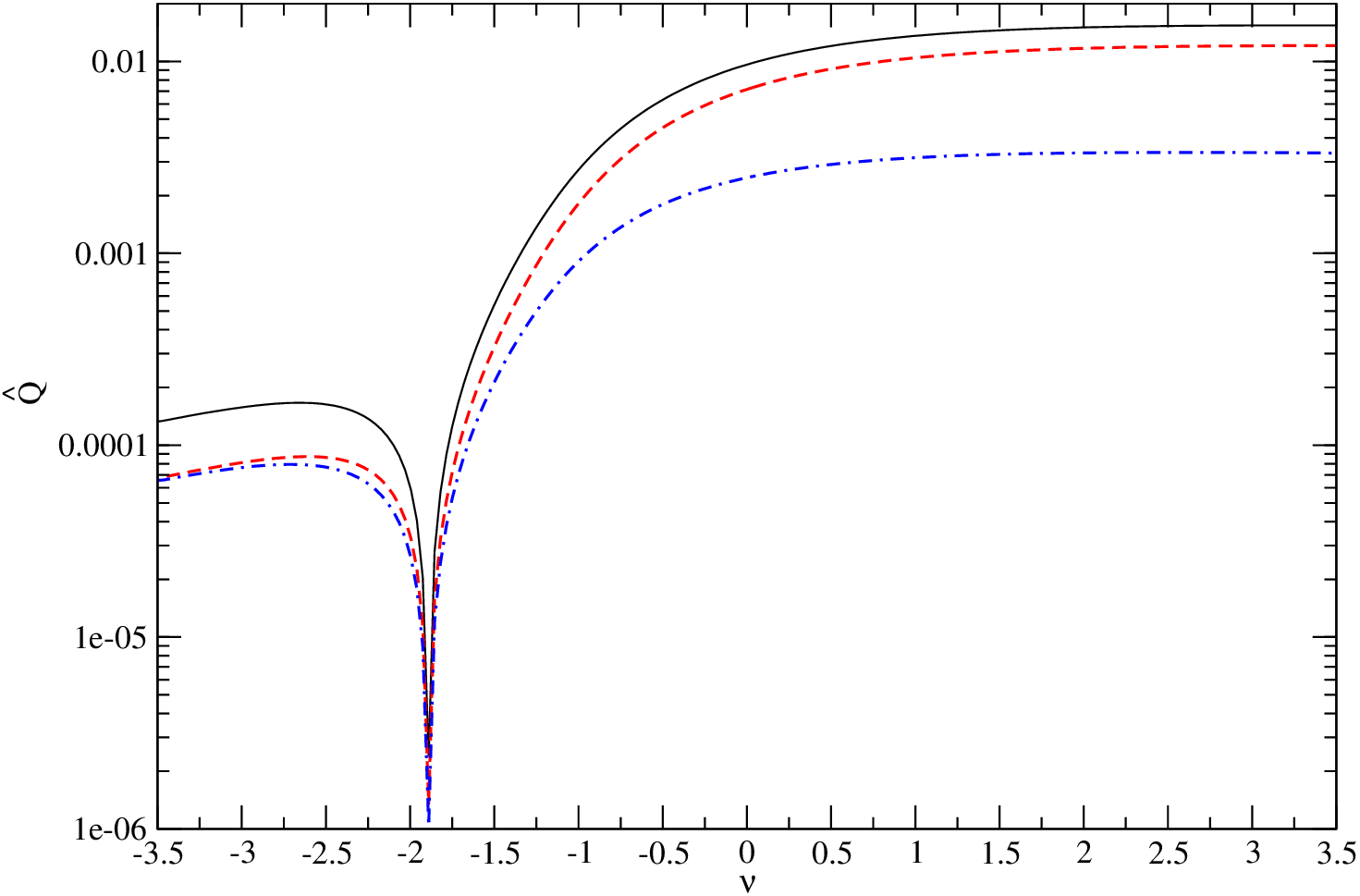}
\end{center}
\vspace{0.5cm} \caption{The value of the overlap integral
(\ref{eq92}) for a fixed eigenfrequency $\omega=1$  plotted as a function
of $\nu$ together with the  contributions to it from the regions just  inside
the radiative core and the convective envelope.  The CD2 fitting model is used. See the text for a
description of the  different curves.} \label{Fig9a}
\end{figure}

\begin{figure}
\begin{center}
\vspace{8cm}\includegraphics{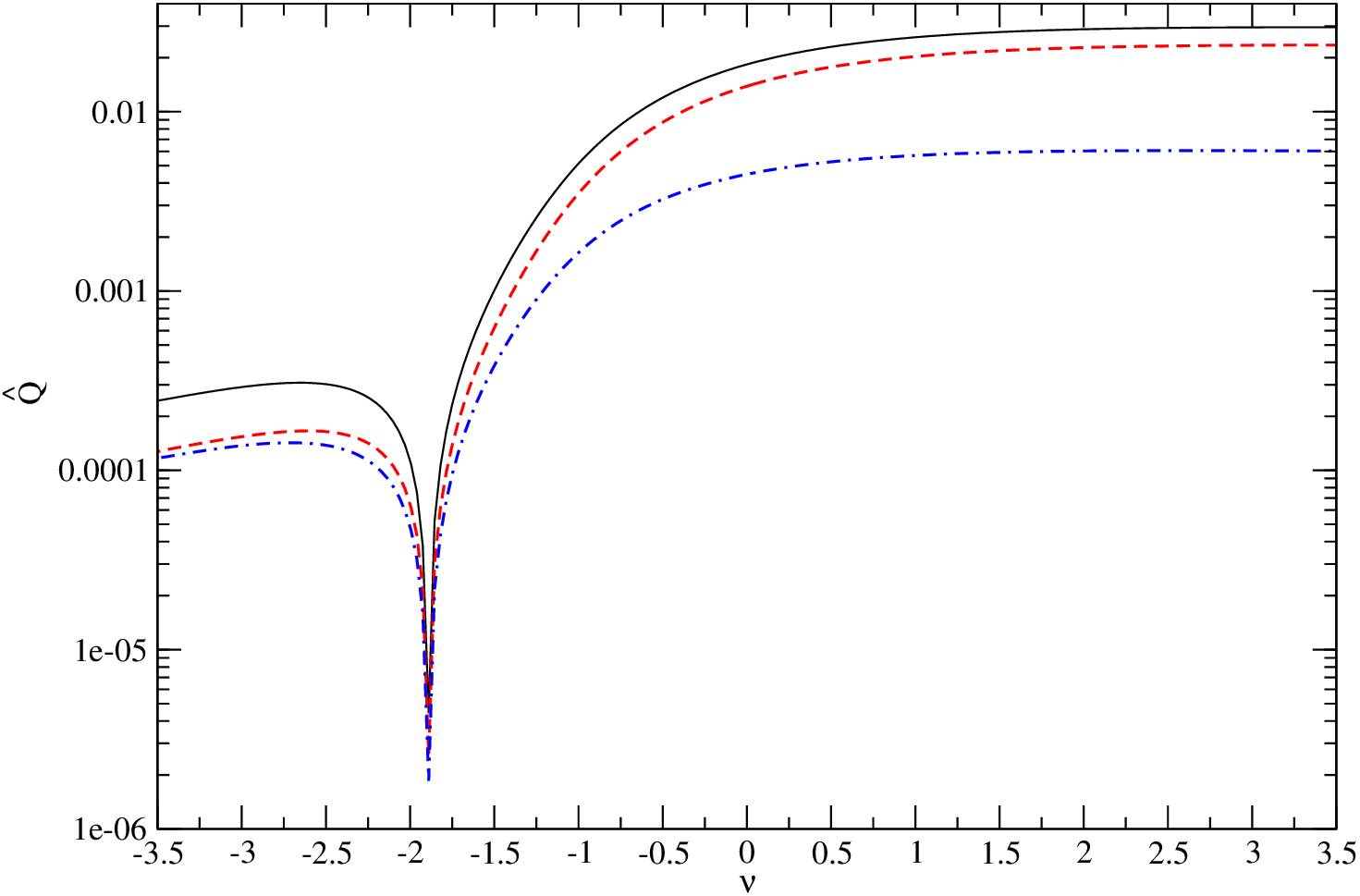}
\end{center}
\vspace{0.5cm} \caption{As for Fig. \ref{Fig9a} but for  the IR2
model.} \label{Fig9b}
\end{figure}

In Figs \ref{Fig9a} and \ref{Fig9b} we show the dependence of the
overlap integral defined through  (\ref{eq92}) (the index $j$ being suppressed)  with  $\omega$ set equal to  $1$ in (\ref{eq86})
and (\ref{eq90}) on $\nu,$  for the  CD2 and IR2 models, respectively. Note that
the overlap integral given by (\ref{eq92}) includes the factor contributed by the
angular overlap integral $\alpha(\nu)$. The solid curves show  $\hat Q,$
the dashed curves show $\hat Q$  calculated with the contribution from the convective envelope,
$Q_{r,c}$ defined in (\ref{eq90}) set to zero and the
dot-dashed curves show $\hat Q$ calculated with the contribution from the radiative interior,
$Q_{r,r}$ defined in (\ref{eq86}) set to zero. One can see from
 Figs \ref{Fig9a} and \ref{Fig9b}  that at negative values of $\nu,$ $\hat Q$  becomes quite
small. The minimum at $\nu\approx =1.9$ is determined by the fact
that $\Lambda \approx 30$ there and the integral is strongly
suppressed as discussed above.

As follows from equations (\ref{eq86}) and (\ref{eq90}) the
contributions to the overlap integral coming from the radiative interior near  base of
convection zone and the convective envelope are, to a good approximation, proportional to
powers of $\omega$. This can be used to extend the results
presented in Figs. \ref{Fig9a} and \ref{Fig9b} to other values of
 $\omega$ by making appropriate scaling. Of course, when
using such results, one has to take into account that the values of
$\omega,$  the mode order  and $\Omega$ are linked together through equation (\ref{eq79}) and also  the definition of $\nu.$

\section{Tidal evolution timescales for a  binary
with a small eccentricity and with orbital and spin angular momenta aligned
in the moderately large
dissipation regime}\label{Tidalapp}

As a simple illustration of application of our formalism we
 calculate the  timescales for the
evolution of the orbital elements of a binary system
induced by tidal interaction.
 We consider  a binary consisting of a primary star  of mass $M_*$ on which
tides are exerted by the secondary of mass $M_2$
that is approximated as being a point mass. For
simplicity, we assume below that the orbit has a small
eccentricity $,e,$ and the stellar rotation frequency is small in
comparison with the characteristic frequency $\Omega_*.$

We  shall assume that the action of dissipation on the eigenmodes
is sufficiently strong for the system to be in the regime of
moderately strong viscosity discussed above. Then  we can use
(\ref{eq34}) to calculate the  orbital energy dissipation rate due
to tides. In the case of Sun-like stars considered here, this may
require the excited waves to be of large enough amplitude to break
as they propagate towards the stellar centre (see Barker  2011 and references therein
for a discussion). Although such a condition may not be satisfied,
assuming that it is should provide the most optimistic estimate of
the potential long term tidal evolution due to the excitation of these
modes. Note too that the action of turbulence in convection zones
is likely to be ineffective for the short orbital frequencies we
consider on account of its long characteristic time scale.

This problem has been discussed in a similar setting  by e.g. Zahn
(1977) and Goodman $\&$ Dickson (1998). However,  they did not include the effects
of Coriolis forces, taking account of rotation only through
a shift in tidal forcing frequency. Our formalism includes the effects
of Coriolis forces under the traditional approximation
and so allows us to assess their
significance. Moreover, the formalism of Zahn (1977) has been mainly
applied to more massive stars having a convective core and
radiative envelope in contrast to the solar type stars
explicitly considered here (see also Papaloizou \& Savonije 1997).
 Although our general approach is applicable
to this case,
the  asymptotic calculation of the overlap integrals
needs to be considered separately.

In sections \ref{Basiceq} - \ref{modgm}, the stellar angular velocity was taken to be positive
so that prograde normal modes had positive frequencies while retrograde normal modes had negative%
frequencies. However, for the discussion of normal modes and overlap integrals from section \ref{modgm} onwards,
it was convenient to exploit the symmetry that enabled retrograde modes to have a positive frequency,  with the sign of
$\Omega$  being reversed to become negative.
In this section we consider the tidal interactions of a star with a definite  positive angular velocity, accordingly
from now on we revert to the convention of section \ref{Basiceq} - \ref{modgm} for which retrograde modes
have negative frequencies.

\subsection {Equations for the evolution of  the semi-major
axis and eccentricity}

Equations governing the evolution of the semi-major axis $a,$ and
the eccentricity, $e,$ follow from consideration of the
conservation of energy and angular momentum of the system. We
begin by noting that tidal interaction with $M_*$ induces rates of
change of energy, $\dot E_I$, and angular momentum,  $\dot L_c,$
for the orbit given by  equation (\ref{eq34}).  Conservation of
energy and angular momentum implies that these produce changes in
$a$ and $e$ according to
\begin{equation}
 {G\mu M_*^2{\dot a}\over 2 a^2 } =\dot E_I,\quad
 {G\mu M_*^2e{\dot e}\over a } =\dot E_I(1-{e^2\over 2})-\Omega_{orb}\dot L_c.\label{t001}
\end{equation}
Here the eccentricity is assumed to be small so that powers of $e$ higher than second
order may be neglected and  $\mu =M_2/M_*$ is the mass
ratio.
In what follows,  we find it convenient to rewrite these equations in the form
\begin{equation}
{\dot a\over a}=-{2\over T_{a}}, \quad {\dot e\over e}=-{1\over
T_{e}}, \label{t1}
\end{equation}
where
\begin{equation}
T_{e}=-{G\mu M_*^2e^2\over a\dot E_e}, \quad T_a=-{G\mu M_*^2\over a\dot
E_I}\hspace{2mm} {\rm and} \label{t2}
\end{equation}
\begin{equation}
\dot E_e=\dot E_I(1-{e^2\over 2})-\Omega_{orb}\dot L_c. \label{t3}
\end{equation}

When the rotation frequency $\Omega $ is specified we can use
equations (\ref{eq34}) and the expressions for the coefficients
$A_{m,k}$ given in Appendix A to estimate the time scales $T_a$
and $T_e$. We consider below two important cases 1) the case of
synchronous rotation $\Omega=\Omega_{orb}$ and  2) the case of a
non-rotating primary $\Omega=0,$ labelling the corresponding time
scales by indices $(1)$ and $(2)$, respectively. Since we shall
use  dimensionless units,  we write  $\hat Q
=\sqrt{M_*}R_{*}\tilde Q$, where $\tilde Q$  is the dimensionless
overlap integral,  we recall that we have suppressed the  index
$j.$

\subsection{The case of synchronous rotation}

Here we have $\Omega=\Omega_{orb}$.  As indicated in Appendix A it
turns out that to leading order in $e$,  terms of higher order in
$e$ than linear  in the expansion of the tidal potential may be
neglected when substituting  into  the series (\ref{eq4}).  Thus,
the term proportional $e^2/ 2$ in (\ref{t3}) results in   a higher
order contribution   and can be neglected.  We thus  have $\dot
E_e\approx \dot E_c$. The $(m,k)$ pairs that  remain to be
considered are $(0,1), (2,1)$ and $(2,3)$. The  conditions  for
resonance associated with a normal mode (\ref{eq26}) associated
with these are all included in the expression $\omega
=\omega_{m,k} = k\Omega_{orb} - m\Omega =  \pm \Omega_{orb}$. The
corresponding values of the parameter $\nu =\pm 2$. As seen from
Figs. \ref{Fig1} and \ref{Fig1a} when either $\nu=-2$ and $m=2$ or
$\nu=\pm 2$ and $m=0$ we have $\Lambda \approx 34$. This leads to
a strong suppression of the corresponding overlap integrals
determined by the factor $h(r) \propto 1-{30\over \Lambda}$, see
the discussion above, equation (\ref{eq81}) and the discussion of Fig. \ref{Fig8}
for the $m=2$ case. Therefore, their respective contributions to
the energy and angular momentum exchange rates can be neglected
and we take into account below only the dominant $(2,3)$ term.

Equations (\ref{eq34}) and (\ref{t3})  then give to  the leading
order in eccentricity
\begin{eqnarray}
\dot{E}_e&=&-{147\pi^2e^2\over 40}{\left({GM_2\over
a^3}\right)}^2M_*R_*^2\left[{\tilde
Q^2\over |d\omega /dj|}\right]_1, \hspace{2mm}{\rm and}\nonumber \\
 \dot{E}_I&=& 3\dot E_e,
\label{t4}
\end{eqnarray}
where functions of eigenfrequencies $\omega$ enclosed in square
brackets $[..]_{k}$ are assumed to be evaluated for
$\omega=k\Omega_{orb}$  and $m=2$  from now on.

The term $[|d\omega /dj|]_k$ can be evaluated setting $n=j$ in
equation (\ref{eq79}) and adopting the limit $n \gg 1,$ noting
that $\Lambda$ varies only weakly with $\nu$ in the region of
interest being close to  $4.5$ (~see Fig.~\ref{Fig1}~), with the
result that
\begin{equation}
[|d\omega /dj|]_k={\pi k^2\over \sqrt{\Lambda }
I}{\Omega^2_{orb}\over \Omega_*}, \label{t5}
\end{equation}
where the integral $I$ is expressed in  natural units.
Substituting (\ref{t5}) in (\ref{t4})  we obtain
\begin{eqnarray}
\dot {E_I}&=& -{441\pi e^2{\dot E}_*\over 40}\left[{\tilde
Q}^2\right ]_1,
 \hspace{1mm}{\rm and}\hspace{1mm}
\dot{E_e}=-{147\pi e^2{\dot E}_*\over 40}\left[{\tilde
Q}^2\right]_1,
\nonumber
\end{eqnarray}
where
\begin{equation}
{\dot E}_{*}=\sqrt{\Lambda}I{G^2\mu^2\Omega_*M_*^3R_*^2\over
\Omega^2_{orb}a^{6}}\label{t7}.
\end{equation}
Now we use  (\ref{t7}) to obtain the tidal evolution time scales
given by  (\ref{t2}) thus
obtaining
\begin{eqnarray}
T^{(1)}_e&=&{40 T_*\over 147\pi}\left [{\tilde Q}^{-2}\right ]_1, \nonumber \\
T^{(1)}_a&=&{1\over 3}e^{-2}T^{(1)}_e, \label{t8}
\end{eqnarray}
where
\begin{equation}
T_*={\left(1+\mu \right)^{5/3}\over \sqrt{\Lambda} \mu
I}\left({P_{orb}\Omega_*\over 2\pi}\right)^{4/3}\Omega_*^{-1},
\label{t9}
\end{equation}
and  the orbital period $P_{orb}=2\pi\Omega_{orb}^{-1}$. Thus,
for this configuration we have $T_a \gg T_e$ when $e\ll 1$.
\begin{figure}
\begin{center}
\vspace{8cm}\includegraphics{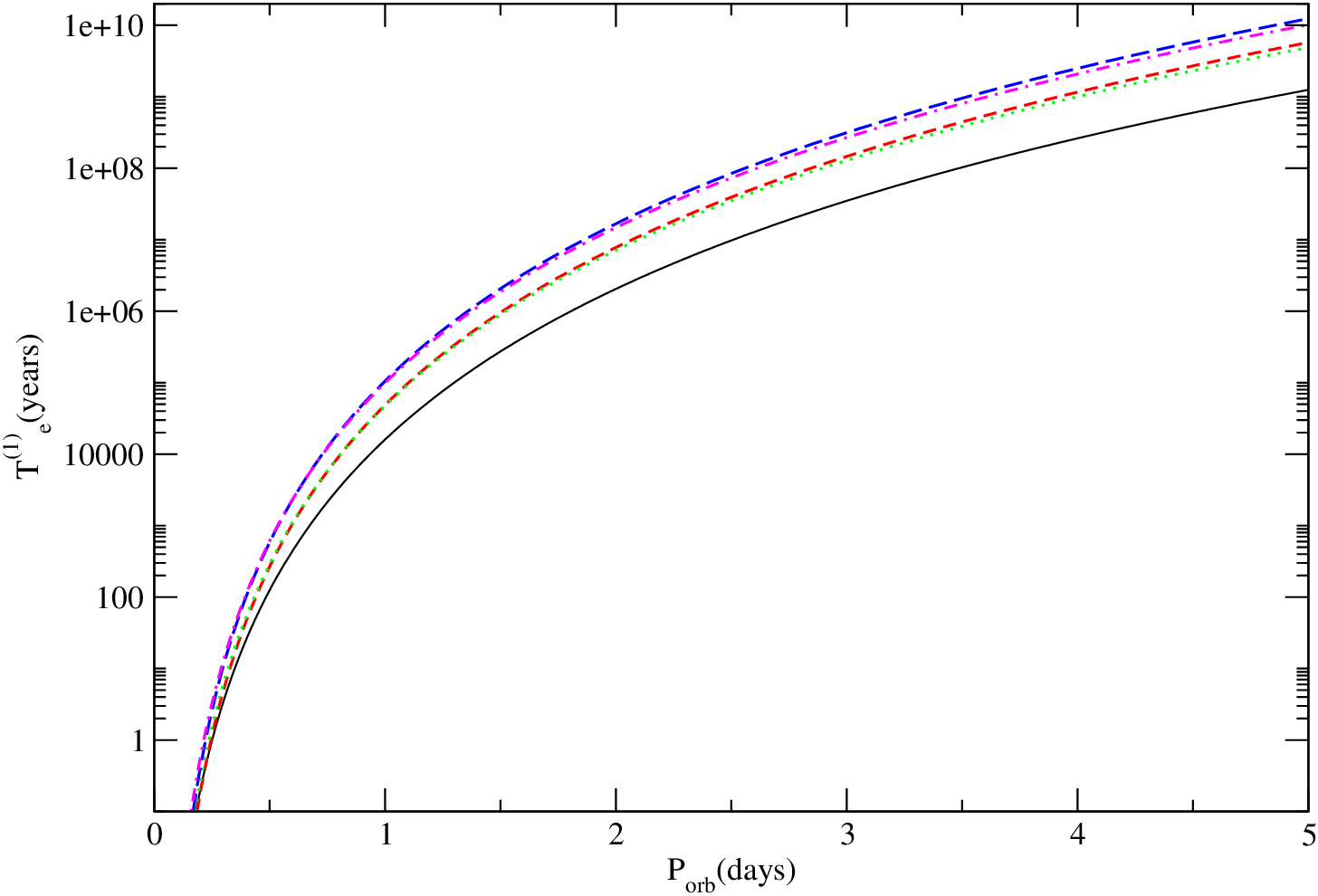}
\end{center}
\vspace{0.5cm} \caption{The evolution time scale $T^{(1)}_{e}$
calculated for the CD and IR models  compared to  the result
of Goodman $\&$ Dickson (1998), for $\mu=1$. See the text for the
description of different curves.} \label{Fig10}
\end{figure}

In order to calculate the overlap integral entering (\ref{t8}) for
$m=2$ we use the results of preceding  section setting values of
parameters determining the overlap integral to correspond to
$\nu=2$. Explicitly, we use $\alpha_1\approx 1$, $\Lambda =4.5$,
$I_{\theta}=-0.1155$, $B_c=6.2$ and $I_c=1.4\cdot 10^{-2}$.

We show the evolution timescales $T^{(1)}_e$ calculated for the CD
and  IR models in Fig. \ref{Fig10}. The short dashed, dotted, long
dashed and dot dashed curves are for the CD2, CDL, IR2 and IRL
fitting models.  One can see from  Fig. \ref{Fig10} that  there is
little difference between the curves corresponding to different fitting
models. The solid line plots  a minor modification of the expression of Goodman $\&$
Dickson (1998)  who give $T_{e}^{GD}=8\cdot 10^3P_{orb}^7,$ with $P_{orb}$
in days,  for a
system of two tidally interacting Sun-like stars of equal mass.

Since we assume that the tides are raised only on one component,
we plot $T_{e}^{GD}=1.6\cdot 10^4P_{orb}^7/\mu$ in  Fig.
\ref{Fig10} and others below. Our results are seen to give  larger
values of $T_{e}$ for a given $P_{orb},$ e.g.  for $P_{orb} = 4 $
days,  the results differ by  a factor of five to ten.  Thus we
have $T_{e}\approx 2.7\cdot 10^8yr$, $1.15\cdot 10^9yr$ and
$2.35\cdot 10^9yr$ from the modified  Goodman $\&$ Dickson (1998)
expression and for our calculations corresponding to the CD and IR
models, respectively. We remark that  Ogilvie \& Lin (2004)
indicate a possible reason why
 the Goodman \& Dickson expression  gives such an underestimate
 for $T_{e}$  and we recall that   the effects of Coriolis forces
are not included.
 Note that the IR model gives
larger values of $T_{e}$ irrespective of the fact that its overlap
integrals are larger than those  of the CD model, for a given
forcing frequency. The is due to a smaller radius of the early
Sun  compared to its present value. The younger
star is more compact and, accordingly, less susceptible to tidal
influence from  a perturber.

Finally  let us recall that
for large values of $P_{orb},$ the appropriate normalised  overlap integrals corresponding
to the CDL and IRL fitting models are proportional to
$P_{orb}^{-17/6}$. From equation (\ref{t8}) it follows that in
this case $T_{e}\propto P_{orb}^{7}$ as for the Goodman $\&$
Dickson (1998) expression.

\begin{figure}
\begin{center}
\vspace{8cm}\includegraphics{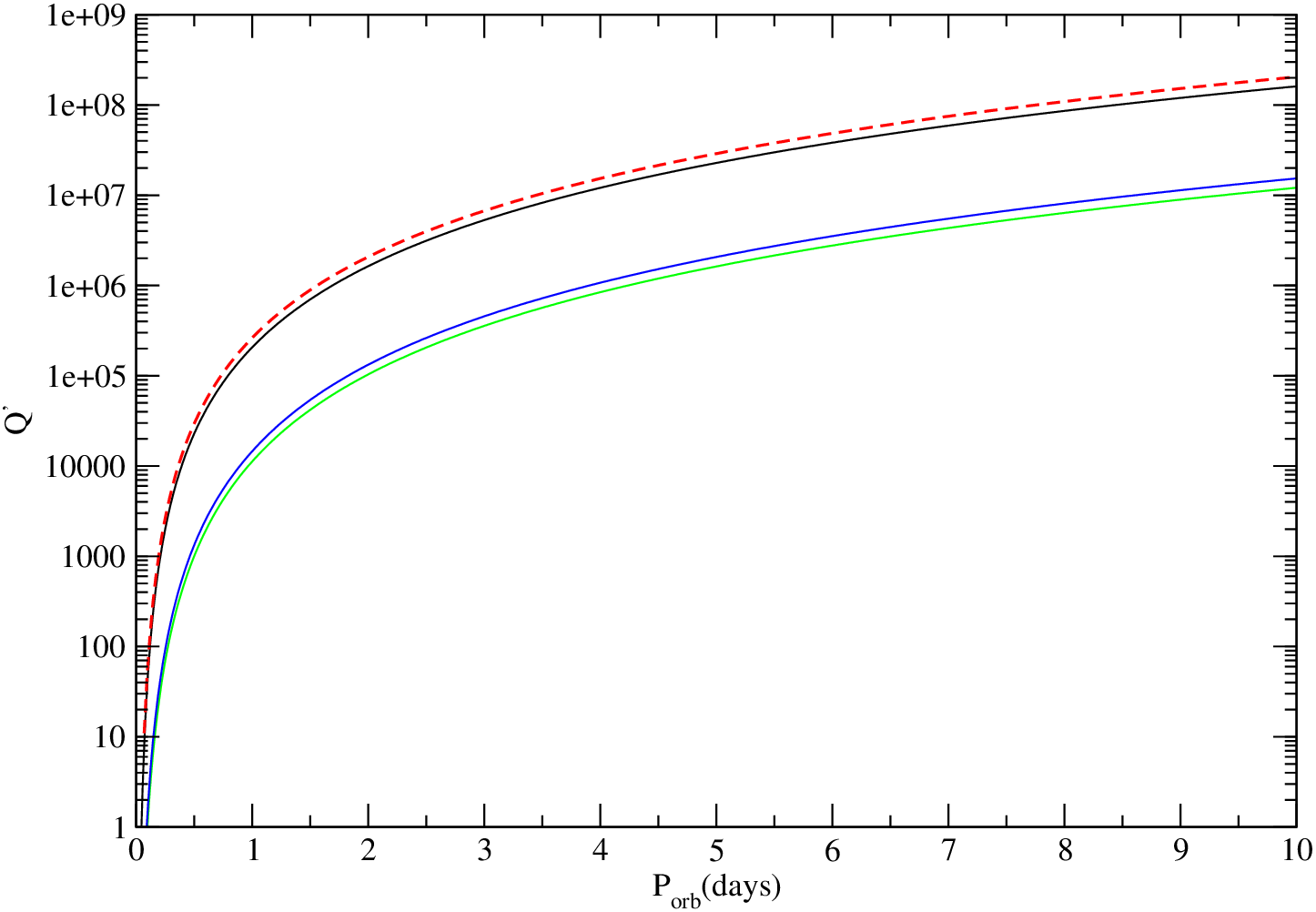}
\end{center}
\vspace{0.5cm} \caption{ The  tidal $Q'$ parameter for orbital circularisation.
The solid and dashed curves are respectively for CD2 and IR2 models in the case of
synchronous rotation. the
dotted and dot-dashed curves  are respectively
for the CD2 and IR2 models  with no rotation.} \label{FigQT1}
\end{figure}

Finally we calculate the effective  tidal $Q'$ for the star by comparing our circularisation rate
to that of Goldreich \& Soter (1966).  This comparison leads to the expression
\begin{eqnarray}
Q^{'}=\frac{285}{98\pi\sqrt{\Lambda}I[\tilde{Q}^2]_1(1+\mu)^{1/2}}\left (\frac{2\pi}{P_{orb}\Omega_\ast}\right)^3.
 \end{eqnarray}
We plot $Q^{'}$ as a function of orbital period for $\mu=1$ in
Fig. \ref {FigQT1}.
This quantity is often assumed to be a constant.  However,
Fig. \ref {FigQT1}
shows that this varies between $10^6$ and  $10^8$ as the orbital period
varies from $1 - 10$ days.

\subsection{The case of non-rotating primary}

\begin{figure}
\begin{center}
\vspace{8cm}\includegraphics{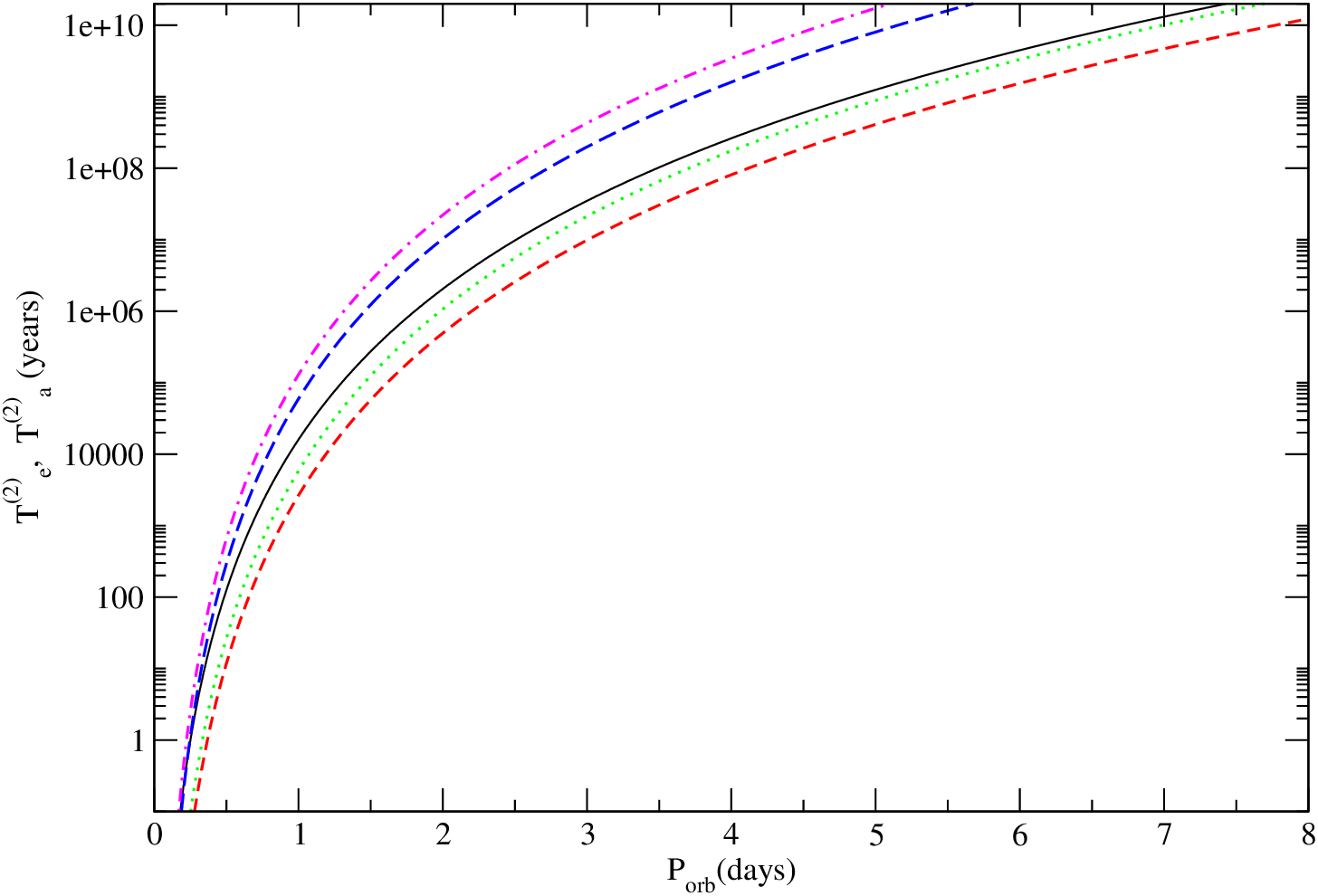}
\end{center}
\vspace{0.5cm} \caption{The evolution time scales $T^{(2)}_a$ and
$T^{(2)}_{e}$ calculated for the CD and IR models  compared
the modified expression of Goodman $\&$ Dickson (1998). See the text for the
description of the different curves.} \label{Fig11}
\end{figure}

\begin{figure}
\begin{center}
\vspace{8cm}\includegraphics{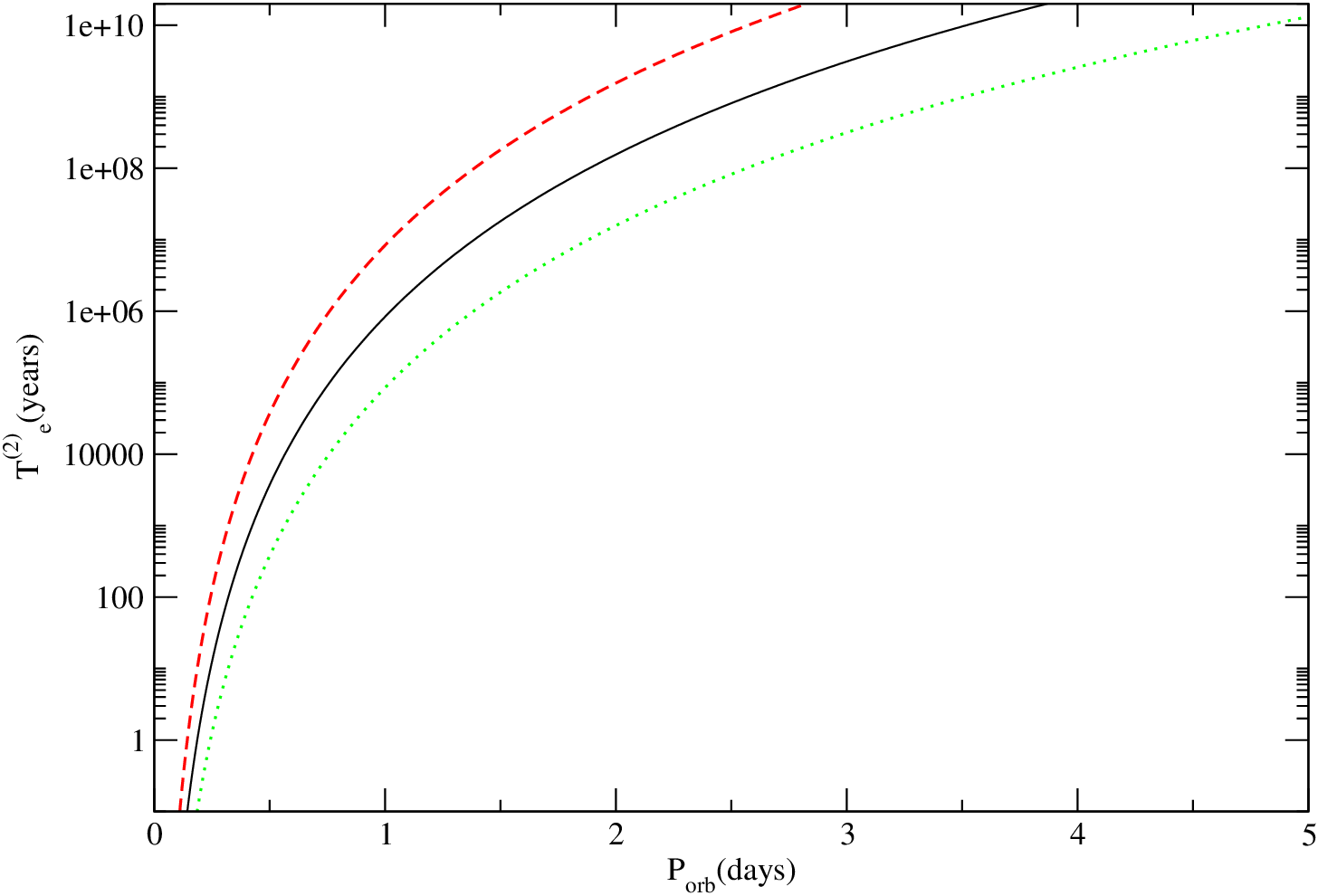}
\end{center}
\vspace{0.5cm} \caption{The evolution time scales $T^{(2)}_e$
plotted for $q=10^{-3}$, $q=10^{-4}$ and $q=10^{-2}$ using
the solid, dashed and dotted curves, respectively. } \label{Fig12}
\end{figure}

\begin{figure}
\begin{center}
\vspace{8cm}\includegraphics{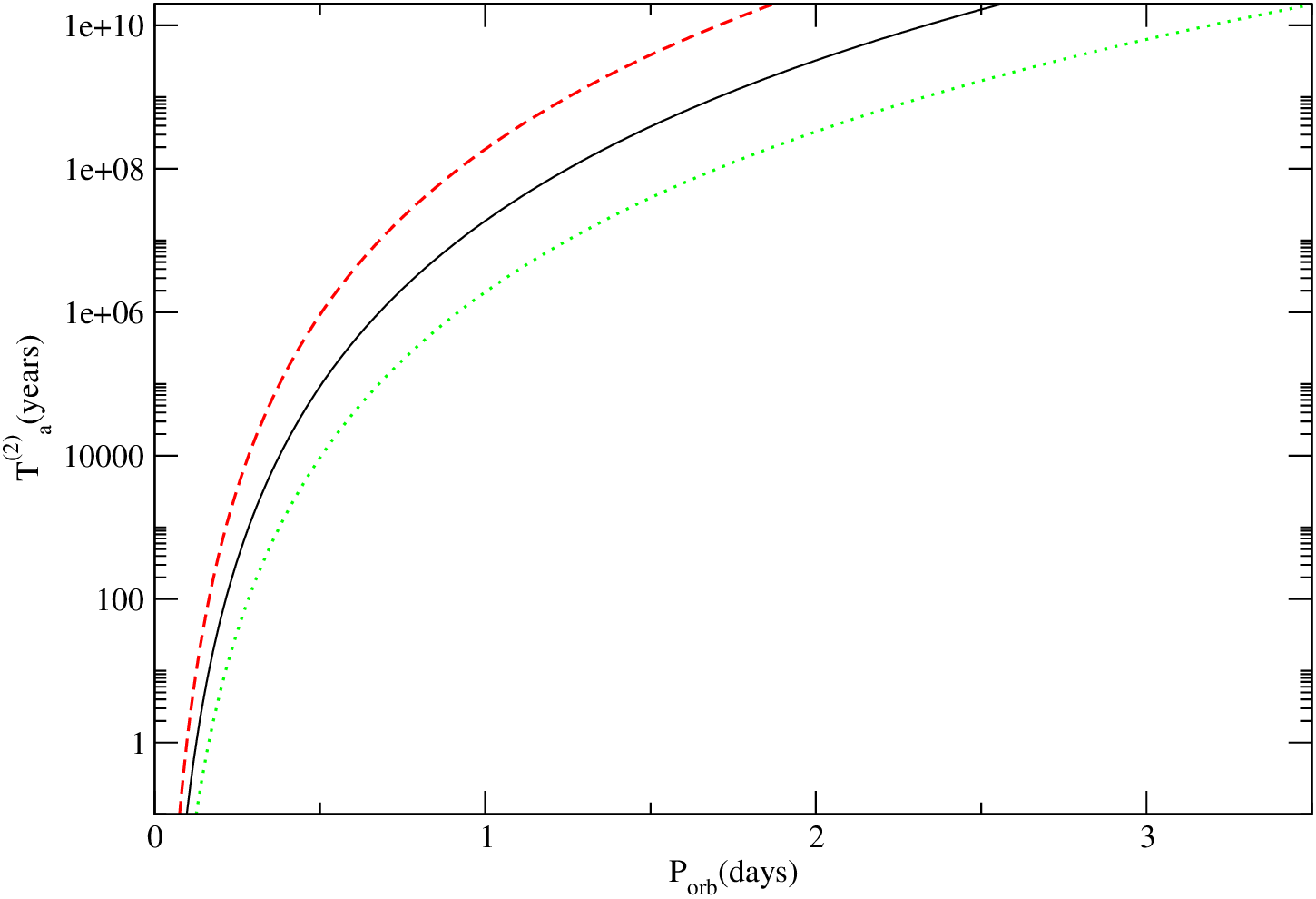}
\end{center}
\vspace{0.5cm} \caption{Same as Fig. \ref{Fig12} but $T^{(2)}_a$
is shown.} \label{Fig13}
\end{figure}

Calculation of the time scales $T^{(2)}_a$ and $T^{(2)}_e$
proceeds analogously to the previous case. Since $\Omega=0,$ the
forcing frequencies are of the form $\omega_{m,k}=k\Omega_{orb}$.

In order to calculate $T^{(2)}_a$ we can set $e=0$ in the
expressions for the  potential expansion coefficients given in
Appendix A. Then  only the $A_{2,2}$ term with corresponding
forcing frequency $\omega_{2,2} =2\Omega_{orb}$
 contributes
to the series. The calculation is straightforward with the result that
\begin{equation}
\dot E_{I}=-{3\pi \dot E_{*} \over 40}\left [{\tilde Q}^{2}\right
]_2, \label{t10}
\end{equation}
and, accordingly,
\begin{equation}
T^{(2)}_{a}={40 T_* \over 3\pi}\left [{\tilde Q}^{-2}\right
]_2\label{t11}
\end{equation}

It is convenient to separate the calculation of ${\dot E}_e$, and,
accordingly, $T^{(2)}_e$ into two stages.  First we evaluate
${\dot E}_I-\Omega_{orb} {\dot L}_c$. This quantity is determined by
terms in the potential expansion with forcing frequencies
$\omega_{0,1} = \Omega_{orb}$, $\omega_{2,1} =\Omega_{orb}$ and
$\omega_{2,3} =3\Omega_{orb}.$  The  respective coefficients
$A_{m,k}$ are given in Appendix A. Substituting these into
(\ref{eq34}), noting that in the non rotating case, the overlap integrals are independent of $m,$ we obtain
\begin{eqnarray}
&&\hspace{-6mm}{\dot E}_I-\Omega_{orb} {\dot L}_c=\nonumber \\
& &\hspace{-6mm}-{\left ({GM_2\over a^3}\right )}^{2}e^2\left \lbrace
{3\pi^2\over 20}\left [{{\hat Q}^2\over
|d\omega/dj|}\right]_{1}+{49\pi^2\over 40}\left [{{\hat Q}^2\over
|d\omega/dj|}\right ]_{3}\right \rbrace . \label{t12}
\end{eqnarray}
Note that the term associated with the forcing frequency
$3\Omega_{orb}$ gives the leading contribution since the normalised overlap
integrals  strongly decrease as $\omega \rightarrow 0.$

 Now we
calculate $e^2{\dot E_{I}}/2$.  To   leading order in $e^2$ it is
sufficient to take into account only the term  in the potential
expansion with forcing frequency  $\omega_{2,2}=2\Omega_{orb},$
 thus from equation (\ref{t10}) we obtain
\begin{equation}
{e^2\over 2}{\dot E_{I}}=-{3\pi^2\over 8}{\left ({GM_2\over
a^3}\right )}^{2}\left [{{\hat Q}^2\over |d\omega/dj|}\right
]_{2}e^2.  \label{t13}
\end{equation}
From (\ref{t3}) and (\ref{t12})  we then find that
\begin{equation}
{\dot E}_e=-{\pi^2\over 20}e^2M_*R_*^2
 {\left({GM_2\over
a^3}\right)}^{2}F(\Omega_{orb}), \label{t14}
\end{equation}
where
\begin{equation}
F(\Omega_{orb})=  {49\over 2}\left [{{\tilde
Q}^2\over |d\omega/dj|}\right ]_{3}-3\left [{{\tilde Q}^2\over
|d\omega/dj|}\right ]_{2}+\frac{15}{2}\left [{{\tilde Q}^2\over
|d\omega/dj|}\right ]_{1} \label{t15}
\end{equation}
Now we make use of (\ref{t5}), applicable in this case,  to eliminate the frequency derivatives
with respect to $j$ in (\ref{t15}) finally obtaining
\begin{equation}
{\dot E}_e=-{\pi \over 20}e^2{\dot E}_* {\cal F}(\Omega_{orb}),
 \label{t16}
\end{equation}
where
\begin{equation}
{\cal{ F}} (\Omega_{orb})= \left \lbrace {49\over 18}\left
[{\tilde Q}^2\right ]_{3}-{3\over 4}\left [{\tilde Q}^2\right
]_{2}+\frac{15}{2}\left [{\tilde Q}^2\right ]_{1}\right \rbrace,
 \label{t17}
\end{equation}
and, accordingly,
\begin{equation}
T^{(2)}_e={20\over \pi}T_*/{\cal{ F}}(\Omega_{orb}).
 \label{t18}
\end{equation}

In Fig. \ref{Fig11} we show the
characteristic times $T^{(2)}_a$ and $T^{(2)}_{e}$, for $\mu=1$. The
short dashed and dotted curves give $T^{(2)}_e$  for the CD2
and IR2 models  respectively. The long dashed and dot dashed
curves give $T^{(2)}_{a}$, respectively for the CD2 and IR2 models.
The solid curves correspond to the modified Goodman $\&$ Dickson (1998)
expression.
 We see that $T^{(2)}_a$ is much larger than
$T^{(2)}_e$ in all cases. This is obviously due to the higher forcing
frequency $3\Omega_{orb}$ effective in the latter case. It is
interesting to note that the numerical values of $T^{(2)}_e$ are
 either smaller than,  or close to the result of Goodman $\&$ Dickson (1998)
 regardless of the fact that our results were calculated for
a  non-rotating primary,   while the Goodman $\&$ Dickson (1998)
result corresponds to a primary in a state of synchronous
rotation.

In Figs. \ref{Fig12} and \ref{Fig13} we show the evolution
timescales $T^{2}_e$ and $T^{2}_a$  for  mass ratios appropriate
for exoplanetary systems. The solid, dashed and dotted curves
correspond to $M_2=M_{J}$, $0.1M_{J}$ and $10M_{J}$. As seen from
these  plots,  dynamical  tides in the regime of moderately large
viscosity are potentially  efficient only when orbital periods are
rather small. Thus, the eccentricity is damped in  a  time less
than $10^9yrs$ only for  $P_{orb} < 1.85$, $2.56$ and $3.48days$,
for $M_2=0.1M_{J}$, $M_{J}$ and $10M_{J}$, respectively. Similarly
the semi-major axis  can decay  in a  time less than $10^9yrs$
only for $P_{orb} < 1.24$, $1.69$ and $2.3days$  respectively.

We obtain $Q'$ for the star by comparing our
circularisation rates $\dot e/e$ to that of Goldreich \& Soter
(1966). In this case we find
\begin{eqnarray}
Q^{'}=\frac{285}{98\pi\sqrt{\Lambda}I[\tilde{Q}^2]_1(1+\mu)^{1/2}}\left
(\frac{2\pi}{P_{orb}\Omega_\ast}\right)^3,
 \end{eqnarray}
for the case of synchronous rotation and
\begin{eqnarray}
Q^{'}=\frac{855}{4\pi {\cal{F}}(\Omega_{orb})(1+\mu)^{1/2}}\left
(\frac{2\pi}{P_{orb}\Omega_\ast}\right)^3
 \end{eqnarray}
 in the non-rotating case.
We plot $Q^{'}$ as a function of orbital period for $\mu=1$ in Fig. \ref{FigQT1}.
This quantity is found to  vary between $10^4$ and  $10^7$ as the orbital period
varies from $1 - 10$ days.

 Finally we make a rough estimate of the effect of inertial waves
that could be excited in the convective envelope in the case of synchronous
rotation. We  note that Ogilvie \& Lin (2007) calculate the  $Q'$ associated
with inertial waves, for a solar model with a rotation period of ten days
as a function of tidal forcing frequency as viewed in the rotating frame,
appropriate to forcing with a $m=2$ quadrupole potential (see their Fig.3).
For the relevant forcing frequencies $\omega =\pm \Omega_{orb},$
$Q'\sim 10^{8},$ implying that the effect of inertial modes
may be comparable to those due to rotationally modified gravity modes
at the longest orbital periods considered $\sim 10$ days.
But note that if the tidal dissipation due to these waves scales as $\Omega^2$
as indicated in Ogilvie (2013), their effect may be somewhat less important
at shorter orbital periods.

\section{Conclusions and Discussion}\label{Discuss}
 In this paper we have
 calculated  the general  tidal response of a uniformly rotating star
and used  it to find the
rate of transfer of energy and angular momentum between the star and the orbit,
The  results  are expressed as a sum of contributions from  normal modes
 each being proportional to the square of the associated  overlap integral.

We  considered the response due to an identifiable  regular dense spectrum of normal modes
such as one provided by the low frequency rotationally modified gravity modes.
We obtained expressions (\ref{eq34}) for the energy and angular momentum transfer rates
appropriate to the moderately viscous case for which the waves associated with the modes are
damped before reaching an appropriate boundary (the centre for a radiative core
and the surface for a radiative envelope)  or  understood in the
  averaged sense as described in section \ref{MLV}. In that case they are independent of the details of the dissipation process.
In addition we obtained  expressions  (\ref{eq35}) applicable to the case of
very weak dissipation where resonant responses may occur.
We also gave corresponding expressions  (\ref{eq37})  that can be used to evaluate the
energy and angular momentum  exchanged as a result of a flyby in a parabolic orbit.
These can be applied to either the tidal capture problem
or the problem of tidal circularisation at large eccentricity (see Ivanov \& Papaloizou 2011).

Our expressions may  be applied to  either Sun-like stars with
radiative cores and convective envelopes or more massive stars
with convective cores and radiative envelopes. However, in this
paper we considered  only  Sun-like stars in detail. We developed
expressions for the overlap integrals in the low frequency limit,
under the traditional approximation.  These were evaluated
analytically following a WKBJ approach that yielded
eigenfrequencies and eigenfunctions for the rotationally modified
gravity modes. Good agreement was found between analytic estimates
of the overlap integrals,  obtained from equation (\ref{eq92})
making use of  equations (\ref{eq86}) and (\ref{eq90}), and values
obtained through direct numerical integration. These showed that
for prograde forcing  the overlap integrals increased as the
rotation rate of the star increased and accordingly tidal
interaction strengthened. The opposite, but more pronounced trend,
was found for retrograde forcing (Figs \ref{Fig8}-\ref{Fig9b}).
Knowledge  of the overlap integrals and mode spectrum enables
calculation of tidal energy and angular momentum exchange rates
with the orbit as a function of tidal forcing frequency in the
moderately viscous regime.

For Sun-like stars with inner radiative regions,
our formalism  allows one to find the tidal response
of low frequency gravity modes without numerical evaluation of
eigenfrequencies and overlap integrals,  and thus find the orbital evolution
of tidally interacting stars of this type.
In order to  implement it, it
suffices to know the behaviour of the Brunt - V$\ddot {\rm a}$is$\ddot
{\rm a}$l$\ddot {\rm a}$ frequency close to the base of convection
zone, the  radius and the value of the  density at that location, as well
as the integrals $I=\int^{r_c}_0{dr\over r}N$, and, for rotating
stars, the quantities determining the angular properties of the
overlap integrals and given in Fig \ref{Fig1}.
In general, it is
also necessary to know the decay rate  of the WKBJ  due to
either dissipation or non-linear processes. Provided this
information is given as a function of stellar age, one can construct
more realistic models of tidal interaction, that allow for
finite values of the eccentricity, a non zero inclination of the stellar rotation axis
to the orbital plane , and stellar evolution  by semi-analytic methods.
It is important to
generalise the analytic evaluation of the overlap integrals to
apply to stars with different  structure, eg. more massive stars  with
convective cores.
This will be undertaken
in future  work.

It is interesting to note  that in our formalism,  the dependence
of energy and angular momentum transfer rates  on semi-major axis is
determined by the dependence of the normalised overlap integrals
$\hat Q$ on the forcing  frequency $\omega $. The standard dependence (Zahn
1977) is recovered only when $\hat Q\propto \omega^{17/6}$, which
may, in general, not be satisfied. Although in the models we
consider in this paper,  the dependence of $\hat Q$ on $\omega $ is
very close to a power law with  index $17/6$ in the limit $\omega
\rightarrow 0, $ it differs from this for  short period orbits. It
may also  differ for  stellar models with a different  radial
structure.
This   emphasises the necessity of
analytical and numerical studies of the behaviour of the  overlap integrals  for
stellar models with different masses and ages.

We used  our formalism to  determine the tidal evolution
timescales for a  binary consisting of a primary Sun-like star in
which the tidal response was excited in the  moderately large
dissipation regime. Assuming synchronous rotation and that
inwardly propagating rotationally modified gravity waves are
dissipated close to the centre, the quantity,
$T_e^{(1)}((1+\mu)/2)^{5/3}/\mu,$  where $T_e^{(1)}$ is the
circularisation time, was found to vary between $10^5 - 10^{10}
yr.$ as the orbital period varied through  the range $1-5 d.$ The
corresponding value of $Q'$ defined by Goldreich \& Soter (1966)
varied through the range $3\times 10^5 - 10^8.$

Similarly when the primary was not rotating,
$T_e^{(2)}((1+\mu)/2)^{5/3}/\mu ,$ $T_e^{(2)}$ being the
circularisation time in this case, was found to vary between
$10^4 - 10^{9} yr.$ as the orbital period varied through  the
range $1-5 d.$ The quantity $Q'$ has a similar range of variation
as is found in the synchronous case.
 In this non synchronous case,  the  quantity
$T_a^{(2)}((1+\mu)/2)^{5/3}/\mu $ where the time scale for
inspiral is $T_a^{(2)}$ varied  between $10^6 - 10^{10} yr$ as the
orbital period varied through the range $1-5 d$. Thus the latter
process is unlikely to have the possibility of being  effective
for masses in the planetary regime ($\mu < 10^{-2}$)  and  orbital
periods exceeding $2.5 d$.

In this paper when tackling the problem with periodic forcing
potential we assume that the mode energy doesn't grow with time
and the amount of energy supplied by tidal interaction is equal to
the amount of dissipated energy. This condition may however be
broken either very close to the resonances or in the case when the
distance between two eigen frequencies is smaller than the inverse
of a characteristic orbital evolution time. In the latter case the
approach based on the Fourier transform may be more adequate as
discussed in eg. Ivanov $\&$ Papaloizou (2004a).

\section*{Acknowledgments}

We are grateful to I. W. Roxburgh and G. I. Ogilvie for help and
fruitful discussions and to A. A. Lutovinov and the referee
for useful comments.

PBI and SVCh were supported in part by Federal programme
"Scientific personnel" contract 8422, by RFBR grant 11-02-00244-a,
by grant no. NSh 2915.2012.2 from the President of Russia and by
programme 22 of the Presidium of Russian Academy of Sciences.

Additionally, PBI was supported in part by the Dynasty Foundation
and thanks DAMTP, University of Cambridge for hospitality.

\begin{appendix}
\section{The coefficients $A_{\lowercase{m},\lowercase{k}}$ for a coplanar slightly eccentric orbit}
Equation (\ref{eq4}) gives the perturbing tidal potential acting on the primary component in the form
\begin{equation}
\Psi=r^{2}\sum_{m,k} \left( A_{m,k}e^{-i\omega_{m,k}t} Y^{m}_{2} +
cc \right) .
\end{equation}
An expansion of the gravitational potential due to the perturbing
secondary in spherical harmonics yields the coefficients
$A_{m,k}.$ We consider the case when the eccentricity of the
binary orbit is small,  such that powers of $e$ higher than the
second may be neglected,  and when the rotational axis is
perpendicular to the orbital plane. A convenient  expression for
the tidal potential that can be used to  determine the $A_{m,k}$
under these conditions can for example be found in Zahn (1977).
The non-zero coefficients are found to be
\begin{eqnarray}
 A_{0,0}\!\!\!&=&\!\!\!\frac{GM_2}{2a^3}\sqrt{\frac{\pi}{5}}\left(1+\frac{3}{2}e^2\right),\quad
 A_{0,1}=\frac{3GM_2}{2a^3}\sqrt{\frac{\pi}{5}}e,\nonumber\\
 A_{0,2}\!\!\!&=&\!\!\!\frac{9GM_2}{4a^3}\sqrt{\frac{\pi}{5}}e^2,\qquad
 A_{2,1}=\frac{GM_2}{4a^3}\sqrt{\frac{6\pi}{5}}e,\nonumber\\
 A_{2,2}\!\!\!&=&\!\!\!-\frac{GM_2}{2a^3}\sqrt{\frac{6\pi}{5}}\left(1-\frac{5}{2}e^2\right),\quad
 A_{2,3}=-\frac{7GM_2}{4a^3}\sqrt{\frac{6\pi}{5}}e,\nonumber\\
 A_{2,4}\!\!\!&=&\!\!\!-\frac{17GM_2}{4a^3}\sqrt{\frac{6\pi}{5}}e^2.
\label{a1}
\end{eqnarray}
Here $M_2$ is the mass
of the secondary and $a$ is the orbital semi-major axis. Note that
for the purpose of evaluating  the complex conjugate in (\ref{eq4}) we have
$A_{m,k}^* = A_{m,k}=A_{-m,-k}$. We also remark that in order to
determine the rate of change of the semi-major axis correct to
zero order in eccentricity and the rate of change of eccentricity
correct to first order in eccentricity, terms $\propto e^2$ may be
dropped when working with these coefficients.

\section{Comparison of asymptotic expressions for torques
and energy dissipation rates
given by normal mode sums with those obtained using
radiation boundary conditions}

Here we consider the response of a spherically symmetric stellar model
with rotation  under the traditional approximation.
The governing equation
for the linear  response to a forcing potential
$\Psi_{m,k}\exp({-i\omega_{m,k} t+im\phi}) $ is
\begin{equation}
-\omega_{m,k}^2\mbox{\boldmath$\xi$}_{m,k}  -2i\omega_{m,k} \left({\bmth {\Omega}}\cdot {\hat {\bf r}}\right){\bf {\hat r}}\times {\bmth{\xi}_{m,k}}
+{\bmth {\cal C}}\mbox{\boldmath$\xi$}_{m,k} = -\nabla \Psi_{m,k},
\label{eom}
\end{equation}
This follows from equation (\ref{eq p3}) provided
${\bmth{\Omega}}$ is replaced by its radial component  ${\bmth
{\Omega}}\cdot {\hat {\bf r}}.$ Making this replacement amounts to
making the traditional approximation described in section
\ref{modgm}. Of course when ${\bmth{\Omega}}$ is set to zero, the
problem reduces to the problem of finding the response of  a
spherically symmetric non rotating star. Thus considering equation
(\ref{eom}) enables us to consider both rotating stars with the
adoption of the traditional approximation and non rotating
spherical stars.

The action of the operators ${\bmth {\cal C}}$ and $\nabla$
respectively  on $\mbox{\boldmath$\xi$}_{m,k}$ and  $\Psi_{m,k}$
is as indicated by equation (\ref{opspec}). It is taken as read
that in addition to time dependence through a factor $\exp(-{\rm
i}\omega_{m,k} t),$ each of $\Psi_{m,k}\exp({{\rm i}m\phi})$ and
$\mbox{\boldmath$\xi$}_{m.k}\exp({{\rm i}m\phi}) $ have separable
angular dependencies related to the Hough functions for the
specified  $m$  and with an appropriate  value of $\Lambda$ (see
section \ref{modgm}). These dependencies  will  not be indicated
below. We  go on to discuss the response due to  a single forcing term.

When the response has been
 obtained for a range of
 assumed angular dependencies  corresponding to related
 values of $\Lambda,$    a total response can be found
 using the orthogonality of the Hough functions (see section \ref{Laplace}).
 Total torques or energy dissipation rates are then obtained by summing
 contributions associated with each value of   $\Lambda$
 involved in the representation of the perturbing potential
 in terms of Hough functions. Note also that the values of $\Lambda$
 that are expected to dominate are associated with the potentially
 resonant mode spectrum, namely that associated with rotationally modified gravity modes,
 or simply $g$ modes in the case of a non rotating star.

For simplicity of notation, the subscripts ${m,k}$ on all
quantities will be suppressed from now on. In addition we adopt
spherical polar  coordinates $(r, \theta, \phi)$ for the purpose
of discussion in this appendix.

\subsection{Torque evaluated with radiation  boundary conditions}
The equation to be solved is  equation (\ref{eom}).
Noting the angular  dependence  indicated above,
A solution  can be found by the method of variation of parameters where we write
\begin{equation}
\mbox{\boldmath$\xi$}= A(r)\mbox{\boldmath$\xi$}_1 +B(r)\mbox{\boldmath$\xi$}_2.
\label{forced}\end{equation}
In the above $\mbox{\boldmath$\xi$}_1$ is a solution of the homogeneous form of
equation (\ref{eom}) (${\Psi }$ set to be zero)
that satisfies the correct boundary conditions  enforcing regularity at $r=0$ and $\mbox{\boldmath$\xi$}_2$ is a solution
of the homogeneous form of  equation (\ref{eom}) that satisfies the appropriate  boundary conditions
at the surface $r\rightarrow R_s.$ Here we take the latter solution to have to correspond to  an outgoing wave.
In this case physical dissipation is supposed to take place for $r\sim R_s.$
But note that we expect a corresponding parallel analysis to hold when $\mbox{\boldmath$\xi$}_1$
is taken to correspond to  an ingoing wave while regularity conditions are applied to  $\mbox{\boldmath$\xi$}_2$
for  $r\rightarrow R_s.$
In that  case physical dissipation is supposed to take place for $r\sim 0.$
In any case the function $A(r)$ is chosen to vanish for $r\rightarrow R_s$
and $B(r)$ is chosen to vanish for $r\rightarrow 0.$
In addition the functions $A(r)$ and $B(r)$ satisfy the constraint
\begin{equation}
\mbox{\boldmath$\xi$}_1 \cdot \nabla A(r) +\mbox{\boldmath$\xi$}_2\cdot \nabla B(r)=0.
\end{equation}
This ensures that derivatives of $A$ and $B$ do not appear when the pressure
perturbation $P'$ is evaluated (see below).

We remark that for the  case when dissipation occurs
near the surface considered here,  when $\omega$ corresponds to a free oscillation eigenvalue,
$\mbox{\boldmath$\xi$}_1$ will correspond to the free normal mode.
As we consider the asymptotic low frequency limit, where the spectrum is dense,
and we assume a degree smoothness in the quantities of interest,
$\mbox{\boldmath$\xi$}_1$ may be assumed to correspond to the
normal mode with eigenfrequency closest to the  forcing frequency.  Thus for comparison
with the modal summation approach described above, we identify
$\mbox{\boldmath$\xi$}_1$ with $\mbox{\boldmath$\xi$}_{j_0}.$
We further comment that the equilibrium  tide  is implicitly included in the decomposition (\ref{forced}).
But  the important contribution
for computing the net torque is the outward going wave component  and the equilibrium tide does not affect this.

We make use of the expressions for ${\bmath {\cal C}}\mbox{\boldmath$\xi$},$  $P'(\mbox{\boldmath$\xi$})$
and $N^2$ given by equations (\ref{eomt}), (\ref{linop}) and (\ref{ep1})
and  we take the  adiabatic exponent $\Gamma_1$ to be constant, for convenience,  as in the main text.
 It follows from these  that for any pair of solutions of the homogeneous form of equation (\ref{eom}) such as $\mbox{\boldmath$\xi$}_1$
and  $\mbox{\boldmath$\xi$}_2$  the latter equation implies that
\begin{equation}
W_{12} = \int_{S_r}\left(P'(\mbox{\boldmath$\xi$}_2)\mbox{\boldmath$\xi$}^*_1 - P'(\mbox{\boldmath$\xi$}_1^*)\mbox{\boldmath$\xi$}_2\right)\cdot d{\bf S},
\label{Wr}\end{equation}
where $S_r$ is an interior spherical surface of radius $r,$ is constant.
In addition for the solution of the inhomogeneous equation of interest, we have
\begin{equation}
\int_{S \rightarrow S_{R_s}}\left(P'(\mbox{\boldmath$\xi$})\mbox{\boldmath$\xi$}^*_1 - P'(\mbox{\boldmath$\xi$}_1^*)\mbox{\boldmath$\xi$}\right)\cdot d{\bf S}
-{\int}_V\rho\mbox{\boldmath$\xi$}_1^*\cdot{\nabla \Psi}d\tau =0.\label{forcedint}
\end{equation}
The expression (\ref{Wr}) is associated with the conservation of wave action
for freely propagating disturbances.

 From equation (\ref{eom}) we find
that the total torque communicated to the forced body
is  given by
\begin{equation}
T_0(\mbox{\boldmath$\xi$}) =-{m}{\cal{I{\it m}}}\left[{{\int}}_{S \rightarrow S_{R_s}} P'(\mbox{\boldmath$\xi$})\mbox{\boldmath$\xi$}^{*}\cdot d{\bf S}\right].\label{action}
\end{equation}
Using equation (\ref{action}) with the fact that  the forced solution (\ref{forced}) becomes $B\mbox{\boldmath$\xi$}_2$
 as the surface of the star is approached,  the angular momentum transfer rate
to oscillatory  disturbances as a result of the tidal interaction may be evaluated as
\begin{equation}
T_0(\mbox{\boldmath$\xi$}) = |B(R_s)|^2  T_0(\mbox{\boldmath$\xi$}_2)\label{faction0}
\end{equation}
Furthermore, using
(\ref{forcedint}) we can express $B(R_s)$ in the form
\begin{equation}
B(R_s)=\frac{{\int}_V\rho\mbox{\boldmath$\xi$}_1^*\cdot{\nabla \Psi}d\tau}{W_{12}} \label{coeff}
\end{equation}
Thus the tidally induced torque may be written as
\begin{equation}
T_0(\mbox{\boldmath$\xi$}) =\frac{ |{\int}_V\rho\mbox{\boldmath$\xi$}_1^*\cdot{\nabla \Psi}d\tau|^2  T_0(\mbox{\boldmath$\xi$}_2)}{|W_{12}|^2 }\label{faction}
\end{equation}
To relate this to expressions given in the main text, we  now use the WKBJ approximation to find alternative
expressions  for  some of the quantities  in (\ref{faction}).
\subsection*{WKBJ solutions}
As follows from equation (\ref{eq58}) in the main text after neglecting
 the $1$ as compared to $N^2/\omega^2$  in the last bracket on the left hand side,
in the small $\omega^2$ limit, free oscillations
are governed by the  following equation for $v=r^2P^{1/\Gamma_1}\xi_r,$
 $\xi_{r}$ being the radial component of the displacement,
\begin{equation}
\frac{\rho}{P^{2/\Gamma_1}}\frac{d}{dr}\left(\frac{\rho}{P^{2/\Gamma_1}}\frac{dv}{dr}\right)=-\chi^2 v
\end{equation}
where
\begin{equation}
\chi^2 = \frac{N^2\Lambda\rho^2}{\omega^2r^2 P^{4/\Gamma_1}}.
\end{equation}
We remark that the pressure disturbance is given by
\begin{equation}
P' = \frac{\omega^2\rho }{ \Lambda P^{1/\Gamma_1} }\frac{dv}{dr}
\end{equation}
As follows from equation (\ref{eq59}), away from turning points,  the WKBJ solution for
$v_1=r^2P^{1/\Gamma_1}\xi_{r,1},$ where
$\xi_{r,1}$  is the radial component of  ${\bmath{\xi}}_1,$    can be written in the form
\begin{equation}
v_1 = \frac{\cos\left( \int _{r_i}^r \Phi  dr+ \eta_1 \right )}{\chi^{1/2}},
\label{WKBJ1}\end{equation}
where $\Phi=N\sqrt{\Lambda} /(r \omega),$
 $r_i$ is the inner boundary radius of the wave propagation zone
 and $\eta_1$ is a constant phase determined by the  boundary condition
 at $r=0.$

The WKBJ solution for
$v_2=r^2P^{1/\Gamma_1}\xi_{r,2},$
 where $\xi_{r,2}$  is the radial component of the outgoing wave solution  ${\bmath{\xi}}_2,$ takes the form
\begin{equation}
v_2 = \frac{\exp i\left(- \int _{r_i}^r \Phi  dr+ \eta_2  \right )}{\chi^{1/2}},
\end{equation}
where $\eta_2$ is another constant phase factor.
We may use the above WKBJ solutions to evaluate
the constant quantities  $T_0(\mbox{\boldmath$\xi$}_2)$
and $W_{12}$ in the WKBJ approximation in the  form
\begin{equation}
T_0(\mbox{\boldmath$\xi$}_2) = \frac{ m \omega|\omega|}{\Lambda}.
\end{equation}
and
\begin{equation}
W_{12} = \frac{ \omega|\omega|}{\Lambda}\exp i(\eta_2-\eta_1-\pi/2).
\end{equation}
We remark that here we have  assumed  without loss of generality
that the integration of the square of the angular
 factors over the spherical polar angles is normalised to
unity. Using the above expressions in (\ref{faction})
we obtain
\begin{equation}
T_0(\mbox{\boldmath$\xi$}) =\frac{ m\Lambda |{\int}_V\rho\mbox{\boldmath$\xi$}_1^*\cdot{\nabla \Psi}d\tau|^2 }{\omega|\omega|}.\label{faction1}
\end{equation}
\subsection*{The WKBJ spectrum}
To proceed further  we recall that
 from the free and undamped  form of (\ref{eom})
\begin{equation}
\omega_j^2{\bmath \xi}_j~-~{\bmath{{\cal C}\xi}_j}~-~\omega_j~{\bmth{\cal B}}{\bmath \xi}_j=0
\label{freqrel}
\end{equation}
We use the fact that in the WKBJ  limit
 the spectrum is given in the large $j$ limit
by the phase integral relation
\begin{equation}
\omega_j =\pm \frac{\sqrt{\Lambda}{\int}_{r_i}^{R_s} Nr^{-1}dr}{j\pi}\equiv \omega_0/j ,
\label{freqre2}
\end{equation}
where the $\pm$ alternative corresponds to the positive and negative frequency modes respectively,
and  see equation (\ref{eq76}) of the main text with $ n\equiv  j > 0.$
Thus,  noting that  $\Lambda$   is frequency dependent, we have
\begin{equation}
d\omega_j/dj\left(1-  \frac{\omega_j}{2\Lambda}\frac{\partial \Lambda}{\partial \omega_j}\right)=-\omega^2_j/\omega_0.
\label{intd0}\end{equation}
In order to re-express the above expression in terms of the mode norm, we begin by noting that
if we replace $N$ by $N(1+\epsilon)$ in (\ref{freqre2}) and regard $\omega_j$ to be a function of $\epsilon,$
we have
\begin{equation}
\left. d\omega_j/d\epsilon \right |_{\epsilon=0}\left(1-  \frac{\omega_j}{2\Lambda}\frac{\partial \Lambda}{\partial \omega_j}\right)
=\omega_j |_{\epsilon=0}.\label{intd}
\end{equation}
However, the same result  as (\ref{intd}) can be obtained from   (\ref{freqrel}) by  multiplying by  $\rho {\bmath \xi}_j^*$
and integrating over the fluid volume and then treating the introduction of $\epsilon$
from the point of view of perturbation theory.
Noting that the operators are self adjoint,  we find from this that
\begin{eqnarray}
&&\left. d\omega_j/d\epsilon \right |_{\epsilon=0} {\int}_V \rho \left(
\omega_j^2|{\bmath \xi}_j|^2+~{\bmath \xi}_j^*{\bmath{{\cal C}\xi}_j}\right)d\tau\nonumber \\
&&\hspace{2cm}=2\omega_j{\int}_V\rho N^2 |\mbox{\boldmath$\xi$}_j\cdot {\hat{\bf r}}|^2d\tau \label{intd1}
\end{eqnarray}
Using the WKBJ solution (\ref{WKBJ1}) to evaluate the right hand side of
(\ref{intd1}),
identifying $\omega$ and   $\omega_j$ with $\omega_{j_0}$ and
$\mbox{\boldmath$\xi$}_1$ with $\mbox{\boldmath$\xi$}_{j_0}$ and making use of
(\ref{intd0}), (\ref{intd})  and (\ref{intd1})
we obtain
\begin{equation}
d\omega_j/dj|_{j_{0}} = -\frac{\omega_{j_0}^3}{2N_{j_0}\Lambda}
\label{freqre3}
\end{equation}
where we recall the definitions of the inner product (\ref{eq p2}) and norm (\ref{norm}).

Using (\ref{freqre3}) together with (\ref{faction1}) enables the latter  to be reduced to the form
\begin{equation}
T_0(\mbox{\boldmath$\xi$})= m\omega_{j_0}
\frac {|{\int}_V\rho\mbox{\boldmath$\xi$}^*_{j_0}\cdot{\nabla \Psi}d\tau|^2}
{N_{j_0} |d\omega_j(j)/dj|_{j=j_0}},
\label{torqueapproxrad}
\end{equation}
Recalling again the definition of the inner product
this can be written in the alternative form
\begin{equation}
T_0(\mbox{\boldmath$\xi$}) =\frac{2\pi^2 m\omega_{j_0}
{({\mbox{\boldmath$\xi$}}
^*_{j_0}|
{\nabla \Psi})^2}}
{N_{j0} |d\omega_j(j)/dj|_{j=j_0}},
\label{torqueapproxrad1}
\end{equation}
The rate of energy dissipation is then
\begin{equation}
{\dot E_c}=-\omega_{j_0}T_0(\mbox{\boldmath$\xi$})/m
=-\frac{2\pi^2 \omega_{j_0}^2
{({\mbox{\boldmath$\xi$}}
^*_{j_0}|
{\nabla \Psi})^2}}
{N_{j0} |d\omega_j(j)/dj|_{j=j_0}},
\end{equation}

This should be multiplied by a factor of two
to account for the response to the complex conjugate of the applied  forcing term
as specified by for example by
equation (\ref{eq4}). This is seen from the form of (\ref{action}) to give an identical contribution.
The result  then becomes identical to  a term corresponding to single values of $m$ and $k$
in the summation given by
(\ref{eq29}).

\end{appendix}

\label{lastpage}

\end{document}